\let\oldvec\vec
\documentclass[graybox]{svmult}
\let\vec\oldvec

\usepackage[utf8]{inputenc}
\usepackage{mathptmx}       
\usepackage{helvet}         
\usepackage{courier}        
\usepackage{type1cm}        
\usepackage{makeidx}         
\usepackage{graphicx}        
\usepackage{multicol}        
\usepackage[bottom]{footmisc}
\usepackage[centertags]{amsmath}
\usepackage{amsfonts}
\usepackage{amssymb}
\usepackage{amsmath}
\usepackage{graphicx,color}
\usepackage[normalem]{ulem}
\usepackage{feynmf}
\usepackage{youngtab}
\usepackage{multirow}
\usepackage{cancel}
\usepackage{cite}

\newcommand{\bea}{\begin{eqnarray}}
\newcommand{\eea}{\end{eqnarray}}

\newcommand{\de}{\partial}

\newcommand{\p}[1]{(\ref{#1})}
\newcommand{\be}{\begin{equation}}
\newcommand{\beq}{\begin{equation}}
\newcommand{\ee}{\end{equation}}
\newcommand{\eeq}{\end{equation}}
\newcommand{\bex}{\begin{exercise}}
\newcommand{\eex}{\end{exercise}}
\newcommand{\bal}{\begin{align}}
\newcommand{\eal}{\end{align}}

\newcommand{\lb}{\left(}
\newcommand{\rb}{\right)}

\newcommand{\ad}{\dot{\alpha}}
\newcommand{\bd}{\dot{\beta}}
\newcommand{\tln}{\,\text{ln}\,}

\def\a{\alpha}
\def\b{\beta}
\def\g{\gamma}
\def\G{\Gamma}
\def\d{\delta}

\def\D{\Delta}
\def\e{\epsilon}

\def\h{\eta}

\def\k{\kappa}
\def\l{\lambda}
\def\L{\Lambda}
\def\m{\mu}
\def\n{\nu}

\def\p{\pi}
\def\P{\Pi}

\def\r{\rho}

\def\s{\sigma}
\def\S{\Sigma}
\def\t{\tau}

\def\f{\phi}
\def\F{\Phi}
\def\vf{\varphi}
\def\ps{\psi}

\def\O{\Omega}

\newcommand{\mm}{{\ensuremath{\underline{m}}}}
\newcommand{\nnn}{{\ensuremath{\underline{n}}}}
\newcommand{\rr}{{\ensuremath{\underline{r}}}}
\newcommand{\sss}{{\ensuremath{\underline{s}}}}

\newcommand{\gad}{{\dot{\alpha}}}
\newcommand{\gbd}{{\dot{\beta}}}

\newcommand{\gnd}{{\dot{\nu}}}

\newcommand{\ga}{\alpha}
\newcommand{\gb}{\beta}

\newcommand{\adD}{{\mathsf{D}}}
\newcommand{\tadD}{\widetilde{\mathsf{D}}}

\newcommand{\pl}{{\partial}}
\newcommand{\bpl}{{{\bar{\pl}}}}

\newcommand{\bry}{{{\bar{y}}}}

\newcommand{\brxi}{{{\bar{\xi}}}}

\newcommand{\bvarpi}{{{\bar{\varpi}}}}

\newcommand{\besubeqs}{\begin{subequations}}
\newcommand{\esubeqs}{\end{subequations}}

\newcommand{\brN}{\bar{N}}

\newcommand{\Fron}{{\Phi}}

\newcommand{\twoR}[2]{
\begin{picture}(75,70)(-20,2)
\multiframe(0,0)(13.5,0){1}(15,10){}\put(20,0){$\lambda_r$}
\multiframe(0,10.5)(13.5,0){1}(20,10){}\put(25,10.5){$\lambda_{r-1}$}
\end{picture}}

\makeindex             

\newcommand{\vac}{|E_0,\mathbf{s}\rangle}
\newcommand{\rep}{D(E_0,\mathbf{s})}

\begin{document}

\title*{From Higher Spins to Strings: A Primer}
\author{Rakibur Rahman and Massimo Taronna}
\institute{Rakibur Rahman \at Max-Planck-Institut f\"ur Gravitationsphsyik (Albert-Einstein-Institut),\\
Am M\"uhlenberg 1, D-14476 Potsdam-Golm, Germany\\ \email{rakibur.rahman@aei.mpg.de}
\and Massimo Taronna \at Universit\'e Libre de Bruxelles,\\
ULB-Campus Plaine C.P. 231, B-1050 Bruxelles, Belgium\\ \email{massimo.taronna@ulb.ac.de}}
%
%

\maketitle
\abstract{A contribution to the collection of reviews \textit{Introduction to Higher Spin Theory} edited by S.~Fredenhagen, this introductory article is a pedagogical account
of higher-spin fields and their connections with String Theory. We start with the motivations for and a brief historical overview of the subject. We discuss the Wigner
classifications of unitary irreducible Poincar\'e-modules, write down covariant field equations for totally symmetric massive and massless representations in flat space,
and consider their Lagrangian formulation. After an elementary exposition of the AdS unitary representations, we review the key no-go and yes-go results concerning higher-spin
interactions, e.g., the Velo-Zwanziger acausality and its string-theoretic resolution among others. The unfolded formalism, which underlies Vasiliev's equations, is then
introduced to reformulate the flat-space Bargmann-Wigner equations and the AdS massive-scalar Klein-Gordon equation, and to state the ``central on-mass-shell theorem''.
These techniques are used for deriving the unfolded form of the boundary-to-bulk propagator in $AdS_4$, which in turn discloses the asymptotic symmetries of (supersymmetric)
higher-spin theories. The implications for string-higher-spin dualities revealed by this analysis are then elaborated.}

\numberwithin{equation}{section}
\section{Introduction}\label{sec:Intro}

The study of higher-spin (HS) fields and their interactions has generated a lot of interest in recent years. The subject however has a long history, dating back
to the 1930's, marked by rather slow progress for over half a century owing to its difficult and subtle nature. The purpose of this review article is to make the
reader familiar with the basics of the subject and its intriguing connections with String Theory. Our attempt is a humble addition to a number of excellent reviews
and lecture notes already existing in the literature \cite{Vasiliev:1999ba,Bianchi:2005yh,Francia:2005bv,Bouatta:2004kk,Bekaert:2005vh,Sagnotti:2005ns,Sorokin:2004ie,
Bekaert:2006py,Francia:2006hp,Fotopoulos:2008ka,Campoleoni:2009je,Sagnotti:2010jt,Bekaert:2010hw,Taronna:2012gb,Rahman:2013sta,Gomez:2013sfb,Sagnotti:2013bha,Giombi:2012ms,Gaberdiel:2012uj,
Didenko:2014dwa,Vasiliev:2014vwa,Sezgin:2012ag}. In this introductory section, we will motivate the study of HS theories and give a brief historical overview of the subject.

\subsection{Motivations}\label{sec:Motivations}

The physics of elementary particles is described by Quantum Field Theory, which associates them in Minkowski space with unitary irreducible representations of the
Poincar\'{e} group. Unitarity is required for compatibility with the principles of Quantum Mechanics, while irreducibility reflects the particles' elementary nature.
In this regard, as long as free propagation is concerned, the spin of a particle can in principle take arbitrary integer or half-integer values. When interactions
are turned on, though, the consistency of the field theory is not at all automatic.

Indeed, theories of interacting massless particles with spin $s$ greater than 2 have serious consistency issues in flat space. If these particles have anything to do with Nature, they must couple to gravity, which on the other hand interacts universally with all matter in the soft limit. However, a set of powerful no-go
theorems together with classification of interactions \cite{Weinberg:1964ew,Grisaru:1977kk,Grisaru:1976vm,Aragone:1979hx,Weinberg:1980kq,Porrati:2008rm,Boulanger:2008tg,Taronna:2011kt,Porrati:2012rd,Joung:2013nma} leads to the unequivocal conclusion that
such particles cannot have gravitational interactions in Minkowski space. The phrase ``higher spin'' conventionally refers to particles with $s>2$, for which
the no-go theorems apply. In the context of electromagnetic interactions in flat space, such no-go theorems rule out a non-zero charge for massless particles
with $s>1$ \cite{Weinberg:1980kq,Porrati:2008rm}.

To construct consistent interacting theories for massless HS gauge fields, it is imperative to find some yes-go results. One of the motivations comes from considering the
high-energy behavior of string scattering amplitudes \cite{Gross:1987kza,Gross:1987ar,Amati:1987wq,Amati:1987uf}. Note that massive HS excitations show up in
the string spectra, and they play a crucial role in some of the spectacular features of String Theory: planar duality, modular invariance and open-closed duality,
for example. In the tensionless limit of String Theory, i.e. when $\a'\rightarrow\infty$, the mass of all these particles tend to zero\footnote{In curved backgrounds, this is expected to be true for a huge part of the spectrum, not all of it.}.
Moreover, there appear an infinite number of linear relations among the tree-level scattering amplitudes of different string states that involve particles of
arbitrary spins \cite{Gross:1988ue,Moeller:2005ez} (see also \cite{Taronna:2010qq,Sagnotti:2010at,Taronna:2011kt} for more recent progress and puzzles). This points towards the restoration of a larger symmetry at high energies. More specifically, the linear
relations are generated by symmetry transformations of the ultra high-energy theory that have conserved charges with arbitrary spin. Such a symmetry may exist
despite the Coleman-Mandula theorem \cite{Coleman:1967ad} that rules out HS symmetries for nontrivial field theories in flat space. This is because the no-go argument relies on the assumption of a Minkowski background, but the introduction of a non-vanishing curvature, for instance, does not alter the short-distance properties of the amplitudes and opens up a possibility to evade the no-go theorem.

In a sense, HS gauge symmetry is the maximal relativistic symmetry, and as such it cannot result from the spontaneous breakdown of a larger one \cite{Fradkin:1986ka}.
Given this, the emergence of HS symmetries at ultrahigh energies above the Planck scale may not come as a surprise. HS gauge theories may therefore
shed light on Quantum Gravity. Because lower-spin symmetries are subgroups of HS ones, lower-spin theories can be realized as low-energy limits of HS theories with the HS gauge group spontaneously broken. It is quite natural to conjecture that String Theory describes the broken phase of a HS gauge theory,
in which the fields with $s>2$ acquire mass. Conversely, the tensionless limit of String Theory, if well defined at all, may be a theory of HS gauge fields.

Remarkably, the inclusion of a cosmological constant enables one to evade the no-go theorems and write down consistent (gravitationally) interacting theories of
HS gauge fields \cite{Fradkin:1987ks,Fradkin:1986qy}. The limit of zero cosmological constant is singular, in accordance with the
no-go results. In Anti-de Sitter (AdS) space, indeed, one can construct a set of non-linear
equations for an infinite number of massless interacting symmetric tensor fields of arbitrary spins:
the Vasiliev's equations \cite{Vasiliev:1990en,Vasiliev:1990vu,Vasiliev:1990bu,Vasiliev:1992av,Vasiliev:2003ev}.

That AdS space is the natural setup for HS gauge theories is very interesting from the point of view of the AdS/CFT correspondence
\cite{Maldacena:1997re,Gubser:1998bc,Witten:1998qj}. In this context, the interconnections between HS gauge theories and tensionless strings
become more concrete \cite{Sundborg:2000wp,Mikhailov:2002bp,Sezgin:2002rt,Bianchi:2003wx,Beisert:2004di,Bianchi:2004ww}. In fact, a new class
of HS holographic dualities emerge. The conjectured dualities between $\text{AdS}_4$ Vasiliev theories and 3d vectorial conformal
models \cite{Klebanov:2002ja,Sezgin:2003pt}, for example, find key evidence from the comparison of 3-point functions \cite{Giombi:2009wh}
and the analysis of HS symmetries \cite{Konstein:1989ij,Giombi:2011rz,Giombi:2011ya,Maldacena:2011jn,Maldacena:2012sf,Boulanger:2013zza,Stanev:2013qra,Alba:2013yda,Friedan:2015xea} (see also the recent study of HS partition functions \cite{Giombi:2013yva,Giombi:2013fka,Tseytlin:2013jya,Giombi:2014iua,Giombi:2014yra,Beccaria:2014jxa,Beccaria:2014xda,Beccaria:2015vaa,Beccaria:2016tqy} and the bulk reconstruction analysis of \cite{Bekaert:2014cea,Bekaert:2015tva}). Similarly, there appear $\text{AdS}_3/\text{CFT}_2$
vector-like dualities \cite{Henneaux:2010xg,Campoleoni:2010zq,Henneaux:2012ny,Gaberdiel:2010pz,Gaberdiel:2013vva}. The HS holography has
some remarkable features that may help us understand the very origin of AdS/CFT \cite{Vasiliev:2012vf}. It does not rely on supersymmetry, and can
be checked on both the sides because of their weak-weak nature.

Most interestingly, the vector-like $\text{AdS}_4/\text{CFT}_3$ dualities, with supersymmetry and Chern-Simons interactions \cite{Aharony:2011jz,Giombi:2011kc},
can actually be embedded within the ABJ duality of String Theory \cite{Aharony:2008ug}, that thereby form a triality \cite{Chang:2012kt}. Among others, it suggests
a concrete way of embedding Vasiliev's theory into String Theory. Also realized in $\text{AdS}_3/\text{CFT}_2$ \cite{Gaberdiel:2012ku}, such a triality takes us a
step closer to finding the actual connection of HS gauge theories and the tensionless limit of String Theory \cite{Gaberdiel:2014cha,Gaberdiel:2015mra,Gaberdiel:2015uca}.

Last but not the least, a more down-to-earth reason for studying HS fields is the existence of massive HS particles in Nature in the form of hadronic resonances,
e.g., $\pi_2$(1670), $\rho_3$(1690), $a_4$(2040) etc. The interactions of these composite particles are described by complicated form factors. At any rate,
their dynamics should be captured by some consistent local actions in a regime where the exchanged momenta are small compared to their masses.
However, massive HS Lagrangians minimally coupled to electromagnetism (EM) suffer from the Velo-Zwanziger problem \cite{Velo:1969bt,Velo:1970ur,Velo:1972rt}:
in nontrivial EM backgrounds one cannot guarantee that no unphysical degrees of freedom start propagating nor that the physical ones propagate only within the light cone.
This pathology manifests itself in general for all charged massive HS particles with $s>1$. It is quite challenging, field theoretically, to construct consistent
interactions of massive HS particles since this problem persists for a wide class of non-minimal generalizations of the theory and also for other interactions \cite{Shamaly:1972zu,Hortacsu:1974bm,Prabhakaran:1975yp,Seetharaman:1975dq,Kobayashi:1978xd,Kobayashi:1978mv,Deser:2000dz,Deser:2001dt}.
The good news is that a judicious set of non-minimal couplings and/or additional dynamical degrees of freedom can actually come to the rescue
\cite{Deser:2001dt,Argyres:1989cu,Porrati:2010hm,Porrati:2009bs,Rahman:2011ik}. The theory of charged open strings proves to be strikingly instructive in this regard:
in a uniform EM background it spells out complicated and highly non-linear contact terms that make the string fields propagate
causally \cite{Argyres:1989cu,Porrati:2010hm}. Of course, the minimal EM coupling disappears in the tensionless limit in accordance with the no-go theorems.
Thus String Theory meets higher spins once again in the context of the Velo-Zwanziger problem for massive fields.

\subsection{Historical Overview}\label{sec:History}

The study of HS fields dates back to a 1932 paper by Majorana \cite{Majorana:1932rj}, who was led to introduce unitary representations of the Lorentz group in an attempt
to get rid of the negative-energy solutions of a Dirac-like equation (see Refs.~\cite{Casalbuoni:2006fa,Bekaert:2009pt,Esposito:2011wf} for a historical account). Then,
in 1936 Dirac considered relativistic wave equations in order to generalize his celebrated spin-$\tfrac{1}{2}$ equation \cite{Dirac:1936tg}.

However, it was not until 1939 that a systematic study of HS particles began, when Fierz and Pauli took a field theoretic approach to the problem and focused on the
physical requirements of Lorentz invariance and positivity of energy \cite{Fierz:1939zz,Fierz:1939ix}. For particles of arbitrary integer or half-integer spin,
they wrote down a set of physical state conditions comprising dynamical equations and constraints. It was noted, rather unexpectedly, that turning on interactions for massive
HS fields by naively covariantizing the derivatives in this set of conditions results in inconsistencies because of the non-commuting nature of the covariant derivatives.
To avoid such difficulties, they suggested that one take recourse to the Lagrangian formulation, which would automatically render the resulting equations of motion (EoM)
and constraints mutually compatible. This is possible only at the cost of introducing lower-spin auxiliary fields, which must vanish on shell when interactions are
switched off. Fierz and Pauli themselves were also able to write down a linearized Lagrangian for a massive graviton (spin-2 field) \cite{Fierz:1939ix}.

At the same time, Wigner published his ground-breaking work on unitary representations of the Poincar\'{e} group~\cite{Wigner:1939cj}. This and later studies by
Bargmann and Wigner on relativistic wave equations \cite{Bargmann:1948ck} made it clear that the requirement of positivity of energy could be replaced by another one,
namely, a one-particle state carries a unitary irreducible representation of the Poincar\'{e} group. Accordingly, a spin-$s$ Bose field can be described by a
symmetric traceless rank-$s$ Lorentz tensor as in \cite{Fierz:1939zz,Fierz:1939ix}, whereas a Fermi field of spin $s=n+\tfrac{1}{2}$ by a symmetric $\g$-traceless rank-$n$
tensor-spinor as first shown in \cite{Rarita:1941mf}. The latter is a 1941 paper by Rarita and Schwinger, who, using the vector-spinor formalism, were able to construct
therein a Lagrangian for a spin-$\tfrac{3}{2}$ field.

After a long hiatus of at least two decades, the study of HS fields resumed in the 1960's only to find further unexpected difficulties. These negative results, although
in hindsight were actually revealing key features of HS interactions, kept most of the community away from the search of consistent HS theories. It started with Johnson
and Sudarshan (1961), who noticed that the Rarita-Schwinger Lagrangian minimally coupled to an external EM background is quantum mechanically inconsistent: the equal-time
commutation relations are not compatible with Lorentz invariance \cite{Johnson:1960vt}. Later, Velo and Zwanziger went on to show that the classical theory itself suffers from
pathologies \cite{Velo:1969bt}. As already mentioned, this Velo-Zwanziger problem is very generic for interactions of massive HS fields
\cite{Velo:1970ur,Velo:1972rt,Shamaly:1972zu,Hortacsu:1974bm,Prabhakaran:1975yp,Seetharaman:1975dq,Kobayashi:1978xd,Kobayashi:1978mv,Deser:2000dz,Deser:2001dt}.
It shows, contrary to what Fierz and Pauli might have anticipated, that neither the Lagrangian formulation does make the consistency of HS interactions automatic.

Furthermore, there appeared some classic no-go results based on $S$-matrix arguments. On the one hand, Weinberg (1964) showed that massless bosons with $s>2$ cannot mediate long-range
interactions \cite{Weinberg:1964ew}. Similar arguments were made for HS fermions as well \cite{Grisaru:1977kk,Grisaru:1976vm}. On the other hand, Coleman and
Mandula (1967) proved that a nontrivial field theory in flat space cannot have a HS conserved charges \cite{Coleman:1967ad}. In addition to these, the 1980 Weinberg-Witten
theorem \cite{Weinberg:1980kq} and its generalizations \cite{Porrati:2008rm,Porrati:2012rd} completely forbid gravitational interactions of HS gauge fields in flat
space. Arguments based on local Lagrangians, due to Aragone and Deser (1979), also rule out minimal gravitational coupling of massless particles with $s\geq\tfrac{5}{2}$.

Despite all the no-go results, some major developments took place during the 1970's that resulted in explicit Lagrangians for massive and massless HS fields of arbitrary spin.
Based on earlier works of Fronsdal \cite{Fronsdal:1958xx} and Chang \cite{Chang:1967zzc} that had spelled out a systematic procedure for introducing the required auxiliary
fields, Singh and Hagen in 1974 completed the Fierz-Pauli program by writing down Lagrangians for free massive fields of arbitrary spin
\cite{Singh:1974qz,Singh:1974rc}. The feat of constructing Lagrangians for HS gauge fields was achieved in 1978 by Fronsdal and Fang \cite{Fronsdal:1978rb,Fang:1978wz}
upon considering the massless limit of the Singh-Hagen ones.

The yes-go results for HS interactions began to show up in the 1980's. The first glimpse of consistent (non-gravitational) interactions of HS gauge fields in flat space
appeared in a couple of 1983 papers by Bengtsson, Bengtsson and Brink \cite{Bengtsson:1983pd,Bengtsson:1983pg}, who used light-cone formulation to construct cubic
self-couplings of HS bosons. In 1987, the full list of cubic interaction vertices for these fields became available \cite{Bengtsson:1986kh}. Later, Metsaev generalized
this to arbitrary dimensions and to fermions, thereby completing the light-cone classification of  flat-space cubic HS vertices \cite{Metsaev:2005ar}. These results have later been promoted to covariant form in \cite{Buchbinder:2006eq,Boulanger:2008tg,Manvelyan:2010jr,Sagnotti:2010at,Fotopoulos:2010ay} and extended to constant curvature backgrounds and to partially-massless fields in \cite{Vasiliev:2001wa,Vasilev:2011xf,Joung:2011ww,Joung:2012rv,Joung:2012hz,Boulanger:2012dx,Joung:2013doa}. For massive fields,
on the other hand, Argyres and Nappi showed in 1989 that the Velo-Zwanziger problem has a string-theoretic remedy for $s=2$ \cite{Argyres:1989cu}. The arbitrary-spin
generalization of this result appeared later in Ref.~\cite{Porrati:2010hm}.

A breakthrough came in 1987 when Fradkin and Vasiliev noticed that gravitational interactions of HS gauge fields are allowed in (A)dS space
\cite{Fradkin:1987ks,Fradkin:1986qy}. They constructed in AdS$_4$ an explicit cubic Lagrangian for all massless HS bosons with $s\geq1$. In 1990, Vasiliev found a
non-linear set of equations of motion to all orders in the coupling constant for HS bosonic gauge fields propagating in AdS$_4$ \cite{Vasiliev:1990en}. Subsequently, Vasiliev's equations were generalized to arbitrary dimension for totally symmetric massless HS fields \cite{Vasiliev:2003ev} and to the supersymmetric
case \cite{Sezgin:1998gg,Engquist:2002vr,Engquist:2002gy}.

With the advent of the AdS/CFT correspondence, HS gauge theories acquired even more interest. The work of Sezgin and Sundell and of Klebanov and Polyakov \cite{Sezgin:2002rt,Klebanov:2002ja,Sezgin:2003pt} in the early 2000's conjectured on a new class of HS holographic dualities between $\text{AdS}_4$ Vasiliev theories and 3d
vectorial conformal models. The first check of these dualities is due to Giombi and Yin in 2009 \cite{Giombi:2009wh}, which generated great interest
in the study of HS gauge theories. In 2010, Gaberdiel and Gopakumar found lower-dimensional versions of such dualities: $\text{AdS}_3/\text{CFT}_2$ vector-like
dualities \cite{Gaberdiel:2010pz}. Remarkably, in 2012 it was realized that when supersymmetry and Chern-Simons interactions are included, HS holographic dualities can be
embedded within stringy dualities to actually make a triality of different theories \cite{Chang:2012kt,Gaberdiel:2012ku}. This makes the connection
between HS gauge theories and the tensionless limit of String Theory ever-more intriguing.

\subsection{Outline}\label{sec:outline}

The outline of this review is as follows. We spell out our conventions and some prerequisites in Section \ref{sec:conventions}. In Section \ref{sec:Flat}, we consider free HS theories in flat space.
Discussions about the Poincar\'e group and the Wigner classifications, the Bargmann-Wigner program, and the Lagrangian formulation appear respectively in Sections
\ref{sec:Wigner}--\ref{sec:Lagrangian}. Section \ref{sec:AdS}, on the other hand, deals with AdS space and free HS fields therein. The issues of HS interactions in flat
space and the yes-go results are considered in Section \ref{sec:Beyond}. We discuss in Section \ref{sec:no-go} the various no-go theorems for interacting HS gauge
fields, and in Section \ref{sec:yes-go} the different ways to bypass them in order to be able to construct HS interactions. The Velo-Zwanziger acausality problem of
massive HS fields and its remedy offered by String Theory are discussed respectively in Sections \ref{sec:vz} and \ref{sec:open}.
Section \ref{sec:HS0} gives an introduction to the unfolded formulation and various other tools which play a key role in HS theories. In particular, Section \ref{sec:Unfolding} is devoted to the so-called
Lopatin-Vasiliev formulation of HS theories in flat space, and Section \ref{sec:AdS_Unf} to the AdS extension of the corresponding systems of equations. Finally, in Section \ref{sec:fromHS} we first derive expressions for boundary-to-bulk propagators as solutions to the unfolded equations of 4d HS theories on AdS. Then we use these tools to have a glimpse of the asymptotic symmetries of (supersymmetric) HS theories, and highlight their role in making interesting conjectures about dualities between Vasiliev's equations and String Theory.

\subsection{Conventions \& Some Prerequisites}\label{sec:conventions}

In this review, we will use the mostly-positive convention for the metric. The expression $(i_1\cdots i_n)$  denotes a totally symmetric one in all the indices
$i_1,\cdots,i_n$ with \underline{no} normalization factor, e.g., $(i_1i_2)=i_1i_2+i_2i_1$ etc. The totally antisymmetric expression $[i_1\cdots i_n]$ comes with the same
normalization. The number of terms included in the symmetrization is assumed to be the minimum possible one.
A prime will denote a trace w.r.t. the Minkowski metric: $\Phi'\equiv\eta^{\mu\nu}\Phi_{\mu\nu}$, and $\partial\cdot$~will denote a divergence:
$\partial\cdot\Phi\equiv\partial^\mu\Phi_\mu$. The Clifford algebra is $\{\g^\m,\g^\n\} \equiv +2\h^{\m\n}$, and the $\g$-matrices obey $\g^{\m\,\dagger}
\equiv \h^{\m\m}\g^\m$. The Dirac adjoint is defined as $\bar{\ps}_\m=\ps^\dagger_\m\g^0$.

Sometimes, for ease of notation, groups of symmetrized or antisymmetrized indices are also denoted as
$T_{\cdots\,a(k)\,\cdots}$~or~$T_{\cdots\,a[k]\,\cdots}$~respectively. More generally, for Young tableaux in the symmetric and antisymmetric conventions we respectively use the notations $(s_1,\cdots,s_k)$ and $[s_1,\cdots,s_k]$, where the $s_i$'s label the length of the $i$-th row or column, depending on the manifest symmetry of a given tableau.

Although we haven chosen \underline{not} to use unit-weight (anti-)symmetrization, we assume that repeated indices
with the same name are (anti-)symmetrized with the minimum number of terms. This results in the following
rules for repeated indices:
\begin{subequations}
\begin{align}
a(k)a&=(k+1) a(k+1)\,,\\[5pt]
a(k)a\,a&=(k+1)\,a(k+1)a=(k+1)(k+2)\,a(k+2)\,,\\[2pt]
a(k)a(2)&=\binom{k+2}{2}\,a(k+2)\,,\\
a(k) a(q)&=\binom{k+q}{k}\,a(k+q)\,,
\end{align}
\end{subequations}
where $a(k)$ has a unit weight by convention, and so the proportionality coefficient gives the weight of the right hand side. This convention simplifies formulae given that one keeps in mind a few binomial coefficients. A general suggestion for working with such symmetrized indices is that one keeps track of the number of terms present in a given expression. A prototypical example is the Young symmetry projection:
\be
T_{a(k),\,ab(l-1)}=0\,.\label{irred}
\ee
For an irreducible tensor with the index configuration $T_{a(k-1)c,\,b(l-1)a}$\,, Eq.~\eqref{irred} enables one to move all the symmetrized ``$a$'' indices to the first group. To see this, let us count the number of ways one can pick an index, say $a_1$, from the $k+1$ terms Eq.~\eqref{irred} contains. The index $a_1$ will appear in $k$ terms in the first row, and only in one term in the second. Then we get:
\be
T_{a_1\,a(k-1),\,a b(l-1)}+T_{a(k),\,a_1b(l-1)}=0\,.
\ee
The first term is of weight $k$, and takes into account the number of times the index $a_1$ appears in the first row of Eq.~\eqref{irred}. Then, we can replace the index $a_1$ with $c$:
\be
T_{a(k-1)c,\,b(l-1)a}=- \, T_{a(k),\,b(l-1)c}\,.\ee
This can be iterated to move all indices in the first row. In general, symmetrized indices can always be moved to the previous row (not to the subsequent ones!).

\bex
In the irreducibility identity: $T_{\cdots,\,a(k),\,a(q)b(l),\cdots}=0$,
pick up $q$ indices from the symmetrized $a$-type ones, and relabel them as $c_1,\cdots,c_q$
to prove, by making use of the identity $\sum_{i=0}^{q-1}(-1)^i\binom{q}{i}=(-1)^q\,$, that
\be
T_{\cdots,\,a(k-q)c(q),\,b(l)a(q),\,\cdots}=(-1)^q T_{\cdots,\,a(k),\,b(l)c(q),\,\cdots}\,.
\ee
\eex

\bex\label{Ex2}
Starting from the relation definition:
\be
T_{a(k),\,b(l)|\,mn}=T^{(k+1,\,l+1)}_{a(k)m,\,b(l)n}-T^{(k+1,\,l+1)}_{a(k)n,\,b(l)m}~,
\ee
and using irreducibility of the tensors appearing on the right-hand side, show that:
\begin{itemize}
\item $T_{a(k),\,b(l)|\,mn}$ is an irreducible tensor of type $(k,l)$ in the symmetric indices:
\be
T_{a(k),\,ab(l-1)|\,mn}=0\,.
\ee
\item After symmetrizing the index $m$ back with the group $a(k)$ one gets:
\be
T_{a(k),\,b(l)|an}=(k+2)\,T_{a(k),\,b(l)n}\,.
\ee
\item Iterative application of the above formula gives
\be
T_{a(k),\,b(l)|an_1|\,\cdots\,|\,an_q}=(k+2)\cdots(k+q+1)\,T_{a(k+q),\,b(l)n(q)}\,.
\ee
\end{itemize}
Note that by convention the left-hand side has weight $(k+1)\cdots(k+q)$.
\eex

For Anti-de Sitter space, underlined lowercase roman letters, $\mm,\nnn,\cdots$, denote
covariant indices, while regular lowercase roman letters, $a,b,\cdots$, (but not $i, j$), denote tangent indices at each point. Uppercase roman letters, $A,B,\cdots$,
denote ambient-space indices only in Section \ref{sec:AdS}, but $sp(4)$ indices later on.
(Un)dotted indices: $\ga,\gb(\gad,\gbd)=1,2$, label the (anti-)fundamental representations of the 4d Lorentz algebra $sl(2,\mathbb{C})$. Symplectic indices are lowered and raised with the antisymmetric metric $\epsilon^{\ga\gb}=-\epsilon^{\gb\ga}$ where $\epsilon^{12}=1$, or with $C_{AB}=\begin{pmatrix}\epsilon_{\ga\gb}&0\\0&\epsilon_{\gad\gbd}\end{pmatrix}$ in the $sp(4)$ case.

For differential forms in 4d, we use the spinorial language. $H^{\ga\ga}=h^{\ga}{}_{\gnd}\wedge h^{\ga\gnd}$ and~$H^{\gad\gad}=h_{\gamma}{}^{\gad}\wedge h^{\gamma\gad}$ are the basis of two forms. For a given vector $y^\mm$, the inner product, denoted by $i_y$, is:\;~$i_y(\omega)[x]=y^\mm\,\omega_{\mm\nnn\,[n-1]}(x)\,dx^{\nnn[n-1]}\,,~ i_y^2=0$.
Note that the de-Rham differential is inverted, up to solutions of the homogeneous equation parameterized by $\epsilon$, by the standard homotopy integral for the de Rham operator:
\beq df_n=g_{n+1}\rightarrow f_n=\underbrace{\int_0^1 dt\,i_x (g_{n+1})[xt]}_{\Gamma_n(g)}+d\epsilon_{n-1}+\delta_{n,0}\epsilon_0\,.\nonumber\eeq

\newpage
\section{Free HS in Flat Space}\label{sec:Flat}

In this Section we will consider free propagation of HS fields in Minkowski space. When far from each other, elementary
particles are essentially free since their interactions become negligible. The study of free fields is therefore important
in understanding the elementary building blocks of interacting theories. In Section \ref{sec:Wigner}, we discuss the Wigner
classification of unitary irreducible representations (UIRs) of the $d$-dimensional Poincar\'e group. These UIRs describe
all possible types of elementary particles in flat space. Their association to linear spaces where Poincar\'e symmetry acts
simply reflects the fact that elementary particles are described by linear relativistic equations. These are asymptotic
states, separated from one another, which would otherwise interact through non-linear corrections to the field equations.
Of particular interest among the UIRs of the Poincar\'e group are the massive and massless totally symmetric tensor(-spinor) representations of arbitrary rank (spin), which we consider in Section \ref{sec:BW} to write down covariant wave equations
that give rise to these UIRs as their solution space. It is desirable that the field equations follow from an action
principle, since non-linear deformations can be introduced relatively easily at the Lagrangian level without running into
immediate difficulties. The Lagrangian formulation of free HS fields is therefore considered in Section \ref{sec:Lagrangian}.

\subsection{Poincar\'e Group \& Wigner Classification}\label{sec:Wigner}

The isometries of $d$-dimensional Minkowski space are captured by the Poincar\'e group $ISO(d-1,1)$. The Lie algebra
$iso(d-1,1)$ of the Poincar\'e group is a semi-direct sum of two algebras: the Lie algebra $so(d-1,1)$ of the Lorentz
generators $\left\{M_{\m\n}\right\}$ and the Lie algebra $\mathbb{R}^d$ of the momentum generators $\left\{p_{\m}\right\}$,
the latter forming an ideal of the Poincar\'e algebra. The commutation relations read:
\bea &\left[M_{\m\n},\,M_{\r\s}\right]=i\left(\h_{\m\r}\,M_{\n\s}-\h_{\n\r}\,M_{\m\s}-\h_{\m\s}\,M_{\n\r}+\h_{\n\s}\,
M_{\m\r}\right),&\label{wc1}\\[3pt]&\left[P_\m,\,M_{\r\s}\right]=-i\left(\h_{\m\r}P_\s-\h_{\m\s}P_\r\right),&\label{wc2}\\
[3pt]&\left[P_\m,\,P_\n\right]=0.&\label{wc3}\eea

A Poincar\'e transformation maps one solution of a relativistic equation to another. In other words, the space of solutions of a particular relativistic equation forms a Poincar\'e-module\footnote{A representation of a group or
algebra by some operators, along with the space where these operators act, is called a module.}. Elementary particles
are described by relativistic equations, and thus associated with various irreducible unitary Poincar\'e-modules
\cite{Wigner:1939cj}. In order to describe all possible types of elementary particles one therefore needs the Wigner
classification$-$the classification of the unitary Poincar\'e-modules. In $d=4$, this classification was made in Refs.~\cite{Wigner:1939cj,Bargmann:1948ck}. A detailed account of unitary irreducible Poincar\'e-modules for any
space-time dimensions is given, e.g., in Ref.~\cite{Bekaert:2006py}.

Let us consider a vector space $\mathcal V$ that forms a unitary Poincar\'e-module. The momentum operators $P_\mu$,
being Hermitian and commuting with each other, are simultaneously diagonalizable. Let the subspace $\mathcal V_p$
be defined by the eigenvalue equation: $P_\m\mathcal V_p=p_\m\mathcal V_p$. Now we note that
the quadratic Casimir operator, defined as \beq\mathcal C_2\equiv\h^{\a\b}P_\a P_\b,\label{wc4}\eeq
commutes with all the Poincar\'e generators; it commutes with $P_\m$ because of the commuting nature of the momenta,
and with the Lorentz generators $M_{\m\n}$ because $\mathcal C_2$ is a Lorentz scalar. Poincar\'e transformations will
relate only those subspaces $\mathcal V_p$ that have the same eigenvalue of $\mathcal C_2$.
Any irreducible Poincar\'e-module will therefore be characterized by some number $-m^2$ such that
\beq\mathcal C_2=-m^2\mathbb I.\label{wc5}\eeq

This irreducibility condition is nothing but a representation-theoretic incarnation of the mass-shell condition:
$p^2=-m^2$. It means that every elementary particle should obey some equation that sets the value of the quadratic
Casimir$-$the Klein-Gordon equation (to be discussed in Section \ref{sec:BW}). Note that $-m^2$ is real-valued
because the representation is unitary, and so three different possibilities emerge:
\begin{itemize}\setlength\itemsep{0.5em}
  \item massive particles: $m^2 > 0$,
  \item massless particles: $m^2 = 0$,
  \item tachyons: $m^2 < 0$.
\end{itemize}

The cases of massive and massless particles are of physical interest\footnote{Tachyons, corresponding to the case $m^2 < 0$,
are unstable since they have upside-down effective potentials. Another not-so-interesting case is $p_\m=0$, which is a subcase
of $m^2 = 0$; it corresponds to constant tensors that do not depend on space-time coordinates.}. Before going into their
discussions, let us note that Poincar\'e transformations cannot change the sign of $p_0$, so that it suffices to consider
only the positive-frequency branch $p_0>0$. Given a particular $p_\m$, the structure of $\mathcal V_p$ is determined by
the subgroup of the Pioncar\'e group that leaves $p_\m$ invariant. This subgroup is called the \textit{Wigner little group}, and its
Lie algebra the  \textit{Wigner little algebra}. It turns out that any $\mathcal V_p$ that forms a unitary module of the little algebra
leads to a UIR of the Poincar\'e algebra. The problem of classifying elementary particles as UIRs of the Poincar\'e algebra
is therefore reduced to problem of classifying unitary modules of the little algebra.

\subsubsection{Massive Case}\label{sec:massiveWLA}

First, we would like to find the possible structure of $\mathcal V_p$ at a given $p_\m$. Because all subspaces
$\mathcal V_p$ with different $p_\m$ on the mass shell $p^2=-m^2$ are related by Lorentz transformations it suffices to analyze the problem in the
\textit{rest frame}, where \beq p_\m=\left(m,\,0,\cdots,0\right).\label{wc6}\eeq
The little algebra in this case is $o(d-1)$$-$the part of the Lorentz algebra that does not act on the
time-like component. Since $\mathcal V_p$ forms an $o(d-1)$-module, elementary massive particles in $d$ dimensions
are classified as UIRs of $o(d-1)$.

The simplest UIR is the \textit{trivial} module, which does not transform at all under the little group. Denoted by\,
$\bullet$\, in the notation of Young diagrams, this corresponds to a scalar field. Next comes the \textit{vector} module,
denoted by the Young diagram {\tiny$\yng(1)$}\,, for which elements of $\mathcal V_p$ are $o(d-1)$-vectors. This corresponds
to a massive vector field. An arbitrary rank-2 \textit{tensor} module decomposes into three UIRs of $o(d-1)$: the symmetric
traceless part\, {\tiny$\yng(2)$}\,, the antisymmetric part\, {\tiny$\yng(1,1)$}\,, and the trace part\, $\bullet$\,.
Their symmetry and tracelessness properties do not change under the little group. More complicated tensor modules will correspond to massive HS fields.

Let us consider an $o(d-1)$-tensor $\f_{i_1,\,i_2,\,\cdots,\,i_r}$ of arbitrary rank $r$, where each index runs as
$1,\,2,\cdots,\,d-1$. The module is irreducible if it is not possible to single out a submodule by imposing further
conditions. This will be the case if two conditions are satisfied. First, the tensor should be traceless;
contraction of any two of its indices should give zero, e.g., $\d^{i_1i_2}\f_{i_1,\,i_2,\,\cdots,\,i_r}=0$. Second, $\f_{i_1,\,i_2,\,\cdots,\,i_r}$ should have some definite symmetry properties. Different possible symmetry properties
of the tensor can be represented by various Young diagrams, each of which consists of $r$ cells {\tiny$\yng(1)$}
arranged in such a way that the length of any given row does not exceed that of any row above. A Young diagram
$Y_{l_1,\,l_2,\cdots,\,l_m}$ has a total of $m$ rows such that there are $l_i$ cells in the $i$-th row
($i=1,\,2,\cdots,\,m$), and $l_1\geq l_2\geq\cdots\geq l_m$\,:
\beq Y_{l_1,\,l_2,\cdots,\,l_m}~=~
\begin{aligned}
&\begin{tabular}{|c|c|c|c|c|}\hline
   $\phantom{a1}$&$\phantom{a1}$&\multicolumn{2}{|c|}{$~~~~\cdots~~~~\cdots~~~~\cdots~~~~$}&\phantom{a1}\\\hline
\end{tabular}~\,l_1\\[-4pt]
&\begin{tabular}{|c|c|c|c|c|}
   $\phantom{a1}$&$\phantom{a1}$&\multicolumn{2}{|c|}{$~~~~\cdots~~~~$}&$\phantom{a1}$\\\hline
\end{tabular}~\,l_2\\[-4.3pt]
&\begin{tabular}{|c|}
   $~~~~\vdots~~~~\vdots~~~~\vdots~~~~$\\\hline
\end{tabular}\\[-4pt]
&\begin{tabular}{|c|c|c|}
   $\phantom{a1}$&$\cdots$&$\phantom{a1}$\\\hline
\end{tabular}~\,l_m
\end{aligned}~~.\label{wc7}\eeq

A tensor $\f_{i_{11}i_{12}\,\cdots\, i_{1l_1},\;i_{21}i_{22}\,\cdots\, i_{2l_2},\cdots~\cdots,\,i_{m1}i_{m2}\,\cdots\,
i_{ml_m}}$ is said to represent the Young diagram~(\ref{wc7}) in the \textit{symmetric basis} if
it is totally symmetric w.r.t. the $l_1$ indices of the type $i_{1j}$, w.r.t. the $l_2$ indices of the type $i_{2j}$,
and so on. Furthermore, symmetrization of \underline{any} index of one type with \underline{all} indices of another type gives zero.
For example, \beq \f_{\left(i_{11}i_{12}\,\cdots\,i_{1l_1},\;i_{21}\right)\,i_{22}\,\cdots\, i_{2l_2},\cdots~\cdots,\,
i_{m1}i_{m2}\,\cdots\,i_{ml_m}}=0.\nonumber\eeq
On the other hand, if $h_j$ is the height of the $j$-th column of the Young diagram~(\ref{wc7}), where
$j=1,\,2,\cdots,\,n$, one has $h_1\geq h_2\geq\cdots\geq h_n$\,. Of course, $n=l_1$ and $m=h_1$.
A tensor $\f_{i_{11}i_{21}\,\cdots\, i_{h_11}|\,i_{12}i_{22}\,\cdots\, i_{h_22}|\,\cdots~\cdots
|\,i_{1n}i_{2n}\,\cdots\,i_{h_nn}}$ is said to represent the Young diagram in the \textit{antisymmetric basis} if
it is totally antisymmetric w.r.t. the $h_1$ indices of the type $i_{j1}$, w.r.t. the $h_2$ indices of the type $i_{j2}$,
and so on. Furthermore, antisymmetrization of \underline{any} index of one type with \underline{all} indices of another type gives
zero. For example, \beq \f_{\left[i_{11}i_{21}\,\cdots\, i_{h_11}|\,i_{12}\right]\,i_{22}\,\cdots\, i_{h_22}|\,
\cdots~\cdots|\,i_{1n}i_{2n}\,\cdots\,i_{h_nn}}=0.\nonumber\eeq

The antisymmetric basis makes manifest the antisymmetry associated with each column. Clearly, the maximum height of a column
can be $d-1$. However, it is possible to restrict one's consideration to a smaller set of Young diagrams. This is because the
$o(d-1)$-epsilon symbol enables one to dualize any set of totally antisymmetric indices. Thus one can set an upper bound on the height (or the total number of rows) of Young diagrams representing tensor modules of $o(d-1)$. \bex Show that in a generic dimension $d$ this upper bound is given by \beq n=h_1\leq\tfrac14\left(2d-3-(-1)^d\right).\label{wc8}\eeq\eex

One can easily convince oneself about the following facts:\vspace{2px}
\begin{itemize}
  \item \textit{Totally symmetric} tensors are described by single-row Young diagrams {\tiny$\young(\null\cdot\null)$}\;.
  \item \textit{Totally antisymmetric} tensors are described by single-column diagrams {\tiny$\young(\null,:,\null)$}\;.
  \vspace{8px}
  \item \textit{Mixed symmetry} tensors, which are neither totally symmetric nor totally antisymmetric, are described by
  diagrams with multiple rows and multiple columns.
\end{itemize}
\vspace{2px}

Note that massive fermions can be described in an analogous manner. For fermions, the unit transformation is in fact a
rotation of $4\p$$-$twice as much as the usual rotation. This means that fermions actually transform under the double
covering group of the Lorentz group: $Spin(d-1,1)$. However, the two groups have the same Lie algebra of $o(d-1,1)$ .
Let us consider an $o(d-1)$-tensor-spinor $\psi^{\,a}_{i_1,\,i_2,\,\cdots,\,i_r}$ of arbitrary rank $r$, where ``$a$''
is the spinor index. In order for it to be irreducible the tensor-spinor should be $\g$-traceless, e.g., $\left(\g^{\,i_1}\right)^a{}_b\psi^{\,b}_{i_1,\,i_2,\,\cdots,\,i_r}=0$. This also imposes the $o(d-1)$-tracelessness
through the Clifford algebra. Besides, the $o(d-1)$-vector indices should have some definite symmetry properties
as in the bosonic case.
\vspace{5px}

One can now analyze the possible types of massive fields in various dimensions.
\begin{itemize}
  \item In $d=1$ or $d=2$, the upper bound~(\ref{wc8}) on the height of possible Young diagrams is zero. This leaves room
   only for the trivial diagram\, $\bullet$\, which corresponds to a scalar or a spinor field. These are the only two possible
   types of massive particles that can propagate in these dimensions.\vspace{5px}
  \item In $d=3$ or $d=4$, the upper bound~(\ref{wc8}) on the height of possible Young diagrams is unity.  Apart from the
  trivial-module, this allows totally symmetric tensors or tensor-spinors. Massive fields in 3d or 4d Minkowski space are
  therefore characterized by a single number $l$$-$the length of a single-row Young diagram. This number is identified with
  the \textit{spin} $s$ for bosons and with $s-\tfrac12$ for fermions.\vspace{5px}
  \item In $d>4$, the height of possible Young diagrams can be greater than one, and so mixed symmetry tensor fields
  show up as well. A generic Young diagram is characterized by multiple numbers, but the word ``spin'' will denote the number
  of columns it possesses. Note that the appearance of massive mixed-symmetry tensor(-spinor) fields in String Theory is
  quite natural since it works in higher space-time dimensions$-$$d=26$ or $d=10$.
\end{itemize}

\subsubsection{Massless Case}

In order to find the possible structure of $\mathcal V_p$ with light-like momenta $p_\m$, which satisfy $p^2=0$
and $p_0>0$, it is convenient to use the \textit{light-cone coordinates}:
\beq x^\pm\equiv\tfrac{1}{\sqrt{2}}\left(x^0\pm x^{d-1}\right),\quad \text{and}~~x^i~~\text{with}~~i=1,\,2,
\cdots,\,d-2,\label{wc10}\eeq in which the metric reads: $\h_{++}=0=\h_{--}$, $\h_{+-}=-1=\h_{-+}$, and
$\h_{ij}=\d_{ij}$ with $i,j=1,\,2,\cdots,\,d-2$. Then, one can write
\beq p^2=-2p^+p^-+\sum_{i=1}^{d-2}p^ip^i=0,\qquad p^\pm\equiv\tfrac{1}{\sqrt{2}}\left(p^0\pm p^{d-1}\right).\label{wc11}\eeq
It is possible to consider the momentum $p_\m$ such that
\beq p_+\equiv-p^-=0,\qquad p_{-}\equiv-p^+=\sqrt{2}p_0>0,\qquad p_i\equiv p^i=0.\label{wc12}\eeq

We would like to find the part of the Lorentz algebra that does not affect the momentum~(\ref{wc12}).
For this purpose, we will use the light-cone Lorentz generators that include $M_{ij},\,M_{+i},\,M_{-i}$ and
$M_{+-}$ with $i,j=1,\,2,\cdots,\,d-2$.
From the commutation relation~(\ref{wc2}), it is easy to see that the generators $M_{ij}$ and $M_{+i}$
do not change the momentum, while $M_{-i}$ and $M_{+-}$ do have an affect.
\bex Prove the above statement by computing the following commutators: $\left[P_\m,\,M_{ij}\right]$,
\,$\left[P_\m,\,M_{+i}\right]$,\,$\left[P_\m,\,M_{-i}\right]$ and $\left[P_\m,\,M_{+-}\right]$.\eex

Thus we are left with the set of generators $\{M_{ij},\,M_{+i}\}$, which form the Lie algebra $iso(d-2)$.
To see this, let us first define $\p_i\equiv P_-M_{+i}$\,.
\bex Derive from Eq.~(\ref{wc1})--(\ref{wc3}) the following commutation relations:
\bea &\left[M_{ij},\,M_{kl}\right]=i\left(\d_{ik}\,M_{jl}-\d_{jk}\,M_{il}-\d_{il}\,M_{jk}+\d_{jl}\,
M_{ik}\right),&\label{wc13}\\[3pt]&\left[\p_i,\,M_{jk}\right]=-i\left(\d_{ij}\p_k-\d_{ik}\p_j\right),&
\label{wc14}\\[3pt]&\left[\p_i,\,\p_j\right]=0.&\label{wc15}\eea\eex
Eqs.~(\ref{wc13})--(\ref{wc15}) indeed define the Lie algebra $iso(d-2)$. The massless little
algebra is therefore $iso(d-2)$, and $\mathcal V_p$ has to form a unitary irreducible module thereof.

However, because the Lie algebra $iso(d-2)$ non-compact, its unitary representations does not admit faithful
finite-dimensional representations. On the other hand, finite dimensionality of a representation is necessary as it
ensures that the corresponding field contains a finite number of components. For $iso(d-2)$, this is possible
only when the quasi-momentum generators $\{\p_i\}$ act trivially, i.e., they are zero. This reduces the algebra
to a compact one defined just by Eq.~(\ref{wc13}), which is nothing but $o(d-2)$. Therefore, massless
elementary particles described by fields with a finite number of components are characterized by UIRs of $o(d -2)$.
With an abuse of terminology, $o(d-2)$ is customarily called the massless Wigner little algebra.

The structure of the modules of $o(d-2)$ is completely analogous to that of the massive little algebra $o(d-1)$.
Modulo this unit shift in space-time dimensionality, the discussions of Section \ref{sec:massiveWLA}
hold verbatim in the massless case. In Sections \ref{sec:BW} and \ref{sec:Lagrangian}, we will see further manifestations
of this simple connection between massive and massless representations of the Poincar\'e group, e.g., in the degrees-of-freedom
count and in the Lagrangian formulation.
\vspace{5px}

The Wigner classification allows one to analyze possible types of massless fields in flat space. These will
be analogous to their massive counterparts with the shift of $d\rightarrow d-1$. First, we write down the upper bound
on the height (total number of rows) of possible Young diagrams, i.e., the massless counterpart of Eq.~(\ref{wc8}). It is:
\beq n=h_1\leq\tfrac14\left(2d-5+(-1)^d\right).\label{wc18}\eeq
The possibilities in various dimensions are discussed below.
\begin{itemize}
  \item For $d\leq3$, Eq.~(\ref{wc18}) says that Young diagrams cannot a non-zero height.
  Therefore, scalars and spinors are the only two possible types of massless particles that can propagate
  in these lower dimensions. Consequently, there are no propagating gravitational waves nor any
  local degrees of freedom carried by HS fields.\vspace{5px}
  \item In $d=4$ or $d=5$, the upper bound~(\ref{wc18}) on the height of possible Young diagrams is unity. This allows also
  for totally symmetric tensors or tensor-spinors. So, massless fields in these dimensions are characterized by a single
  number $l$$-$the length of a single-row Young diagram$-$identified with
  the \textit{spin} $s$ for bosons and with $s-\tfrac12$ for fermions.\vspace{5px}
  \item For $d>5$, Young diagrams can have heights greater than unity, and so massless mixed symmetry tensor fields
  appear as well. There are multiple numbers characterizing a given Young diagram, but again ``spin'' will denote only the
  total number of columns appearing in the diagram.
\end{itemize}

We parenthetically remark that \underline{faithful} representations of the massless little group $iso(d-2)$, which are known as
\textit{infinite-} or \textit{continuous-spin representations}, have also been discussed in the literature \cite{Wigner:1939cj}
(see also \cite{Schuster:2014hca} for a recent discussion). Analysis of the unitary $iso(d-2)$-modules can be carried out
in the same way as that of the unitary Poincar\'e-modules.
It turns out that continuous-spin fields are classified by various finite-dimensional $o(d-3)$-modules. It is however not clear
if such fields are of any relevance for the physical world.\vspace{5px}

To conclude this discussion, let us recapitulate the following points: \vspace{-3px}
\begin{svgraybox}\vspace{-15px}
\begin{itemize}
  \item Elementary massive particles in $d$ dimensions are classified as unitary irreducible representations of $o(d-1)$.\vspace{5px}
  \item Massless elementary particles, described by fields with a finite number of components, are characterize
  by unitary irreducible $o(d -2)$-modules.\vspace{-5px}
\end{itemize} \end{svgraybox}
\subsection{Bargmann-Wigner Program}\label{sec:BW}

The Bargmann-Wigner program consists in associating, with each UIR of the Poincar\'e group, a covariant wave equation whose solution
space carries the original representation.
The program was originally completed in $d=4$ \cite{Bargmann:1948ck}, in which case only totally symmetric representations suffice.
Because of their simplicity, these single-row Young diagrams continue to receive most of the attention in $d>4$ as well, where there
also appear more complicated diagrams describing mixed symmetry tensors. Here we will focus mainly on the totally symmetric representations.

\subsubsection{Totally Symmetric Massive Fields:}

In the massive case, the little algebra is $o(d-1)$. We consider the $o(d-1)$-module $\mathcal V_p$, which is the space of symmetric traceless tensors
$\f_{i_1\cdots i_s}$ with the indices taking the values $1,\cdots,d-1$. We would like to write down Poincar\'e invariant field equations that
reproduce $\mathcal V_p$ at a given momentum $p$. It is sufficient to take a totally symmetric traceless Lorentz tensor field $\Phi_{\m_1\cdots\m_s}$,
with $\m_i=0,\cdots,d-1$, which obey
\begin{svgraybox}\vspace{-17px}\bea (\square-m^2)\Phi_{\m_1\cdots\m_s}&=&0,\label{kgb}\\
\partial\cdot\Phi_{\m_1\cdots\m_{s-1}}&=&0,\label{divb}\\
\Phi^\prime_{\m_1\cdots\m_{s-2}}&=&0.\label{trb}\eea\vspace{-22px}\end{svgraybox}
These equations are known as the Fierz-Pauli conditions for a symmetric bosonic field of mass $m$ and spin $s$.
The Klein-Gordon equation~(\ref{kgb})
sets the value of the quadratic Casimir $\mathcal C_2$ of the Poincar\'e group. In the rest frame, where the momentum is $p_\m=(m,0,\cdots,0)$,
the transversality condition~(\ref{divb}) renders  vanishing the components $\Phi_{0\m_2\cdots\m_s}$. The Lorentz-trace condition~(\ref{trb})
thereafter boils down to the $o(d-1)$-tracelessness. As a result, at the rest-frame momentum we indeed have an irreducible $o(d-1)$-tensor to
represent the $o(d-1)$-module $\f_{i_1\cdots i_s}$.

To count the number of degrees of freedom (DoFs) the field $\Phi_{\m_1\cdots\m_s}$ propagate, note that the number
of independent components of symmetric rank-$s$ tensor is $\binom{d-1+s}{s}\equiv\frac{(d-1+s)!}{s!(d-1)!}$\,.
%
%
The tracelessness condition is a symmetric rank-$(s-2)$ tensor that removes $\binom{d-3+s}{s-2}$ of them. Similarly, the
divergence constraint~(\ref{divb}) should eliminate $\binom{d-2+s}{s-1}$ many, but its trace part
has already been incorporated in the tracelessness of the field itself, so that the actual number
is less by $\binom{d-4+s}{s-3}$. The total number of DoFs for a massive symmetric spin-$s$ boson is therefore
\begin{svgraybox}\vspace{-5px}
\beq \mathfrak{D}_{boson}^{(m\neq0)}~=~2\binom{d-4+s}{s-1}+\binom{d-4+s}{s}.\label{dofBm}\eeq\vspace{-10px}
\end{svgraybox}
\bex Derive this result, and confirm that in $d=4$ it reduces to $2s+1$.\eex

The story is similar for fermions. We consider the space $\mathcal V_p$ of symmetric $\g$-traceless tensor-spinors $\psi^{\,a}_{i_1\cdots i_n}$, where
$i_k=1,\cdots,d-1$, and `$a$' is the $o(d-1)$-spinor index. In order to find Poincar\'e invariant field equations that reproduce $\mathcal V_p$ at a
given momentum $p$, we consider a symmetric $\g$-traceless Lorentz tensor-spinor field $\Psi_{\m_1\cdots\m_n}$, with $\m_i=0,\cdots,d-1$, whose
$o(d-1,\,1)$-spinor index has been suppressed. The desired field equations are:
\begin{svgraybox}\vspace{-15px}
\bea (\not{\!\partial\!}-m)\Psi_{\mu_1\cdots\mu_n}&=&0,\label{kgf}\\
\partial\cdot\Psi_{\mu_1\cdots\mu_{n-1}}&=&0.\label{divf}\\
\gamma^{\,\mu_1}\Psi_{\mu_1\cdots\mu_n}&=&0.\label{trf}\eea\vspace{-20px}
\end{svgraybox}

These are the Fierz-Pauli conditions for a symmetric fermionic field of mass $m$ and spin $s=n+\tfrac{1}{2}$. First of all, it follows directly from
the Dirac equation~(\ref{kgf}) that each component of the tensor-spinor field obeys the Klein-Gordon equation: $(\Box-m^2)\Psi_{\mu_1\cdots\mu_n}=0$.
The latter sets the value of the quadratic Casimir $\mathcal C_2$.
\bex Show that the Dirac equation projects out half of the solutions of the Klein-Gordon equation, and that the momentum-space projectors
reduce in the rest frame to: $\P_{\pm\a}{}^\b=\tfrac12\left(\d_\a^\b\pm i(\g^0)_\a{}^\b\right)$, where $\a,\b$ are $o(d-1,\,1)$-spinor indices.
\eex
Because the projectors $\P_{\pm\a}{}^\b$ do not transform under spatial rotations, they produce $o(d-1)$-spinors out of the
$o(d-1,\,1)$-spinors. Now, the divergence constraint~(\ref{divf}) makes the components $\Psi_{0\m_2\cdots\mu_n}$ vanish in the
rest frame. This, in turn, reduces the $\g$-trace constraint~(\ref{trf}) to the $o(d-1)$-$\g$-trace condition. The resulting
$o(d-1)$-tensor-spinor represents the $o(d-1)$-module $\psi^{\,a}_{i_1\cdots i_n}$.

The counting of DoFs is analogous to that for bosons. In $d$ dimensions, a symmetric rank-$n$ tensor-spinor
has $\binom{d+n-1}{n}\times 2^{[d]/2}$ independent components, where $[d]\equiv d+\tfrac{1}{2}\left[(-1)^d-1\right]$.
The $\gamma$-trace and divergence constraints, being symmetric rank-$(n-1)$ tensor-spinors, each eliminates
$\binom{d+n-2}{n-1}\times 2^{[d]/2}$ of them. But there is an over-counting of $\binom{d+n-3}{n-2}\times 2^{[d]/2}$
constraints, since only the traceless part of the divergence condition~(\ref{divf}) should be counted. On the other hand,
the Dirac equation sets to zero half of the components of the tensor-spinor. Therefore, the number of DoFs for a massive symmetric
spin $s=n+\tfrac12$ fermion is
\begin{svgraybox}\vspace{-5px}
\beq \mathfrak{D}_{fermion}^{(m\neq0)}~=~\binom{d-3+n}{n}\times 2^{[d-2]/2}.\label{dofFm}\eeq\vspace{-10px}
\end{svgraybox}
\bex Produce this result. Show that when $d=4$, the DoF count reduces to $2(n+1)=2s+1$, as expected.\eex

\subsubsection{Symmetric Massless Fields:}\label{sec:BWmassless}

The massless UIRs of the Poincar\'e algebra are associated with finite-dimensional modules of the little algebra $o(d-2)$.
Group theoretical arguments and quantum consistency require that massless HS fields be gauge fields and that their
gauge invariant field strengths (curvatures) satisfy irreducibility conditions which constitute the geometric HS field equations.
The Bargmann-Wigner program for the massless case is a bit subtle, and there exist different approaches to it.

\paragraph{\underline{Fronsdal Approach}:}

Let us consider a totally symmetric rank-$s$ Lorentz tensor, $\vf_{\mu_1\cdots\mu_s}$, which is traceful but \underline{doubly traceless}:
$\vf^{\prime\prime}_{\mu_1\cdots\mu_{s-4}}=0$. Then, the Fronsdal tensor is defined as:
\beq \mathcal F_{\mu_1\cdots\mu_s}\equiv\Box\vf_{\mu_1\cdots\mu_s}-\partial_{(\mu_1}\partial\cdot\vf_{\mu_2\cdots\mu_s)}+\partial_{(\mu_1}\partial_{\mu_2}
\vf'_{\mu_3\cdots\mu_s)}.\;\label{abba1}\eeq
\bex Prove that the Fronsdal tensor is also doubly traceless: $\mathcal F''_{\mu_1\cdots\mu_{s-4}}=0$.\eex
\bex Prove that the Fronsdal tensor~(\ref{abba1}) enjoys a gauge invariance with a symmetric \underline{traceless} rank-$(s-1)$ gauge parameter $\lambda_{\mu_1...\mu_{s-1}}$:
\beq \delta\vf_{\mu_1\cdots\mu_s}=\partial_{(\mu_1}\lambda_{\mu_2\cdots\mu_s)},\qquad\quad \lambda'_{\mu_1\cdots\mu_{s-3}}=0.\label{b01}\eeq\eex
\begin{svgraybox}
The Fronsdal equation for a symmetric massless spin-$s$ boson is given by:
\beq \mathcal F_{\mu_1\cdots\mu_s}\equiv\Box\vf_{\mu_1\cdots\mu_s}-\partial_{(\mu_1}\partial\cdot\vf_{\mu_2\cdots\mu_s)}+\partial_{(\mu_1}\partial_{\mu_2}
\vf'_{\mu_3\cdots\mu_s)}=0.\label{abba2}\eeq\vspace{-15px}\end{svgraybox}
This gauge-theoretic description leads to the correct number of physical DoFs, which is simply the dimension of a rank-$s$ symmetric traceless
$o(d-2)$-module \cite{deWit:1979pe}. Because $\mathcal F_{\mu_1\cdots\mu_s}$ is symmetric and doubly traceless, the Fronsdal equation indeed describes
the dynamics of a symmetric rank-$s$ tensor with
a vanishing double trace, which has $\binom{d-1+s}{s}-\binom{d-5+s}{s-4}$ independent components. Now, the gauge invariance~(\ref{b01}) with its symmetric traceless
rank-$(s-1)$ gauge parameter enables one to remove $\binom{d-2+s}{s-1}-\binom{d-4+s}{s-3}$ of these components by imposing an appropriate covariant gauge
condition (e.g., an arbitrary-spin generalization of the Lorenz gauge for $s=1$ and the de Donder gauge for $s=2$):
\beq \mathcal G_{\mu_1\cdots\mu_{s-1}}\equiv\partial\cdot\vf_{\mu_1\cdots\mu_{s-1}}-\tfrac{1}{2}\partial_{(\mu_1}\vf'_{\mu_2\cdots\mu_{s-1})}=0,\label{abba3}\eeq
where $\mathcal G_{\mu_1\cdots\mu_{s-1}}$ is traceless because $\vf''_{\mu_1\cdots\mu_{s-4}}=0$. This gauge choice reduces Eq.~(\ref{abba2}) to the Klein-Gordon
equation for a massless field:
\beq \Box\vf_{\mu_1\cdots\mu_s}=0.\label{abba4}\eeq
The gauge condition~(\ref{abba3}) involves derivatives, and converts constraints into evolution equations. More precisely, it renders dynamical
the traceless part of $\vf_{0\mu_2\cdots\mu_s}$, which was non-dynamical originally in Eq.~(\ref{abba2}).
The gauge fixing~(\ref{abba3}) however is not complete, since one can still allow for gauge parameters that satisfy $\Box\lambda_{\mu_1\cdots\mu_{s-1}}=0$.
So, one can further gauge away $\binom{d-2+s}{s-1}-\binom{d-4+s}{s-3}$ components. Thus the total number of DoFs for a massless symmetric spin-$s$
bosonic field is:
\begin{svgraybox}\beq \mathfrak{D}_{boson}^{(m=0)}=2\binom{d-5+s}{s-1}+\binom{d-5+s}{s}.\label{dofB0}\eeq\vspace{-10px}\end{svgraybox}
\bex Check that the substitution $d\rightarrow(d-1)$ in formula~(\ref{dofBm}) produces the same DoF count as above. This confirms that the
Fronsdal equation along with its gauge symmetry indeed represents a rank-$s$ symmetric traceless $o(d-2)$-module.\eex
\bex Confirm that in 4d a massless bosonic field with $s\geq1$ carries 2 DoFs.\eex

To describe fermions, let us consider a totally symmetric rank-$n$ Lorentz tensor-spinor, $\psi_{\mu_1\cdots\mu_n}$, which is \underline{triply $\gamma$-traceless}:
$\gamma^{\mu_1}\gamma^{\mu_3}\gamma^{\mu_3}\psi_{\mu_1\cdots\mu_n}=0$. The Fronsdal tensor this fermionic field is given by:
\beq \mathcal S_{\mu_1\cdots\mu_n}\equiv i\left[\displaystyle{\not{\!\partial\,}}\psi_{\mu_1\cdots\mu_n}
-\partial_{(\mu_1}\displaystyle{\not{\!\!\psi}}_{\mu_2\cdots\mu_n)}\right].\label{noname45}\eeq
\bex Prove that $\mathcal S_{\mu_1\cdots\mu_n}$ is triply $\gamma$-traceless, and
enjoys a gauge symmetry with a symmetric \underline{$\gamma$-traceless} rank-$(n-1)$ tensor-spinor gauge parameter
$\varepsilon_{\mu_1\cdots\mu_{n-1}}$: \beq \delta\psi_{\mu_1\cdots\mu_n}=\partial_{(\mu_1}\varepsilon_{\mu_2\cdots\mu_n)},
\qquad\quad\gamma^{\mu_1}\varepsilon_{\mu_1\cdots\mu_{n-1}}=0.\label{f01}\eeq\eex
\begin{svgraybox} The Fronsdal equation for a massless symmetric spin $s=n+\frac12$ fermion is: \beq \mathcal S_{\mu_1\cdots\mu_n}\equiv
i\left[\displaystyle{\not{\!\partial\,}}\psi_{\mu_1\cdots\mu_n}-\partial_{(\mu_1}\displaystyle{\not{\!\!\psi}}_{\mu_2\cdots\mu_n)}\right]=0.
\label{noname46}\eeq\vspace{-15px}\end{svgraybox}

Because $\mathcal S_{\mu_1\cdots\mu_n}$ is symmetric and triply $\gamma$-traceless, Eq.~(\ref{noname46}) contains the appropriate number of independent components,
namely $\left[\binom{d+n-1}{n}-\binom{d+n-4}{n-3}\right]\times 2^{[d]/2}$. However, the $\gamma$-traceless part of
$\psi_{0\mu_2\cdots\mu_n}$ is actually not dynamical, so that there $\left[\binom{d+n-2}{n-1}-\binom{d+n-3}{n-2}\right]\times 2^{[d]/2}$ number of constraints.
Now, the gauge freedom of Eq.~(\ref{f01}) enables us to choose the gauge
\beq \mathcal G_{\mu_1\cdots\mu_{n-1}}\equiv\displaystyle{\not{\!\!\psi}}_{\mu_1\cdots\mu_{n-1}}-\tfrac{1}{d-4+2n}\,\gamma_{(\mu_1}\psi'_{\mu_2\cdots\mu_{n-1})}=0,
\label{abba7}\eeq
where $\mathcal G_{\mu_1\cdots\mu_{n-1}}$ is $\gamma$-traceless. This algebraic gauge condition does not convert constraints into dynamical equations, and
allows further gauge transformations. The residual gauge parameter entails that the following tensor-spinor vanishes:
$\displaystyle{\not{\!\partial\,}}\varepsilon_{\mu_1\cdots\mu_{n-1}}-\tfrac{2}{d-4+2n}\,\gamma_{(\mu_1}\partial\cdot\varepsilon_{\mu_3\cdots\mu_{n-1})}$\,.
The latter quantity is $\gamma$-traceless, thanks to the $\gamma$-tracelessness of $\varepsilon_{\mu_1\cdots\mu_{n-1}}$  itself.
The gauge condition~(\ref{abba7}) and the residual gauge choice will eliminate $\left[\binom{d+n-2}{n-1}-\binom{d+n-3}{n-2}\right]\times 2^{[d]/2}$ components each.

Note that the gauge fixing~(\ref{abba7}) alone does not reduce the EoMs~(\ref{noname46}) to the Dirac form. However, one can exploit part of the residual gauge
invariance to set $\psi'_{\mu_1\cdots\mu_{n-2}}$ to zero. The latter choice reduces the field equations to the form:
\beq \displaystyle{\not{\!\partial\,}}\psi_{\mu_1\cdots\mu_n}=0.\label{amipagol}\eeq
This is nothing but the Dirac equation for a massless fermion, and it projects to zero half of the components of $\psi_{\mu_1\cdots\mu_n}$.
Then, the total number of DoFs for a massless symmetric spin $s=n+\tfrac{1}{2}$ fermionic field is given by:
\begin{svgraybox}\beq \mathfrak{D}_{fermion}^{(m=0)}=\binom{d-4+n}{n}\times 2^{[d-2]/2}.\label{dofF}\eeq\vspace{-10px}\end{svgraybox}
This is indeed the dimension of a rank-$n$ symmetric $\g$-traceless $o(d-2)$-module. Note, in particular, that in $d=4$ this number is 2 for all
half-integer spin as expected.
\vspace{-5px}

The Fronsdal equations~(\ref{abba2}) and~(\ref{noname46}) are \underline{local} second- and first-order differential equations for
massless symmetric bosonic and fermionic HS fields. As we will see in Section~\ref{sec:Lagrangian}, these equations do result from Lagrangians
constructed in \cite{Fronsdal:1978rb,Fang:1978wz}. An unappealing feature of the Fronsdal approach however is the presence of algebraic (trace)
constraints on the fields and gauge parameters, which is somewhat unnatural. Only when the HS fields satisfy the trace constraints
\underline{off shell}, can the Fronsdal equations follow from Lagrangians. One may forego these constraints on the fields and gauge
parameters in three different but related ways:
\begin{itemize}\setlength\itemsep{0.5em}
\item One way is the BRST approach \cite{Pashnev:1997rm,Pashnev:1998ti,Burdik:2001hj,Burdik:2000kj},
which introduces a set of auxiliary fields (whose number grows with the spin) such that the field equations remain Lagrangian
\cite{Pashnev:1997rm,Pashnev:1998ti,Burdik:2001hj,Burdik:2000kj,Buchbinder:2004gp,Buchbinder:2005ua}. A ``minimal'' solution of this
kind for symmetric tensor(-spinor) fields is the compensator equation \cite{Francia:2002aa,Francia:2002pt,Sagnotti:2003qa}; but
it is non-Lagrangian.
\item Another way is to introduce \underline{non-locality} in the theory, as shown in Refs.~\cite{Francia:2002aa,Francia:2002pt}.
The non-local formulation is interesting in that it casts the HS field equations into
``geometric'' forms involving generalized curvatures of de Wit and Freedman \cite{deWit:1979pe} and of Weinberg \cite{Weinberg:1965rz}.
Moreover, the Lagrangian nature of the equations can be preserved by suitable choices of non-locality
\cite{Francia:2005bu,Francia:2007qt,Francia:2008hd,Francia:2010qp}.
\item The third possibility$-$the Bargmann-Wigner approach, which we are going to consider below in details$-$does not invoke
gauge fields to begin with. In $d=4$ these equations were constructed long ago by Bargmann and Wigner
in terms of two-component tensor-spinors \cite{Bargmann:1948ck}. Generalizations to higher dimensions were also made for arbitrary
tensorial UIRs \cite{Bekaert:2002dt,Bekaert:2003az} and for spinorial UIRs \cite{Bandos:2005mb}.
\end{itemize}

\paragraph{\underline{Bargmann-Wigner Approach}:}

Let us consider a mixed-symmetry \underline{traceless} tensor field $R_{\m_1\n_1|\,\cdots\,|\,\m_s\n_s}$, which is a $o(d-1,1)$-module
with the symmetries of the following Young tableau:
\beq \begin{tabular}{|c|c|c|c|c|}\hline
   $\m_1$&$\m_2$&\multicolumn{2}{|c|}{$~~~~\cdots~~~~$}&$\m_s$\\\hline
   $\n_1$&$\n_2$&\multicolumn{2}{|c|}{$~~~~\cdots~~~~$}&$\n_s$\\\hline
\end{tabular}~,\label{w-curvature0}
\eeq
where the manifestly antisymmetric convention is taken for convenience. The tensor $R_{\m_1\n_1|\,\cdots\,|\,\m_s\n_s}$
is antisymmetric under the interchange of ``paired'' indices,
but symmetric under the interchange of any two sets of paired indices, e.g.,
\beq R_{\m_1\n_1|\,\m_2\n_2|\,\cdots\,|\,\m_s\n_s}=-R_{\n_1\m_1|\,\m_2\n_2|\,\cdots\,|\,\m_s\n_s}
=R_{\m_s\n_s|\,\m_2\n_2|\,\cdots\,|\,\m_{s-1}\n_{s-1}|\,\m_1\n_1},\label{w-curvature0.1}\eeq
and vanishes upon complete antisymmetrization of any three indices, e.g.,
\beq R_{[\,\m_1\n_1|\,\m_2]\,\n_2\,|\,\cdots\,|\,\m_s\n_s}=0\,.\label{w-curvature0.2}\eeq
The desired wave equations are the following set of first-order field equations:
\bea \partial_{[\,\r} R_{\m_1\n_1]\,|\,\cdots\,|\,\m_s\n_s}&=&0,\label{w-curvature1}\\
\partial\cdot R_{\m_1\n_1|\,\cdots\,|\,\m_{s-1}{\n_{s-1}}|\,\m_s}&=&0.\label{w-curvature2}\eea
It follows immediately that the tensor obeys the Klein-Gordon equation:
\beq \Box\,R_{\mu_1\nu_1|\,\cdots\,|\,\mu_s\nu_s}=0,\label{w-curvature3}\eeq
which fixes the value of the quadratic Casimir $\mathcal C_2$ of the Poincar\'e group to zero, consistently with the masslessness of the UIR.

We will now see that the solution space of the wave equations~(\ref{w-curvature1})--(\ref{w-curvature3}) indeed carries a spin-$s$ massless
representation of the Poincar\'e group. Let us first note that a generalization of the Poincar\'e lemma states that the differential Bianchi-like
identity~(\ref{w-curvature1}) and the irreducibility conditions of\, $R_{\m_1\n_1|\,\cdots\,|\,\m_s\n_s}$ imply that this tensor is in fact
the $s^{\text{th}}$ derivative of a totally symmetric traceless rank-$s$ tensor \cite{DuboisViolette:2001jk}.
This can be understood in the light-cone frame with the momentum given by Eq.~(\ref{wc12}).
\bex Show that one can solve Eq.~(\ref{w-curvature1}) in terms of a totally symmetric traceless rank-$s$ tensor $\vf_{\n_1\cdots\n_s}$ as
\beq R_{\m_1\n_1|\,\cdots\,|\,\m_s\n_s}=Y_{s,\,s}\left[p^-_{\m_1}\cdots p^-_{\m_s}\vf_{\n_1\cdots\n_s}\right]\,,\qquad p_\m^-\equiv p_-\d_\m^-\,,
\label{w-curvature4}\eeq
where $Y_{s,\,s}$ is the projector in the corresponding Young tableau~(\ref{w-curvature0}).
Check that only the $\vf_{\m_1\cdots\m_s}$-components with no ``$-$'' index give nontrivial solutions.
\eex
Finally, using the scalar product formula: $X\cdot Y=-X^-Y^+-X^+Y^-+\sum_{i=1}^{d-2}X^iY^i$, one can show that the transversality condition~(\ref{w-curvature2})
amounts to setting to zero all components of $\vf_{\m_1\cdots\m_s}$ with at least one ``$+$'' index. The tracelessness of $\vf_{\m_1\cdots\m_s}$ then
boils down to the $o(d-2)$-tracelessness. The resulting irreducible $o(d-2)$-tensor is a module of the little algebra $o(d-2)$, as required.

Similarly, for a spin $s=n+\tfrac12$ fermion one considers an $o(d-1,1)$-module $\mathfrak{R}_{\m_1\n_1|\,\cdots\,|\,\m_n\n_n}$, which is
a \underline{$\g$-traceless} mixed-symmetry tensor-spinor sharing the same properties~(\ref{w-curvature0})--(\ref{w-curvature0.2}) of its bosonic
counterpart. The wave equations are:
\bea \partial_{[\,\r} \mathfrak R_{\m_1\n_1]\,|\,\cdots\,|\,\m_n\n_n}&=&0,\label{w-curvaturef1}\\
\partial\cdot \mathfrak R_{\m_1\n_1|\,\cdots\,|\,\m_{n-1}{\n_{n-1}}|\,\m_n}&=&0.\label{w-curvaturef2}\eea
The Dirac equation then follows from taking a $\g$-trace of Eq.~(\ref{w-curvaturef1}):
\beq  \displaystyle{\not{\!\partial\,}}\mathfrak R_{\mu_1\nu_1|\,\cdots\,|\,\mu_n\nu_n}=0.\label{w-curvaturef3}\eeq
\bex Show that the solution space of the wave equations (\ref{w-curvaturef1})--(\ref{w-curvaturef3}) indeed carries
a spin $s=n+\tfrac12$ massless representation of the Poincar\'e group.
\eex

Note, in view of Eq.~(\ref{w-curvature1}) and the generalized Poincar\'e lemma \cite{DuboisViolette:2001jk}, that
the tensor $R_{\m_1\n_1\,|\,\cdots\,|\,\m_s\n_s}$ generalizes the familiar linear Weyl tensor for $s=2$ and EM field
strength for $s=1$. When the traces of $R_{\m_1\n_1\,|\,\cdots\,|\,\m_s\n_s}$ are included, the resulting irreducible
representation of $GL(d)$ is a generalization of the linear Riemann tensor$-$the Weinberg curvature tensor for a
massless symmetric spin-$s$ boson \cite{Weinberg:1965rz}. A posteriori, the Bargmann-Wigner
equations~(\ref{w-curvature1})--(\ref{w-curvature3}) are a number of conditions imposed on the generalized curvature,
and so the HS gauge fields satisfy manifestly gauge-invariant but higher-order field equations. The requirement of on-shell
tracelessness of the Weinberg tensor then results in a generalization of the linearized Einstein equations. The latter equations
are non-Lagrangian for $s>2$, but offer unconstrained gauge symmetries for the symmetric tensor gauge potentials. In fact,
they result in the compensator equations presented in \cite{Francia:2002aa,Francia:2002pt,Sagnotti:2003qa}, which in turn
give rise to the non-local equations of Refs.~\cite{Francia:2002aa,Francia:2002pt} after algebraic manipulations.
These interconnections and the subsequent rederivation of the Fronsdal equations are
beautifully explained in Refs.~\cite{Sorokin:2004ie,Bandos:2005mb,Bekaert:2006ix,Sagnotti:2003qa}.

A drawback of the Bargmann-Wigner equations for massless Poincar\'e-modules is the lack of explicit appearance of the
gauge fields, which show up only upon solving the differential Bianchi identity \footnote{A related fact is that in $d=4$
the Bargmann-Wigner equations are conformally invariant, while the Fronsdal equations are not. Conformal invariance is
actually lost as one solves the differential Bianchi identities and thereby introduces gauge potentials.}.
This is quite unnatural from the perspective of relativistic Quantum Field Theory, where non-linear Lagrangians for massless
fields require gauge-theoretic description. Historically, this prompted the development of the Fronsdal approach, which
aimed at writing down free Lagrangians for gauge fields and finding non-linear deformations thereof. Vasiliev, on the other hand,
took a different path in which gauge fields are introduced by realizing the Bargmann-Wigner equations in terms of Free Differential
Algebras \cite{Vasiliev:1980as,Vasiliev:1986td,Lopatin:1987hz}. This is the so-called ``\underline{frame-like}'' formulation of
HS fields as opposed to the ``\underline{metric-like}'' formulation which includes all other approaches discussed so far.

\paragraph{Remarks:}
\begin{itemize}\setlength\itemsep{1em}
\item \underline{Mixed-Symmetry Case}: It is not very difficult to extend the analysis to arbitrary irreducible mixed-symmetry
tensor(-spinor) representations. Many qualitative features of the totally symmetric case still persists.
In the massive case, the wave equations are essentially the same as the Fierz-Pauli conditions~(\ref{kgb})--(\ref{trb})
or~(\ref{kgf})--(\ref{trf}). In the massless case, the Bargmann-Wigner program was first completed as an extension of
the Fronsdal formulation by Labastida \cite{Labastida:1986gy,Labastida:1986ft,Labastida:1987kw}. The gauge transformations
take a more complicated shape for mixed-symmetry fields, since each index family comes with one independent gauge parameter.
The algebraic constraints on the fields and gauge parameters are also more involved, since it is possible to consider different
($\g$-)traces involving different index families, and only some of their linear combinations are forced to vanish. The mathematical
structures appearing in the mixed-symmetry case are richer and more complex. Interested readers can find the details in Refs.~\cite{Bandos:2005mb,Bekaert:2006ix,Alkalaev:2008gi,Campoleoni:2008jq,Campoleoni:2009gs,Campoleoni:2009je}.
\item \underline{Triplets \& String Field Theory}: It is instructive to explore the link between String Theory and HS fields
at the free level. As it turns out, String Theory favors reducible representations with no on-shell tracelessness
conditions on the fields. For symmetric bosonic string fields, one obtains in the tensionless limit an interesting system
of three fields: a ``triplet'' \cite{Bengtsson:1986ys,Bengtsson:1987jt,Bonelli:2003kh,Fotopoulos:2008ka,Francia:2002aa,
Francia:2002pt,Pashnev:1997rm,Pashnev:1998ti,Burdik:2000kj,Buchbinder:2001bs,Buchbinder:2002ry,Bekaert:2003uc,Buchbinder:2004gp}.
The triplet system comprises symmetric traceful tensors of rank $s$, $s-1$ and $s-2$: $\vf_{\m_1\cdots\m_s}$, $C_{\m_1\cdots\m_{s-1}}$
and $D_{\m_1\cdots\m_{s-2}}$. They obey the following equations:
\bea \Box\,\vf_{\m_1\cdots\m_s}&=&\de_{(\m_1}C_{\m_2\cdots\m_s)},\nonumber\\[3pt]
C_{\m_1\cdots\m_{s-1}}&=&\de\cdot\vf_{\m_1\cdots\m_{s-1}}-\de_{(\m_1}D_{\m_2\cdots\m_{s-1})},\nonumber\\[3pt]
\Box\,D_{\m_1\cdots\m_{s-2}}&=&\de\cdot C_{\m_1\cdots\m_{s-2}},\nonumber\eea
which are invariant under unconstrained gauge transformations:
\bea \d\vf_{\m_1\cdots\m_s}&=&\de_{(\m_1}\l_{\m_2\cdots\m_s)},\nonumber\\[3pt]
\d C_{\m_1\cdots\m_{s-1}}&=&\Box\,\l_{\m_1\cdots\m_{s-1}},\nonumber\\[3pt]
\d D_{\m_1\cdots\m_{s-2}}&=&\de\cdot\l_{\m_1\cdots\m_{s-2}}.\nonumber\eea

This is a Lagrangian system. It is easy to see that the lower-rank tensors disappear on shell, and so one is left with the
massless Klein-Gordon equation and transversality of $\vf_{\m_1\cdots\m_s}$. These equations therefore propagate modes of
spin $s,\,s-2,\,s-4,\cdots,$ down to zero or one if $s$ is even or odd. Note that
fermionic and mixed-symmetry triplet systems share similar qualitative features.
\end{itemize}

\subsection{Lagrangian Formulation}\label{sec:Lagrangian}

Initiated by Fierz and Pauli \cite{Fierz:1939zz,Fierz:1939ix}, the Lagrangian formulation of HS fields requires
the inclusion of a lower-spin auxiliary fields, which must vanish on shell. Singh and Hagen achieved the feat of
writing down Lagrangians for massive symmetric bosonic and fermionic fields of arbitrary spin \cite{Singh:1974qz,Singh:1974rc}.
These Lagrangians simplify considerably in the massless limit, and therefore Lagrangians for
HS massless fields turn out to be much simpler \cite{Fronsdal:1978rb,Fang:1978wz}. Indeed, the most straightforward way of obtaining
a massive spin-$s$ Lagrangian in $d$ dimensions is to start with a $(d+1)$-dimensional Lagrangian of a massless spin-$s$ field,
and perform a Kaluza-Klein reduction with a single dimension compactified on a circle of radius $1/m$.

\subsubsection{Symmetric Massive Fields:}

The Singh-Hagen Lagrangian for massive fields has the following field contents:
\begin{itemize}\setlength\itemsep{0.5em}
\item For a spin-$s$ boson, symmetric \underline{traceless} tensor fields of rank $s,\,s-2,\,s-3,\cdots,0$.
\item For a spin-$s=n+\tfrac{1}{2}$ fermion, a set of symmetric
\underline{$\gamma$-traceless} tensor-spinors: one of rank $n$, another of rank $n-1$, and doublets of rank $n-2,\,n-3,\cdots,0$.
\end{itemize}
All the lower-spin auxiliary fields are forced to vanish when the Fierz-Pauli conditions~(\ref{kgb})--(\ref{trb}) and~(\ref{kgf})--(\ref{trf})
are satisfied \footnote{Interestingly, by field redefinitions one can package all the Singh-Hagen fields nicely into just a couple of reducible
representations \cite{Singh:1981aw}: for bosons a rank-$s$ and a rank-$(s-3)$ symmetric \underline{traceful} fields, and for fermions
a rank-$n$ and a rank-$(n-2)$ symmetric \underline{$\g$-traceful} tensor-spinors.}.
The salient features of the Singh-Hagen construction can be understood by considering the simple case of a massive spin-2 field $\Phi_{\mu\nu}$.
The Klein-Gordon equation, $(\Box-m^2)\Phi_{\mu\nu}=0$, and the transversality condition, $\partial\cdot\Phi_\mu=0$, may potentially be
derived from a Lagrangian of the form:
\beq \mathcal L_{\text{trial}}=-\tfrac{1}{2}(\partial_\mu\Phi_{\nu\rho})^2 -\tfrac{1}{2}m^2\Phi_{\mu\nu}^2 +\tfrac{1}{2}\l(\partial\cdot\Phi_{\mu})^2,\label{h5}\eeq
where $\l$ is a constant to be determined. While taking variations of the trial action one must take care of the symmetric traceless nature of $\Phi_{\mu\nu}$.
The EoMs read:
\beq (\Box-m^2)\Phi_{\mu\nu}-\tfrac{1}{2}\l\left[\partial_{(\mu}\partial\cdot\Phi_{\nu)}-\tfrac{2}{d}\,\eta_{\mu\nu}\partial\cdot\partial\cdot\Phi\right]=0.\label{h6}\eeq
The divergence of the above equation gives
\beq [(\l-2)\Box+2m^2]\partial\cdot\Phi_{\mu}+\l\left(1-\tfrac{2}{d}\right)\partial_\mu\partial\cdot\partial\cdot\Phi=0.\label{h7}\eeq
The transversality condition is recovered by setting $\l=2$ and requiring $\partial\cdot\partial\cdot\Phi=0$. But the latter
requirement follows from the EoMs only if an auxiliary scalar field $\r$ is introduced. This allows the following
addition to the trail Lagrangian~(\ref{h5}):
\beq \mathcal L_{\text{add}}=\a_1(\partial_\mu\r)^2+\a_2\,\r^2+\r\,\partial\cdot\partial\cdot\Phi,\label{h8}\eeq
where $\a_{1,2}$ are constants. On the one hand, the double divergence of the $\Phi_{\mu\nu}$-EoMs, resulting from the Lagrangian
$\mathcal L=\mathcal L_{\text{trial}}+\mathcal L_{\text{add}}$, now gives
\beq [(2-d)\Box-dm^2]\,\partial\cdot\partial\cdot\Phi+(-1)\Box^2\r=0.\label{h9}\eeq
On the other hand, the auxiliary scalar $\r$ has the EoM:
\beq \partial\cdot\partial\cdot\Phi-2(\a_1\Box-\a_2)\r=0.\label{h10}\eeq
Eqs.~(\ref{h9})--(\ref{h10}) comprise a linear homogeneous system in the variables $\partial\cdot\partial\cdot\Phi$ and $\r$.
The double-divergence condition, $\partial\cdot\partial\cdot\Phi=0$, and the vanishing of the auxiliary scalar, $\r=0$, follow
if the associated determinant $\Delta$ is non-zero. We have
\beq \Delta=[2(d-2)\a_1-(d-1)]\Box^2+2[dm^2\a_1-(d-2)\a_2]\Box-2dm^2\a_2.\label{h11}\eeq
Note, in particular, that $\Delta$ gets rid of differential operators and becomes algebraic, proportional to $m^2$, and hence non-zero if
\beq \a_1=\tfrac{(d-1)}{2(d-2)}, \qquad\quad \a_2=\tfrac{m^2d(d-1)}{2(d-2)^2}, \qquad\quad d>2.\label{h12}\eeq
Thus one ends up with the following Lagrangian for a massive spin-2 field:
\beq \mathcal L=-\tfrac{1}{2}(\partial_\mu\Phi_{\nu\rho})^2-\tfrac{1}{2}m^2\Phi_{\mu\nu}^2+(\partial\cdot\Phi_{\mu})^2+\r\,\partial\cdot\partial\cdot\Phi
+\tfrac{(d-1)}{2(d-2)}\left[(\partial_\mu\r)^2+\tfrac{dm^2}{d-2}\r^2\right].\label{h13}\eeq
\bex Check that the double-divergence condition $\partial\cdot\partial\cdot\Phi=0$ and the vanishing of the auxiliary scalar
and $\r=0$ can indeed be derived from the resulting EoMs, so that the Klein-Gordon equation and the transversality condition follow.\eex

\bex Combine $\Phi_{\mu\nu}$ and $\r$ into a \underline{traceful} field: $\varphi_{\mu\nu}=\Phi_{\mu\nu}
+\tfrac{1}{d-2}\,\eta_{\mu\nu}\r$, to reduce the Singh-Hagen Lagrangian~(\ref{h13}) to the Fierz-Pauli form:
\beq \mathcal L_{\text{FP}}=-\tfrac{1}{2}\,(\partial_\mu\vf_{\nu\rho})^2+(\partial\cdot\vf_{\mu})^2 +\tfrac{1}{2}\,
(\partial_\mu\vf')^2\,-\partial\cdot\vf_{\mu}
\partial^\mu\vf'-\tfrac{1}{2}\, m^2[\vf_{\mu\nu}^2-\vf'^2].\label{f1}\eeq\label{ex:FP}\eex
\vspace{-20px}

The above procedure can be carried out for higher spins. For an arbitrary integer spin $s$, the following pattern emerges:
it is required that one successively obtain the conditions: $\partial^{\mu_1}\cdots\partial^{\mu_k}\Phi_{\mu_1\cdots\mu_s}=0$,
for $k=2,\,3,\cdots,s$. At each value of $k$, an auxiliary symmetric traceless rank-$(s-k)$ tensor field needs to be introduced.
\vspace{5px}

Similarly, for a half-integer spin $s=n+\tfrac{1}{2}$, one obtains the transversality condition, $\partial\cdot\Psi_{\mu_1\cdots\mu_{n-1}}=0$,
by introducing a symmetric $\gamma$-traceless rank-$(n-1)$ tensor-spinor, say $\chi_{\mu_1\cdots\mu_{n-1}}$, provided that the following
conditions are satisfied as well: $\partial\cdot\partial\cdot\Psi_{\mu_1\cdots\mu_{n-2}}=0=\partial\cdot\chi_{\mu_1\cdots\mu_{n-2}}$.
The latter conditions call for a couple of symmetric $\gamma$-traceless rank-$(n-2)$ auxiliary tensor-spinors. Successively, there appear
a pair of rank-($n-k$) conditions for $k=3,\cdots,n$, which require a pair of auxiliary fields that are symmetric $\gamma$-traceless
rank-($n-k$) tensor-spinors. The explicit form of the Lagrangian for an arbitrary spin is rather complicated and not so illuminating.
The interested reader can look it up in the original references \cite{Singh:1974qz,Singh:1974rc}.

\subsubsection{Symmetric Massless Fields:}

For the bosonic case, in the massless limit of the Singh-Hagen Lagrangian, all but one auxiliary fields decouple: the one with the
highest rank $s-2$. The surviving pair of symmetric traceless rank-$s$ and rank-$(s-2)$ tensors can be combined into a single
symmetric tensor, $\vf_{\mu_1\cdots\mu_s}$, which is traceful but \underline{doubly traceless}: $\vf^{\prime\prime}_{\mu_1\cdots\mu_{s-4}}=0$.
\begin{svgraybox}
The Fronsdal Lagrangian for a massless bosonic spin-$s$ field takes the form:
\bea \mathcal L&=&-\tfrac{1}{2}(\partial_\rho\vf_{\mu_1\cdots\mu_s})^2 +\tfrac{1}{2}s(\partial\cdot\vf_{\mu_2\cdots\mu_s})^2+\tfrac{1}{2}s(s-1)
\left(\partial\cdot\partial\cdot\vf_{\mu_3\cdots\mu_s}\right)\vf^{\prime\,\mu_3\cdots\mu_s}\nonumber\\&&+\tfrac{1}{4}s(s-1)(\partial_\rho
\vf'_{\mu_3\cdots\mu_s})^2+\tfrac{1}{8}s(s-1)(s-2)(\partial\cdot\vf'_{\mu_4\cdots\mu_s})^2.\label{b00}\eea\vspace{-17px}
\end{svgraybox}
\bex Prove that the Fronsdal Lagrangian enjoys the gauge invariance~(\ref{b01}).\eex

A very simple illustration of the Fronsdal-Fang formulation is given again by the spin-2 case. Ex.\,\ref{ex:FP} shows that the
Singh-Hagen spin-2 Lagrangian~(\ref{h13}), after a field redefinition, gives the Fierz-Pauli Lagrangian~(\ref{f1}),
whose massless limit is
\beq \mathcal{L}=-\tfrac{1}{2}(\partial_\mu\vf_{\nu\rho})^2\,+\,(\partial\cdot\vf_\mu)^2\,+\,\tfrac{1}{2}(\partial_\mu\vf')^2\,
+\,\vf'\,\partial\cdot\partial\cdot\vf,\label{h15}\eeq
where the field $\vf_{\mu\nu}$ is symmetric and traceful. This is actually the linearized Einstein-Hilbert
action with the massless spin-2 field identified as the metric perturbation around Minkowski background:
$\vf_{\mu\nu}\equiv M_P\left(g_{\m\n}-\eta_{\m\n}\right)$. The corresponding gauge symmetry is just the
infinitesimal version of diffeomorphism invariance:
\beq \delta\vf_{\mu\nu}=\partial_{(\mu}\lambda_{\nu)}.\label{h16}\eeq
\bex Show that the Euler-Lagrange equations resulting from the Lagrangian (\ref{b00}) can be expressed in terms
of the Fronsdal tensor~(\ref{abba1}) as:
\beq \mathcal F_{\m_1\cdots\m_s}-\tfrac12\h_{(\m_1\m_2}\mathcal F'_{\m_3\cdots\m_s)}=0,\label{noname1122}\eeq
from which the Fronsdal equation~(\ref{b01}) follows immediately.
\eex

The Lagrangian~(\ref{b00}) is the unique 2-derivative quadratic one invariant under the gauge transformations~(\ref{b01})
for a canonically normalized field $\vf_{\m_1\cdots\m_s}$.
\bex
Write down the most generic form a Lagrangian of the above kind:
\bea \mathcal L&=&-\tfrac{1}{2}(\partial_\rho\vf_{\mu_1\cdots\mu_s})^2 +\a_1(\partial\cdot\vf_{\mu_2\cdots\mu_s})^2+\a_2
\left(\partial\cdot\partial\cdot\vf_{\mu_3\cdots\mu_s}\right)\vf^{\prime\,\mu_3\cdots\mu_s}\nonumber\\&&+\a_3(\partial_\rho
\vf'_{\mu_3\cdots\mu_s})^2+\a_4(\partial\cdot\vf'_{\mu_4\cdots\mu_s})^2,\nonumber\eea
where the $\a_i$'s are constants, and gauge vary the Lagrangian to show that the variation vanishes only when the coefficients
$\a_i$ are given by: \beq\a_1=\tfrac{1}{2}s,\quad\a_2=\tfrac{1}{2}s(s-1),\quad\a_3=\tfrac{1}{4}s(s-1),\quad
\a_4=\tfrac{1}{8}s(s-1)(s-2).\nonumber\eeq\eex

For a $s=n+\tfrac{1}{2}$ fermion, the Singh-Hagen Lagrangian in the massless limit \cite{Fang:1978wz} keeps only three symmetric
$\gamma$-traceless tensor-spinors: rank-$n$, rank-$(n-1)$ and rank-$(n-2)$, and decouple completely other auxiliary fields.
Furthermore, the surviving triplet can be combined into a single symmetric rank-$n$ tensor-spinor, $\psi_{\mu_1\cdots\mu_n}$.
The latter is \underline{triply $\gamma$-traceless}: $\gamma^{\mu_1}\gamma^{\mu_3}\gamma^{\mu_3}\psi_{\mu_1\cdots\mu_n}=0$.
\begin{svgraybox}
The Lagrangian for a $s=n+\tfrac{1}{2}$ massless fermionic field takes the form:
\bea i\mathcal L &=& \bar{\psi}_{\mu_1\cdots\mu_n}\not{\!\partial\!}\;\psi^{\mu_1\cdots\mu_n}+n\bar{\displaystyle\not{\!\!\psi}}_{\mu_2\cdots\mu_n}\not{\!\partial\!}\,
\displaystyle\not{\!\!\psi}^{\mu_2\cdots\mu_n}-\tfrac{1}{4}n(n-1)\bar{\psi}'_{\mu_3\cdots\mu_n}\not{\!\partial\!}\;\psi'^{\,\mu_3\cdots\mu_n}\nonumber\\&&
-\left[\left(n\bar{\displaystyle\not{\!\!\psi}}_{\mu_2...\mu_n}\partial\cdot\psi^{\mu_2\cdots\mu_n}-\tfrac{1}{2}n(n-1)\bar{\psi}'_{\mu_3\cdots\mu_n}
\partial\cdot\displaystyle\not{\!\!\psi}^{\,\mu_3\cdots\mu_n}\right)-\text{h.c.}\right].\label{f00}\eea\vspace{-17px}\end{svgraybox}

\bex Show that the fermionic Lagrangian enjoys the gauge symmetry~(\ref{f01}).\eex
The EoMs resulting from the Fronsdal Lagrangian~(\ref{f00}) read
\beq \mathcal S_{\mu_1\cdots\mu_n}-\tfrac{1}{2}\gamma_{(\mu_1}\displaystyle\not{\!\!\mathcal S}_{\mu_2\cdots\mu_n)}-\tfrac{1}{2}
\eta_{(\mu_1\mu_2}\mathcal S^\prime_{\mu_3\cdots\mu_n)}=0,\label{noname44}\eeq
which clearly has the equivalent form of the fermionic Fronsdal equation~(\ref{noname46}).
\vspace{10px}

Naturally, the Fronsdal Lagrangians~(\ref{b00}) and~(\ref{f00}) share with their resulting EoMs the same unappealing feature:
the presence of algebraic (trace) constraints on the fields and gauge parameters, already discussed in Section \ref{sec:BWmassless}. The ways to avoid this at the Lagrangian level include, among others, the BRST approach
\cite{Pashnev:1997rm,Pashnev:1998ti,Burdik:2001hj,Burdik:2000kj,Buchbinder:2004gp,Buchbinder:2005ua} and the non-local formulation \cite{Francia:2005bu,Francia:2007qt,Francia:2008hd,Francia:2010qp}.

\subsubsection{Massive Lagrangian from Kaluza-Klein Reduction:}

We have seen in Section \ref{sec:Wigner} that the Wigner little groups for the massless and massive UIRs differ by a unit shift in space-time dimensionality. This fact has a number of important implications. One is the match between the respective DoF counts modulo this shift, as we already noticed in Section \ref{sec:BW}. Another one, which we discuss below, is that the flat-space free Lagrangian for a massive field of mass $m$ and spin $s$ in $d$ dimensions can be obtained by starting with the Lagrangian of a massless spin-$s$ field in $(d+1)$-dimensional Minkowski space, and performing a Kaluza-Klein reduction with a single dimension compactified on a circle of radius $1/m$.

We will clarify this point with the example of a spin-3 field. We start with the spin-3 Fronsdal Lagrangian in $d+1$ dimensions:
\beq \mathcal L_{d+1}=-\tfrac{1}{2}(\partial_Q\vf_{MNP})^2+\tfrac{3}{2}(\partial\cdot\vf_{MN})^2+3\vf'_M\partial\cdot\partial\cdot\vf^{M}
+\tfrac{3}{2}(\partial_M \vf'_N)^2 +\tfrac{3}{4}(\partial^M \vf'_M)^2,\label{t1}\eeq
where $\vf'_N=\vf^M{}_{MN}$. The above Lagrangian has the gauge symmetry:
\beq \delta \vf_{MNP}=\partial_{(M}\l_{NP)},\qquad\l^M{}_M=0.\label{t2}\eeq
As we will see, the tracelessness condition on the gauge parameter has important consequences. Now let us perform a Kaluza-Klein
(KK) reduction by writing
\beq \vf_{MNP}(x^\mu,y)=\left(\frac{m}{2\pi}\right)^{\tfrac12} \frac{1}{\sqrt{2}}\left\{\F_{MNP}(x^\mu)e^{imy}+
\text{c.c.}\right\},\label{t3}\eeq
where the Greek index runs as $0,1,\cdots,d-1$, and the $y$-dimension is compactified on a circle of radius $1/m$.
In $d$ space-time dimensions this procedure gives rise to a quartet: a spin-3 field $\F_{\mu\nu\rho}$, a spin-2
field $W_{\mu\nu}\equiv-i\F_{\mu\nu y}$, a vector field $B_\mu\equiv-\F_{\mu yy}$, and a scalar $\r\equiv i\F_{yyy}$.
We also reduce the gauge parameter $\l_{MN}$ as
\beq \l_{MN}(x^\mu,y) = \left(\frac{m}{2\pi}\right)^{\tfrac12} \frac{1}{\sqrt{2}}\left\{\L_{MN}(x^\mu)e^{imy}
+ \text{c.c.}\right\},\label{t4}\eeq
so that in $d$ dimensions we have three gauge parameters: $\L_{\mu\nu}$, $\L_\mu\equiv-i\L_{\mu y}$, and $\L\equiv-\L_{yy}$.
The higher-dimensional gauge invariance~(\ref{t2}) translates itself in lower dimension into the St\"{u}ckelberg symmetry:
\beq \begin{aligned}
\delta \F_{\mu\nu\rho}&=\partial_{(\mu}\L_{\nu\rho)},\\
\delta W_{\mu\nu}&=\partial_{(\mu}\L_{\nu)}+m\L_{\mu\nu},\\
\delta B_{\mu}&=\partial_{\mu}\L + 2m\L_{\mu},\\
\delta\r&=3m\L.\end{aligned}\label{t8}\eeq
The tracelessness of the higher-dimensional gauge parameter gives rise to the following relation among the $d$-dimensional
gauge parameters: \beq \L^\mu{}_\mu=\L.\label{t9}\eeq

The $d$-dimensional action is obtained by integrating out the compact dimension $y$ in the original action.
We can gauge-fix the KK-reduced Lagrangian by setting to zero the traceless part of the spin-2 field $W_{\mu\nu}$, the vector field
$B_\mu$, and the scalar $\r$. Note that, given the relation~(\ref{t9}), it is possible to set only the traceless part $W_{\mu\nu}$
to zero. The gauge-fixed Lagrangian, which describes a massive spin-3 field, therefore unavoidably contains
an auxiliary scalar field $W'$$-$the trace $W_{\mu\nu}$. We get
\beq\begin{aligned} \mathcal L_d &=-\tfrac{1}{2}(\partial_\sigma\F_{\mu\nu\rho})^2 + \tfrac{3}{2} (\partial\cdot\F^{\mu\nu})^2
+\tfrac{3}{4}(\partial\cdot\F')^2 + \tfrac{3}{2}(\partial_\mu\F'_\nu)^2 + 3\F'_\m\partial\cdot\partial\cdot\F^{\mu}
\\[3pt]&-\tfrac{1}{2}m^2\left(\F_{\mu\nu\rho}^2-3\F^{\prime\,2}_\mu-\tfrac92W^{\prime\,2}\right)+\tfrac{3(d-2)}{2d}
\left[\left(1-\tfrac{1}{d}\right)(\partial_\mu W')^2-mW'\partial\cdot\F'\right],\end{aligned}\label{t10}\eeq
where $\F'_\mu=\F^\n{}_{\n\mu}$ is the trace of the spin-3 field. This is the Lagrangian for a massive spin-3 field with minimal number
of auxiliary fields; it is equivalent to the Singh-Hagen spin-3 Lagrangian \cite{Singh:1974qz} up to some field redefinitions.
\bex Derive the Euler-Lagrange equations from Eq.~(\ref{t10}). Show that they lead to the on-shell vanishing of the auxiliary
scalar: $W'=0$, and the Fierz-Pauli conditions for the spin-3 field: $\left(\Box-m^2\right)\F_{\mu\nu\rho}=0,\,\de\cdot\F_{\m\n}=0$
and $\F'_\m=0$.\eex

\section{Higher Spins in Anti-de Sitter Space}\label{sec:AdS}

Anti-de Sitter space plays a very important role for HS theories, and we already emphasized this point in the Introduction.
This section is devoted the study of free HS fields on AdS. Section \ref{sec:AdSGeometry} gives a lightening
review of the AdS geometry, Section \ref{sec:AdSisometry} is a brief account of the AdS isometry algebra and
its unitary irreducible representations, and Section \ref{sec:BFbound} presents the unitarity bound on the
lowest-energy eigenvalue and the wave equations for totally symmetric representations. We conclude the section
with some remarks, particularly on the singleton representations.

\subsection{The AdS Geometry}\label{sec:AdSGeometry}

A $d$-dimensional anti-de Sitter space $\text{AdS}_d$ is a solution of the Einstein equations with a negative
cosmological constant $\L$: \be G_{\mm\nnn}=-\L g_{\mm\nnn},\qquad \L=-\frac{(d-1)(d-2)}{2L^2}<0.\ee
The length scale $L$ is called the radius of the AdS space. It is a maximally symmetric space, with the Riemann tensor given by
\be R_{\mm\nnn\rr\sss}=-\frac{1}{L^2}\left(g_{\mm\rr}\,g_{\nnn\sss}-g_{\mm\sss}\,g_{\nnn\rr}\right).\ee
It is therefore a conformally flat Einstein manifold with a negative scalar curvature:
\be W_{\mm\nnn\rr\sss}=0,\qquad R_{\mm\nnn}=-\frac{d-1}{L^2}\,g_{\mm\nnn},\qquad R=-\frac{d(d-1)}{L^2}=\frac{2d\L}{d-2}<0.\ee

$\text{AdS}_d$ can be viewed as the $d$-dimensional hyperboloid
\be X_0^2+X_d^2-\sum_{i=1}^{d-1}X_i^2=L^2,\label{hyp}\ee                                                         %
in a $(d+1)$-dimensional flat space with signature $(-\,+\,\cdots\,+\,-)$, called the \textit{ambient space}, with the metric
\be ds^2=-dX_0^2-dX_d^2+\sum_{i=1}^{d-1}dX_i^2.\label{hyp100}\ee Therefore, by construction AdS space is homogeneous and isotropic,
and has the isometry group $SO(d-1,\,2)$.
\bex Check that the hypersurface \eqref{hyp} has the parametric equations \be X_0=L\cosh\r\cos\t,\qquad X_d=L\cosh\r\sin\t,
\qquad X_i=L\sinh\r\,\O_i,\label{hyp1}\ee where $\O_i$'s, with $i=1,\cdots,d-1$, satisfy $\sum\O_i^2=1$ and parameterize the
unit sphere $S^{d-2}$. Show that the $\text{AdS}_d$ metric in these coordinates reduces to                       %
\be ds^2=L^2\left(-\cosh^2\r\,d\t^2+d\r^2+\sinh^2\r\,d\O^2\right),\ee\eex
where $d\O^2$ is the line element on the unit sphere $S^{d-2}$.
$\left(\t,\,\r,\,\O_i\right)$ are called the \textit{global coordinates} of AdS since the solution \eqref{hyp1}, with $\r\geq0$ and $0\leq\t<2\p$,
covers the entire hyperboloid once. Near $\r=0$, the metric behaves like $ds^2\simeq L^2\left(-d\t^2+d\r^2+\r^2\,d\O^2\right)$, so the hyperboloid
has the topology of $S^1\times\mathbb R^{d-1}$. The circle $S^1$ represents closed timelike curves in $\t$-direction. One can simply unwrap the circle,
i.e., take $-\infty<\t<\infty$ with no identifications, and obtain a causal spacetime with no closed time-like curves. The result is a universal covering
space of the hyperboloid, which from now on will be identified as $\text{AdS}_d$.

Another set of coordinates, called the \textit{Poincar\'e coordinates}, is particularly useful in the context of holography. It is defined as:
\bea &X_0=\frac{1}{2z}\left[L^2-t^2+\vec{x}^2+z^2\right],&\nonumber\\&X_{d-1}=\frac{1}{2z}\left[L^2+t^2-\vec{x}^2-z^2\right],&                                      \label{poincarecoord}\\&\left(X_d,\left(X_1,\cdots,X_{d-2}\right)\right)=\frac{L}{z}\left(t,\,\vec{x}\right).&\nonumber\eea
These coordinates make manifest the conformal flatness of AdS space:
\beq ds^2=\frac{L^2}{z^2}\left(dz^2-dt^2+d\vec{x}^2\right),\qquad z\in[0,\,L],~~\text{and}~~(t,\,\vec{x})\in\mathbb R^{d-1}.\eeq
%

\subsection{Isometry Group \& Unitary Representations}\label{sec:AdSisometry}

The isometry group $SO(d-1,\,2)$ of AdS$_d$ has a total of $\tfrac{1}{2}d(d+1)$ many generators $M_{AB}=-M_{BA}$, which obey the commutation relation:
\be [M_{AB},\,M_{CD}]=i\left(\h_{AC}\,M_{BD}-\h_{BC}\,M_{AD}-\h_{AD}\,M_{BC}+\h_{BD}\,M_{AC}\right),\label{algebra}\ee
where $\h_{AB}=\text{diag}(-\,+\,\cdots\,+\,-)$ is the flat metric in the ambient space, and $A, B, C, D=0, 1,\cdots,d$. Since are interested in the
lowest weight \underline{unitary} modules of the isometry algebra $o(d-1,\,2)$ for classifying physically meaningful relativistic fields, the generators
are assumed to be Hermitian: $M_{AB}^\dagger=M_{AB}$. Because the group is non-compact, its unitary modules will necessarily be infinite dimensional.

It is convenient to choose the following basis in the algebra:
\be H=M_{0d},\qquad J_{ij}=iM_{ij},\qquad J_i^\pm=M_{0i}\pm iM_{id},\label{gen}\ee where $i, j=1,\cdots,d-1$.
The Hermiticity condition then gives
\be H^\dagger=H,\qquad J_{ij}^\dagger=-J_{ij},\qquad \left(J_i^\pm\right)^\dagger=J_i^\mp.\label{gen101}\ee
\bex Show that in this basis the commutation relations \eqref{algebra} take the form
\begin{align} [H,\,J_i^\pm]&=\pm J_i^\pm,\label{comm1}\\ [J_i^-,\,J_j^+]&=2(H\d_{ij}-L_{ij}),\label{comm2}\\
[J_{ij},\,J^\pm_k]&=\d_{kj}J_i^\pm-\d_{ki}J_j^\pm,\label{comm3}\\
[J_{ij},\,J_{kl}]&=\d_{jk}J_{il}-\d_{ik}J_{jl}-\d_{jl}J_{ik}+\d_{il}J_{jk},\label{comm4}\end{align}
while the other commutators are all zero.\eex
The $\text{AdS}_d$ isometry group has a maximal compact subgroup $SO(2)\times SO(d-1)$. The compact generators are $H$ and
$J_{ij}$, identified respectively with the energy and angular momenta. The former produces translations in $\t$ (this is called the
\textit{global time coordinate} of AdS, since the time-like Killing vector $\partial_\t$ has a non-vanishing norm everywhere), while
the latter gives rotations in $S^{d-2}$. The remaining non-compact generators $J^\pm_i$ combine AdS translations and Lorentz boosts.
It is clear from the commutation relation \eqref{comm1} that $J_i^+$ and $J_i^-$, respectively, raise and lower by one unit the
energy eigenvalue of the state they act on. Therefore, they are called respectively the \textit{energy boost} and \textit{energy deboost} operators.

To construct lowest weight UIRs, one starts with the vacuum space $\vac$, which forms a unitary module of the maximal compact
subalgebra $o(2)\oplus o(d-1)$. $E_0$ is the lowest eigenvalue of the energy operator:
\be H\vac=E_0\vac,\qquad J^-_i\vac=0,\label{gen102}\ee
while $\mathbf s$ is the generalized spin characterizing the $o(d-1)$-module. The latter is given by a set of $r$ numbers,
$\mathbf{s}=\left(s_1,\cdots,s_r\right)$, for a Young tableau of $r$ rows with $s_i$ cells in the $i$-th row.
For a totally symmetric spin-$s$ representation, in particular, it reduces to the form: $\mathbf{s}=(s,0,\cdots,0)$.

Then, one builds the representation Fock space by acting with the boost operators $J^+_i$ on the vacuum.
The full $o(d-1,\,2)$-module, conventionally denoted as $\rep$, is therefore spanned by vectors of the form
\be J^+_{i_1}\cdots J^+_{i_n}\vac,\qquad n=0,1,\dots\,.\label{gen103}\ee
The states with a given $n$, called \textit{level-$n$ states}, have the energy eigenvalue $E_0+n$.
They constitute a finite-dimensional subspace for any $n$.
\bex Show that any level-$n$ state is orthogonal to any level-$(n\pm2)$ state.\eex
While the vacuum states are assumed to be orthonormal, the norm of any level-$n$ state can be calculated by using the commutation
relations \eqref{comm1}--\eqref{comm4} and the properties \eqref{gen102} of the vacuum. The positivity of the norms at any level
guarantees that the representation will be unitary. This requirement leads to a lower bound on $E_0$, which we  will derive in
Section~\ref{sec:BFbound}. A slick way of understanding the existence of a lower bound is to compute the norm
of the level-1 state $J^+_i\vac$. The norm is $2(d-1)E_0\langle E_0,\mathbf{s}\vac$, which implies that the level-1 state
cannot have a positive norm for a positive-definite vacuum subspace if $E_0$ is negative.

\subsection{Unitarity Bound, Masslessness \& Wave Equations}\label{sec:BFbound}
It is clear that the unitarity region is bounded from below by a positive value
\beq E_0\geq E_0(\mathbf s),~~\text{with}~~E_0(\mathbf s)>0.\label{BF1}\eeq
Below this bound, where $E_0<E_0(\mathbf s)$, some states acquire negative norm and should therefore be excluded from the
physical spectrum. At the boundary of the unitarity region $E_0=E_0(\mathbf s)$, some zero-norm states will appear.
These states have vanishing scalar product with any other state.
\bex Prove the above statement by showing that if the scalar product is non-zero, it is always possible to build a negative-norm state.
This is in contradiction with the assumption of being at the boundary of the unitarity region.\eex
The zero-norm states therefore form an invariant submodule, which can be factored out so that one is left with a ``shorter'' unitary
representation. This is the phenomenon of \textit{multiplet shortening}. The ``short'' unitary representation corresponds either to
a \textit{massless field} or to a novel kind of representation called the \textit{singleton} (see Remarks at the end of this section).
In the case of massless fields, multiplet shortening can be interpreted as enhancement of gauge symmetry. Note, on the other hand,
that above the bound $E_0(\mathbf s)$ unitary representations correspond to \textit{massive fields} on AdS space.

In what follows we will focus only on totally symmetric representations with $\mathbf{s}=(s,0,\cdots,0)$. To derive the unitarity
bound, let us first note that the quadratic Casimir operator of the $o(d-1,2)$ algebra is given by
\beq \mathcal C_2~\equiv~\tfrac12M_{AB}M^{AB}~=~E(E-d+1)-\tfrac12J_{ij}J^{ij}-\d^{ij}J^+_iJ^-_j.\label{ads1}\eeq
\bex Find the eigenvalue of $\mathcal C_2$ for a totally symmetric representation by using the eigenvalue
equation:~$J_{ij}J^{ij}|E_0,s\rangle=-2s(s+d-3)|E_0,s\rangle$. The result is
\beq \langle\mathcal C_2\rangle~=~E_0(E_0-d+1)+s(s+d-3).\label{ads2}\eeq\eex

It is convenient to make a rescaling of the ambient-space coordinates: $y^A=L^{-1}X^A$, so that the hyperboloid that defines
AdS$_d$ is given by
$\h_{AB}y^Ay^B=-1$. Now, let us split the $o(d-1,2)$ generators into orbital part and spin part as follows:
\beq M^{AB}=-i\,y^{[\,A}\mathcal{D}^{B\,]}+\S^{AB},\qquad \mathcal{D}^A\equiv\left(\h^{AB}+y^Ay^B\right)\frac{\partial}
{\partial y^B}\,,\label{ads3}\eeq where $\S^{AB}$ is the spin operator, and the tangent derivative $\mathcal{D}^A$ has the
properties:
\beq y\cdot\mathcal{D}=0,\quad\mathcal{D}\cdot y=d,\quad\left[\mathcal{D}^A,\,y^B\right]=\h^{AB}+y^Ay^B,\quad
\left[\mathcal{D}^A,\,\mathcal{D}^B\right]=-y^{[\,A}\mathcal{D}^{B\,]}.\label{ads3.1}\eeq
The form of $\S^{AB}$ depends on the realization of the representation.
\bex Use the splitting \eqref{ads3} and the properties of the tangent derivative to show that the quadratic
Casimir $\mathcal C_2$ can be written as:
\beq \mathcal C_2=\mathcal{D}^2-2i\,y_A\mathcal{D}_B\S^{AB}+\tfrac12\S_{AB}\S^{AB},
\qquad \mathcal{D}^2\equiv\mathcal{D}_A\mathcal{D}^A.\label{ads6}\eeq
\eex

In the tensor realization, we use as a carrier of $D(E_0,s)$ a totally symmetric \underline{traceless}
$o(d-1,2)$-tensor field $\F_{A_1\cdots\,A_s}$ defined on the hyperboloid \eqref{hyp}.
It is useful to introduce an auxiliary Fock space and work with the generating function:
\beq |\F\rangle=\F_{A_1\cdots\,A_s}\a^{+A_1}\cdots\a^{+A_s}|0\rangle,\label{ads3.2}\eeq
where $|0\rangle$ is the Fock vacuum, and the creation and annihilation operators satisfy
\beq\left[\,\a_A,\,\a^+_B\,\right]=\h_{AB},\qquad \a_A|0\rangle=0.\label{ads3.3}\eeq
Now, the spin operator $\S_{AB}$ takes the form:
\beq \S_{AB}=-i\,\a^+_{[\,A}\a_{B\,]}.\label{ads5}\eeq
\bex Use Eqs.~\eqref{ads6}--\eqref{ads5} to show that one can write
\beq \mathcal C_2|\F\rangle=\left(\,\mathcal{D}^2+s(s+d-3)\,\right)|\F\rangle.\label{ads7}\eeq\eex

Because $|\F\rangle$ is a carrier of $D(E_0,s)$, it obeys the equation:
\beq \left(\,\mathcal C_2-\langle\mathcal C_2\rangle\,\right)\,|\F\rangle=0,\label{ads8}\eeq
with $\langle\mathcal C_2\rangle$ given by Eq.~\eqref{ads2}. By using Eq.~\eqref{ads7}, we can rewrite
this equation as
\beq \left(\,\mathcal{D}^2-E_0(E_0-d+1)\,\right)|\F\rangle=0.\label{ads8.0}\eeq
In addition, $|\F\rangle$ will satisfy the following subsidiary conditions:
\beq \a\cdot\mathcal{D}\,|\F\rangle=0,\qquad y\cdot\a\,|\F\rangle=0.\label{ads8.1}\eeq
The first one is an $SO(d-1,2)$ analog of the usual zero divergence condition, while the second one ensures
that the tensor $\F_{A_1\cdots\,A_s}$, when reduced to the Lorentz subgroup $SO(d-1,1)$, gives rise to a single
tensor of the same rank.

Next, we would like to reformulate Eqs.~\eqref{ads8.0} and \eqref{ads8.1} in the intrinsic AdS$_d$ coordinates.
We set the AdS radius to \underline{unity}: $L=1$. Let the intrinsic AdS$_d$ coordinates be $x^\mm$~with~$\mm=0,1,\cdots,d-1$.
The induced AdS metric is given by
\beq g_{\mm\nnn}=\h_{AB}\,\de_\mm\,y^A \de_\nnn\,y^B,\label{ads9}\eeq
where $y^A=y^A(x)$ is the embedding map, with $y^Ay_A=-1$. The inverse metric and the Christoffel connection,
on the other hand, are given by
\beq g^{\mm\nnn}=\mathcal{D}_A\,x^\mm\,\mathcal{D}^A\,x^\nnn\,,\qquad \G^\rr_{\mm\nnn}=\left(\d^A_B+y^Ay_B\right)
\frac{\de x^\rr}{\de y^A}\frac{\de^2y^B}{\de x^\mm\,\de x^\nnn}\,,\label{ads9.1}\eeq
where $x^\mm=x^{\mm}(y)$ is some representation of the intrinsic coordinates. By making use of the AdS vielbein
$E^\mm_A\equiv g^{\mm\nnn}\,\de_\nnn\,y_A$, it is possible to obtain a tensor on AdS$_d$ from the $o(d-1,2)$-tensor
as follows:
\beq \vf^{\mm_1\cdots\mm_s}(x)=E^{\mm_1}_{A_1}\cdots E^{\mm_s}_{A_s}\F^{A_1\cdots A_s}(y).\label{ads9.2}\eeq
Likewise, we obtain an auxiliary set of creation and annihilation operators on AdS$_d$:
\beq\left(a^\mm,\,a^{+\mm}\right)=E^\mm_A\left(\a^A,\,\a^{+A}\right),\qquad\left[a^\mm,\,a^{+\nnn}\right]
=g^{\mm\nnn}.\label{ads9.3}\eeq
\bex If $\nabla_\mm=\de_\mm+\G^{\cdot}_{\mm\,\cdot}$ is the AdS covariant derivative, prove that
\beq E^{\mm_1}_{A_1}\cdots E^{\mm_s}_{A_s}\,\mathcal{D}^2\F^{A_1\cdots A_s}(y)=\left(\nabla^2+s\right)
\vf^{\mm_1\cdots\mm_s}(x),\qquad \nabla^2\equiv\nabla_\mm\nabla^\mm.\label{ads9.4}\eeq\eex
Let us denote
\beq|\vf\rangle=\vf^{\mm_1\cdots\mm_s}a^+_{\mm_1}\cdots a^+_{\mm_s}|0\rangle.\label{ads9.5}\eeq
Then, under the desired reformulation, Eq.~\eqref{ads8.0} and \eqref{ads8.1} translate into
\bea &\left(\,\nabla^2-E_0(E_0-d+1)+s\,\right)|\vf\rangle=0,&\label{ads10.1}\\[3pt]
&a\cdot\nabla\,|\vf\rangle=0,&\label{ads10.2}\eea
while the irreducibility (tracelessness) condition can be written as:
\be a\cdot a\,|\vf\rangle=0.\label{ads10.3}\ee

Let us recall that the phenomenon of multiplet shortening is associated with the appearance of gauge invariance for
$s\geq1$. In other words, when the quantity $E_0$ in Eq.~\eqref{ads10.1} takes its lowest value $E_0(s)$, the
system of equations \eqref{ads10.1}--\eqref{ads10.3} acquires a gauge invariance of the form:
\beq \delta|\vf\rangle=a^+\cdot\nabla|\l\rangle,\qquad |\l\rangle=\l^{\mm_1\cdots\mm_{s-1}}a^+_{\mm_1}\cdots a^+_{\mm_{s-1}}|0\rangle,\label{ads11}\eeq
where the symmetric traceless rank-$(s-1)$ gauge parameter is \underline{on-shell}:
\beq \left(\,\nabla^2-\m^2\,\right)|\l\rangle=0,\qquad a\cdot\nabla\,|\l\rangle=0,\qquad a\cdot a\,|\l\rangle=0,
\label{ads12}\eeq
with $\m^2$ a parameter to be determined.
\bex Show that the gauge transformation \eqref{ads11} of the system of equations \eqref{ads10.1}--\eqref{ads10.3}
is compatible with the on-shell conditions \eqref{ads12} provided that
\bea &\m^2=(s-1)(s+d-3),&\label{ads13.1}\\&E_0(s)=s+d-3.&\label{ads13.2}\eea
Hint: The following relations, which the reader should derive first, will be useful.
\bea \left[\,a\cdot\nabla,\,a^+\cdot\nabla\,\right]&=&a^+\cdot\;a^+a\cdot a-\left(a^+\cdot a+d-2\right)a^+\cdot a,
\nonumber\\\left[\,\nabla^2,\,a^+\cdot\nabla\,\right]&=&-a^+\cdot\nabla\left(2a^+\cdot a+d-1\right)+2a^+\cdot a^+
a\cdot\nabla.\nonumber\eea\eex

Eq.~\eqref{ads13.2} is our desired result, which gives the unitarity bound on the lowest-energy eigenvalue of a totally
symmetric spin-$s$ representation on AdS:
\begin{svgraybox}\vspace{-7px}\beq E_0\geq s+d-3,\qquad\text{for}~s\geq1.\label{ads14}\eeq\vspace{-17px}\end{svgraybox}
The lower bound is saturated by the massless representation, whose mass $M_0$ can easily be computed from Eq.~\eqref{ads10.1}.
Reintroducing the AdS radius $L$, we can write:
\begin{svgraybox}\vspace{-10px}\beq M_0^2\,L^2=s^2+s(d-6)-2(d-3),\qquad(\text{Mass of massless UIR}).\label{ads15}
\eeq\vspace{-20px}
\end{svgraybox}
Finally, one can rewrite Eqs.~\eqref{ads10.1}--\eqref{ads10.3} in the tensor language with a mass parameter $M^2\geq M_0^2$\,.
The resulting set of equations is a direct analog of the Fierz-Pauli conditions in flat space. Given that the contraction of indices is
made with the AdS metric $g^{\mm\nnn}$, the equations for a totally symmetric massive field on AdS read:
\begin{svgraybox}\vspace{-17px}\bea (\nabla^2-M^2)\vf_{\mm_1\cdots\mm_s}&=&0,\label{adskgb}\\
\nabla\cdot\vf_{\mm_1\cdots\mm_{s-1}}&=&0,\label{adsdivb}\\
\vf^\prime_{\mm_1\cdots\mm_{s-2}}&=&0.\label{adstrb}\eea\vspace{-22px}\end{svgraybox}

\paragraph{Remarks:}
\begin{itemize}\setlength\itemsep{1em}
\item \underline{Lower Spin Case}: The unitarity bound \eqref{ads14} relies on the appearance of gauge symmetries,
and therefore does not apply to lower-spin fields with $s<1$. For these fields, the bound can be derived in
a different way. It is given by:
\be E_0\geq s+\tfrac{1}{2}\left(d-3\right),\qquad\text{for}~s=0,\,\tfrac{1}{2}\,.\ee
For $s=0$, this is the well-known Breitenlohner-Freedman bound \cite{Breitenlohner:1982bm}.
One could find the $d=4$ derivation of this bound, for both $s=0$ and $\tfrac{1}{2}$, in
Ref.~\cite{Nicolai:1984hb}.
\item \underline{Fermions \& Mixed-Symmetry Case}: Similar considerations hold for more general representations of the isometry group
associated with mixed-symmetry Young tableaux or fermions. The corresponding analysis is rather technical; the details can be found
in the literature, for example, in Refs.~\cite{Metsaev:1995re,Metsaev:1997nj,Brink:2000ag,Campoleoni:2008jq,Skvortsov:2008vs,
Campoleoni:2009gs,Campoleoni:2012th,Vasiliev:2012tv}.
\item \underline{Lagrangian Formulation}: The Lagrangian formulation of HS theories in AdS was developed in parallel with that
in flat space (see e.g. \cite{Alkalaev:2003qv,Alkalaev:2003hc,Alkalaev:2009vm,Alkalaev:2011zv}). The Fronsdal action can be extended to AdS
by requiring gauge invariance for massless representations.
The radial reduction trick has played here a key role (see e.g. \cite{Francia:2008hd,Grigoriev:2011gp}). For totally symmetric
massive fields in AdS$_d$, covariant actions were written down in Ref.~\cite{Zinoviev:2001dt}.
\item \underline{The Singletons}: The discussion we carried out for gauge fields involved \underline{a} possible definition of masslessness in AdS background.
In general, however, the notion of masslessness beyond flat space needs to be reconsidered. In AdS space we can indeed distinguish
at least two notions of masslessness: ``composite'' masslessness which we have discussed above, and ``conformal'' masslessness. Both cases correspond
to the appearance of singular vectors. Remarkably, these two concepts coincide only in $d=3,\,4$. The latter definition is tantamount to the possibility
of uplifting a given UIR to a representation of the full conformal group: $o(d-1,2)\rightarrow o(d,2)$. Conversely, it is also possible to require that
irreducibility of the conformal module be preserved upon the reduction $o(d+2)\rightarrow o(d+1)$ in any signature. In this case, the corresponding UIR
is called a singleton. The former definition of masslessness, on the other hand, characterizes massless particles as composite objects in terms of singleton
representations. The only representations of $o(d-1,2)$ which admit an uplift to UIRs of $o(d,2)$ in any dimension are the scalar and the spin-$\tfrac{1}{2}$
fermion. In even dimensions, though, they also include the field strengths described by maximal window diagrams of the type (see e.g. \cite{Bekaert:2009fg,Bekaert:2013zya}):
\be
R_{\mm_1(s),\,\cdots,\,\mm_{d/2}(s)}(x)~
\sim~\tfrac{d}{2}
\left\{\vphantom{\begin{matrix}
a\\[3pt]
b\\[3pt]
c
\end{matrix}}\right.
\underbrace{\begin{aligned}
&\begin{tabular}{|c|c|c|c|c|}\hline
   $\phantom{a1}$&$\phantom{a1}$&\multicolumn{2}{|c|}{$~~~~\cdots~~~~\cdots~~~~\cdots~~~~$}&\phantom{a1}\\\hline
\end{tabular}\\[-4pt]
&\begin{tabular}{|c|}
   $~~~~\vdots~~~~\vdots~~~~\vdots~~~~~\vdots\;~\;~~\;\vdots~~~~\vdots~~~~\vdots~~~~\vdots~~~~$\\
\end{tabular}\\[-4pt]
&\begin{tabular}{|c|c|c|c|c|}\hline
   $\phantom{a1}$&$\phantom{a1}$&\multicolumn{2}{|c|}{$~~~~\cdots~~~~\cdots~~~~\cdots~~~~$}&\phantom{a1}\\\hline
\end{tabular}\\
\end{aligned}}_s,\ee
where $\mm_i$ are $o(d-1,1)$ indices. These field strengths may also satisfy appropriate self-duality
conditions \cite{Govil:2013uta,Govil:2014uwa,Fernando:2015tiu}, with
\be E_0=s+\tfrac{1}{2}d-1\,.\ee
Notice, however, that the uplift to a conformal UIR holds at the level of the Bargmann-Wigner equations in terms of
field-strengths but fails at the gauge-field level, unless the gauge module is trivial ($s=1$). This fact can be appreciated
by noticing that Fronsdal equations in 4d are not conformal \cite{Barnich:2015tma}.
\end{itemize}

\newpage
\section{Beyond Free Theories}\label{sec:Beyond}

Interacting theories of HS fields are generically fraught with difficulties. For massless fields in flat space, interactions are in tension with gauge invariance,
and this leads to various no-go theorems. There are a number of ways to evade these theorems and construct interacting HS gauge theories. Interacting massive
fields, on the other hand, may exhibit superluminal propagation in a non-trivial background. This pathology can be cured by appropriate non-minimal couplings
and/or additional dynamical DoFs. String Theory provides such a remedy for massive HS fields. In this section,
we review these issues of interacting HS fields and their resolutions.

\subsection{Massless HS Fields and No-Go Theorems}\label{sec:no-go}

In Nature we do not observe any massless HS particles. Nor do we know of any String Theory compactification
that gives rise to a Minkowski space wherein
massless particles with $s>2$ exist. In fact, interactions of HS gauge fields in flat space are severely constrained by powerful no-go theorems
\cite{Weinberg:1964ew,Grisaru:1977kk,Grisaru:1976vm,Aragone:1979hx,Weinberg:1980kq,Porrati:2008rm,Porrati:2012rd,Coleman:1967ad}.
Below we give a brief account of these important theorems (see also Ref.~\cite{Bekaert:2010hw} for a nice review).

\subsubsection*{Weinberg (1964)}

There are obstructions to consistent long-range interactions mediated by massless bosonic fields with $s>2$, and this can be understood in a purely $S$-matrix-theoretic
approach \cite{Weinberg:1964ew}. Let us consider the $S$-matrix element of $N$ external particles with momenta $p_i^\mu$, $i=1,...,N$ and a massless spin-$s$ particle
of momentum $q^\mu$. In the soft limit ($q\rightarrow0$) of the spin-$s$ particle, the $S$-matrix element can be expressed in terms of one without the soft particle.
Indeed, the matrix element factorizes as:
\beq S(p_1,\cdots,p_N,q,\epsilon)\rightarrow\sum_{i=1}^N\,g_i\,\left[\frac{p_i^{\mu_1}\cdots p_i^{\mu_s}\epsilon_{\mu_1..\mu_s}(q)}{2q\cdot p_i}\right]
\,S(p_1,\cdots,p_N),\label{2}\eeq
where $\epsilon_{\mu_1\cdots\mu_s}(q)$ is the soft particle's transverse and traceless polarization tensor:
\beq q\cdot\epsilon_{\mu_1\cdots\mu_{s-1}}(q)=0, \qquad \epsilon'_{\mu_1\cdots\mu_{s-2}}(q)=0.\label{3}\eeq
\begin{figure}[ht]
\begin{minipage}[b]{0.47\linewidth}
\centering
\includegraphics[width=1\linewidth,height=1.3\linewidth]{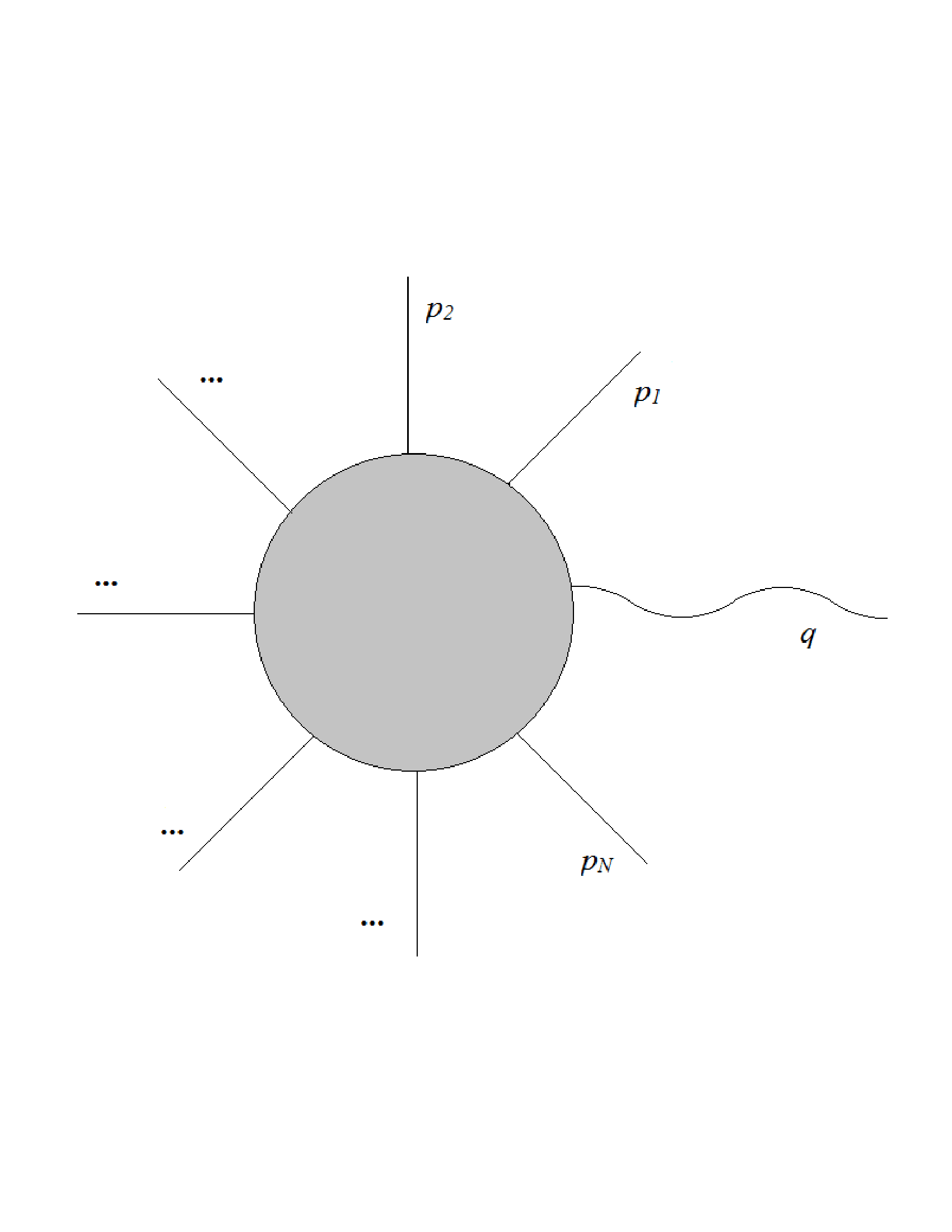}
\vspace{-1.5cm}
\caption{\footnotesize Process involving $N$ external particles of momenta $p_i$, $i=1,\cdots,N$ and one massless spin-$s$ particle of momentum $q$.}
\label{fig:figure1}
\end{minipage}
\hspace{0.6cm}
\begin{minipage}[b]{0.47\linewidth}
\centering
\includegraphics[width=1\linewidth,height=1.3\linewidth]{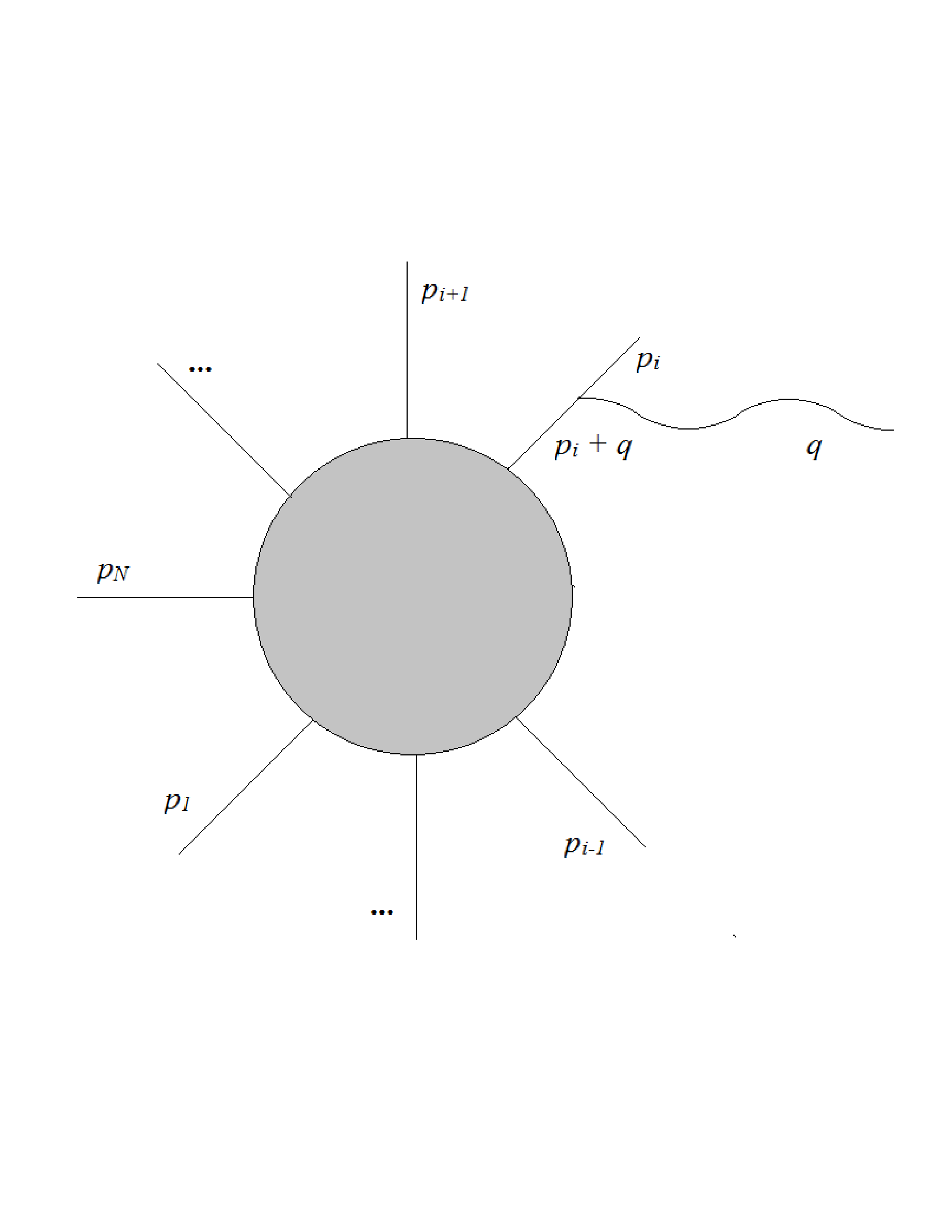}
\vspace{-1.5cm}
\caption{\footnotesize Dominant diagrams for the process when the spin-$s$ particle is soft: $q\rightarrow0$. The soft particle is emitted from an external line.}
\label{fig:figure2}
\end{minipage}
\end{figure}
The polarization tensor is redundant, as it contains more components than the physical polarizations of a massless spin-$s$ particle. This redundancy can be removed
by requiring that the $S$-matrix element vanish for spurious polarizations of the form:
\beq \epsilon^{(\text{spur})}_{\mu_1\cdots\mu_s}(q)\equiv q_{(\mu_1}\l_{\mu_2\cdots\mu_s)}(q),\label{4}\eeq
where $\l_{\mu_1\cdots\mu_{s-1}}$ is also transverse traceless. In order for the spurious polarizations to decouple for generic momenta $p_i^\mu$, given
Eq.~(\ref{2}) one must have
\beq \sum_{i=1}^N g_i p_i^{\mu_1}\cdots p_i^{\mu_{s-1}}=0.\label{5}\eeq
Let us analyse this condition for different values of spin $s$:
\begin{itemize}
\item For $s=1$: The condition requires $\sum g_i=0$, i.e. the conservation of charge.\vspace{5px}
\item For $s=2$: We have $\sum g_ip^\mu_i=0$, and this can be satisfied if (i) $\sum p^\mu_i=0$,
and (ii) $g_i=\kappa=\text{constant}$. The first requirement imposes energy-momentum conservation, while the second demands that all
particles interact with the same strength $\kappa$ with the soft massless spin-2 field (graviton). The latter is simply
the principle of equivalence.\vspace{5px}
\item For $s\geq3$: The condition~(\ref{5}) cannot be fulfilled by generic momenta.
\end{itemize}

Weinberg's argument does not rule out the existence of massless bosons with $s\geq3$, but simply shows that these particles cannot give rise to long-range
interactions. The same argument also applies to half-integer spins \cite{Grisaru:1977kk,Grisaru:1976vm}, and gives a no-go for $s\geq\tfrac{5}{2}$.
These theorems leave open the possibility that HS gauge fields may mediate short-range interactions in flat space.


\subsubsection*{Coleman-Mandula (1967) \& Extension Thereof}

The Coleman-Mandula theorem \cite{Coleman:1967ad} and its supersymmetric extension by Haag, Lopuszanski and Sohnius \cite{Haag:1974qh} pose strong
restrictions on what symmetries the $S$-matrix of an interacting relativistic field theory in 4d flat space may posses. More precisely, under the hypotheses
of non-triviality of the $S$-matrix and finiteness of the spectrum, these theorems show that the maximal extension of the Poincar\'e algebra can only be
the (semi)direct sum of a superalgebra (or a superconformal one in the absence of mass gap) and an internal symmetry algebra that commutes with the Poincar\'e
generators. In particular, they rule out HS conserved charges. Because HS symmetry generators are higher-rank Lorentz tensors, they simply do not commute with
Poincar\'e generators. We shall not go into the details of these rather involved theorems here, and refer the interested reader to the original literature (see also \cite{Fradkin:1986ka,Boulanger:2013zza,Joung:2013nma,Govil:2013uta,Govil:2014uwa,Joung:2014qya,Fernando:2015tiu} for AdS generalizations thereof).

\subsubsection*{Aragone-Deser (1979)}

A byproduct of Weinberg's theorem is that a soft graviton interacts universally with all matter. If a massless HS particle is relevant to our universe, it cannot
avoid gravitational coupling. Therefore, the study of gravitational interactions may completely rule out the existence of such particles in Nature.
Aragone and Deser \cite{Aragone:1979hx} considered the gravitational coupling of a massless spin-$\tfrac{5}{2}$ field, only to discover that the theory is fraught
with grave inconsistencies: The unphysical gauge modes do not decouple unless the theory is free, as we shall demonstrate below.

The theory of a spin-$\tfrac{5}{2}$ gauge field minimally coupled to gravity can be constructed by covariantizing the Fronsdal Lagrangian~(\ref{f00}) for $n=2$.
The result is
\beq i\mathcal L= \sqrt{-g}\,\left[\bar{\psi}_{\mu\nu}\displaystyle{\not{\!\nabla}}\psi^{\mu\nu}
+2\bar{\displaystyle{\not{\!\!\psi}}}_\mu\displaystyle{\not{\!\nabla}}\displaystyle{\not{\!\!\psi}}^\mu
-\tfrac{1}{2}\bar{\psi}'\displaystyle{\not{\!\nabla}}\psi'+\left(\bar{\psi}'\nabla\cdot\displaystyle{
\not{\!\!\psi}}-2\bar{\displaystyle{\not{\!\!\psi}}}_\mu\nabla\cdot\psi^\mu-\text{h.c.}\right)\right],\label{6}\eeq
where the flat-space $\g$-matrices have been replaced by $\G^\mu\equiv e^\m_a\g^a$, with $e^\m_a$ being the vielbein.
The redundancies of the gauge field $\psi_{\mu\nu}$ ought to be eliminated by a gauge invariance of the form:
\beq \delta\psi_{\mu\nu}=\nabla_{(\mu}\epsilon_{\nu)},\qquad\quad \G^\mu\epsilon_\mu=0, \label{8}\eeq
however since the covariant derivatives do not commute, the gauge variation of the Lagrangian leaves one with terms proportional to the Riemann tensor:
\bex Use the commutator formula
\beq [\nabla_\mu, \nabla_\nu]\psi_{\rho\s}=-R_{\mu\nu(\rho}{}^{\l}\psi_{\sigma)\l}+\tfrac{1}{4}R_{\mu\nu\alpha\beta}\G^\a\G^\b\psi_{\rho\s},
\nonumber\eeq to show that the variation of the Lagrangian~(\ref{6}) under the gauge transformation~(\ref{8}) is given, up to a total derivative
and an overall non-zero factor, by \beq \delta\mathcal L\sim\sqrt{-g}\left(i\,\bar{\epsilon}_\mu\G_\nu\psi_{\alpha\beta}R^{\mu\alpha\nu\beta}+
\text{h.c.}\right).\label{9}\eeq\eex

Thus, the Lagrangian is gauge invariant only for a vanishing Riemann tensor. In other words, if the spin-$\tfrac{5}{2}$ gauge field interacts with
gravity, the gauge modes no longer decouple. In contrast to supergravity, here the gauge variation does depend on the off-shell (Weyl) components
of the curvature. Therefore, a ``hypersymmetric'' extension of coordinate invariance with a spin-$\tfrac{3}{2}$ charge is ruled out as a possible
savior. In addition, it is not difficult to see that the addition of local non-minimal couplings, regular in the neighborhood
of flat space, do not alleviate the issue \cite{Aragone:1979hx}. The same problem is expected to show up for any massless field with $s>2$.

This no-go theorem is based on the Lagrangian formulation, and therefore crucially depends on the assumption of locality. It is possible that
some non-locality in the Lagrangian appearing in the guise of gravitational form factors, for example, may remove the difficulties.
This issue can be addressed, again in the language of $S$-matrix, by considering the scattering of a massless HS particle off soft gravitons. This is discussed in the following.

\subsubsection*{Weinberg-Witten (1980) \& Its Generalization}

The Weinberg-Witten theorem \cite{Weinberg:1980kq} states that
``a theory that allows the construction of a conserved Lorentz covariant energy-momentum tensor $\Theta^{\mu\nu}$ for which
$\int \Theta^{0\nu}d^3x$ is the energy-momentum four-vector cannot contain massless particles of spin $s>1$.''
To demonstrate this, let us consider the matrix element for the elastic scattering of a massless spin-$s$ particle off a soft graviton in four dimensions.
Let the initial and final momenta of the spin-$s$ particle respectively be $p$ and $p'$, and the polarizations be identical, say $+s$.
The graviton is off shell with momentum $p'-p$. In the soft limit $p'\rightarrow p$, the matrix element is non-vanishing under the
assumptions stated above, and determined completely by the equivalence principle:
\beq \lim_{p'\rightarrow p} \langle p', +s|\,\Theta_{\mu\nu}\,| p, +s \rangle = p_\mu p_\nu,\label{11}\eeq
where the relativistic normalization of 1-particle states $\langle p| p'\rangle = (2\pi)^32p_0\delta^3(\vec{p} -\vec{p}\,')$ has been used.
The momentum of the off-shell graviton is space-like, so that one can choose the ``brick wall''
frame~\cite{Weinberg:1980kq}, in which $p^\mu= (\,|\vec{p}|,+\vec{p}\,)$ and $p^{\prime\,\mu}=(\,|\vec{p}|,-\vec{p}\,)$.

Now, let us consider a rotation $R(\theta)$ around the direction of $\vec{p}$ by an angle $\theta$. The 1-particle states transform as
\beq R(\theta)|p,+s\rangle = e^{\pm i\theta s}|p,+s\rangle,\qquad\quad R(\theta)|p', +s\rangle = e^{\mp i\theta s}|p',+s\rangle.\label{13}\eeq
The difference of the signs in the exponents arises because $R(\theta)$ amounts to a rotation of $-\theta$ around $\vec{p}'=-\vec{p}$.
Under spatial rotations, the energy-momentum tensor decomposes into a pair of real scalars, a vector and a symmetric traceless rank-2 tensor.
In the standard basis in which the total angular momentum and projection thereof along the $\vec{p}$-direction are the commuting variables,
these fields are represented by spherical tensors. We combine the two real scalars into a complex one, and thereby have three spherical tensors
$\Theta_{l,m}$ with $l=0,1,2$, where $m$ takes integer values in $[-l,+l]$. Then, rotational invariance and the transformations~(\ref{13}) give
\beq \langle p', +s|\,\Theta_{l,m}\,| p, +s \rangle = e^{i\theta(m\pm 2s)}\langle p',+s|\,\Theta_{l,m}\,| p,+s\rangle.\label{14}\eeq
Since $|m|\leq 2$, this equation cannot be satisfied for $s>1$ unless the matrix element vanishes. Because the helicities
are Lorentz invariant and $\Theta_{\mu\nu}$ is a Lorentz tensor, it must vanish in all frames as long as $\left(p'-p\right)^2\neq0$.
However this is in direct contradiction with the non-vanishing result given by Eq.~(\ref{11}).

Note that the Weinberg-Witten theorem does not apply to those theories that lack a conserved Lorentz-covariant energy-momentum tensor:
General Relativity, supergravity, HS gauge theories etc. In these theories gauge invariance of the stress-energy tensor is incompatible
with manifest Lorentz covariance \cite{Deser:2004rr}. The theorem, however, can be generalized to such theories by introducing unphysical
helicity states corresponding to the spurious polarizations \cite{Weinberg:1980kq,Porrati:2008rm}. The unphysical states mix with the physical
ones under Lorentz transformations, such that the matrix elements of $\Theta_{\mu\nu}$
are Lorentz tensors. But the spurious polarizations must decouple from all physical matrix elements. The generalization was made
by Porrati \cite{Porrati:2008rm}, who considered one-graviton matrix
elements to show that for any massless field with $s>2$ coupled to gravity, the decoupling of the unphysical states contradicts the principle
of equivalence. Therefore, HS gauge fields cannot have minimal coupling with gravity in flat space.

In fact, HS gauge fields in flat space cannot have any sort of gravitational interactions whatsoever. This can be seen by combining the theorems
of Weinberg \cite{Weinberg:1964ew} and Porrati \cite{Porrati:2008rm}: Consider the $S$-matrix element of one soft graviton, another graviton of
arbitrary momentum and two other particles with $s>2$. Upon the soft-limit factorization~(\ref{2}), one concludes that the soft graviton
couples with a non-zero strength $\k$ to the other graviton, but with a vanishing strength $\k'=0$ to the HS particles, in accordance with
Porrati's theorem. This contradicts Weinberg's theorem, and so the $S$-matrix element itself must vanish. The argument can be extended to any
number of particles with the same result. So, HS gauge fields in flat space may exist only on their own, completely decoupled from
everything that interacts with gravity.

\subsection{Yes-Go Results for HS Interactions}\label{sec:yes-go}

Having outlined the various no-go theorems in the previous subsection, we now give a brief account of the various loop-holes through which they can be bypassed, 
and the existing yes-go results for constructing HS interactions. Some of these have already been discussed in Ref.~\cite{Bekaert:2010hw}.

First, the no-go theorems do not apply to massive HS particles for a simple reason: Gauge transformations are no longer a symmetry because of the mass term.
After all, such particles appear in Nature in the form of hadronic resonances, which do interact with electromagnetism and gravity. Yet, writing down consistent
interactions for these fields is also quite challenging. We elaborate on this in section \ref{sec:vz}, and then in section \ref{sec:open} we discuss how these 
challenges are overcome in some known examples, particularly in the theory of charged open strings.

Second, the Aragone-Deser obstruction~\cite{Aragone:1979hx} to consistent local interactions of HS gauge fields is due to the appearance of the Weyl tensor
in the gauge variation. In three dimensions such obstructions are non-existent since the Weyl tensor vanishes. Moreover,
massless fields in 3d with $s>1$ do not carry any local DoFs.
\bex Confirm from Eqs.~(\ref{dofB0}) and~(\ref{dofF}) this DoF count in $d=3$.\eex
3d HS fields, therefore, are immune from many consistency issues that may arise in higher dimensions. In fact, in 3d flat space an interacting theory of a
spin-$\tfrac52$ gauge field and gravity$-$hypergravity$-$was constructed long ago \cite{Aragone:1983sz}, which has recently been reformulated as a
Chern-Simons theory of a new extension of the Poincar\'e group with spin-$\tfrac{3}{2}$ fermionic generators \cite{Fuentealba:2015jma}.
Likewise, one can also make a non-trivial extension with a spin-3 conserved charge \cite{Afshar:2013vka,Gonzalez:2013oaa}.
Sure enough, the Coleman-Mandula theorem does not apply to lower dimensions either. 3d HS gauge theories can also be constructed in AdS space making
them interesting for holography. The literature of this field is vast and complex, and we refer to other reviews on the subject for more details.

Third, in flat space with $d\geq4$, one may still construct interactions of massless HS fields, although these ``exotic'' theories will be completely decoupled
from any ``ordinary'' stuff that couples to gravity. After all, cubic interaction vertices for such fields do exist and can be completely classified
in the light-cone formulation \cite{Bengtsson:1983pd,Bengtsson:1983pg,Bengtsson:1986kh,Metsaev:2005ar,Fradkin:1991iy,Metsaev:1993gx,Metsaev:1993mj,
Metsaev:1993ap,Fradkin:1995xy}. The light-cone formulation puts restrictions on the number of derivatives $p$ in a general $s_1-s_2-s_3$ cubic vertex that
involves massless fields and thereby provides a way of classification \cite{Metsaev:2005ar}. For bosonic fields in $d>4$,
there is one vertex for each value of $p$ in the range \beq s_1+s_2-s_3\leq p\leq s_1+s_2+s_3,\label{derivativeb}\eeq
with $s_3$ being the smallest of the three spins. For a vertex containing pair of fermions and a boson, the formula is the same modulo that one uses
$s-\frac{1}{2}$ as ``spin'' for fermions. In $d=4$ however, only two of these vertices exist: Those with the extremal number of derivatives,
$p=s_1+s_2\pm s_3$. Among others, these results constructively reconfirm the non-existence of the low-derivative minimal-like couplings of HS gauge fields. 
Many of the no-go theorems are relatively simple consequences of the cubic analysis (see e.g. \cite{Taronna:2011kt}).

The construction of these vertices in manifestly Poincar\'e-invariant form started with Berends, Burgers and van Dam
\cite{Berends:1984wp,Berends:1984rq,Berends:1985xx}, who systematically employed the Noether procedure to introduce interactions.
More recently, the same procedure has been used in Refs.~\cite{Manvelyan:2010jr,Manvelyan:2010je} to explicitly construct covariant cubic
vertices for bosonic fields. Tensionless limit of String Theory, on the other hand, also gives rise to flat-space cubic vertices
\cite{Polyakov:2009pk,Polyakov:2010qs,Sagnotti:2010at,Taronna:2011kt}, which are completely in accordance with the other results. One can also use
the powerful BRST-antifield formalism \cite{Barnich:1993vg,Henneaux:1997bm,Barnich:1994db,Barnich:1994mt,Barnich:1994cq,Barnich:2000zw} to compute
gauge-invariant and manifestly Lorentz-invariant cubic vertices \cite{Boulanger:2006gr,Boulanger:2008tg,Henneaux:2012wg,Henneaux:2013gba}.
Beyond the cubic order, however, interactions of HS gauge fields in flat space exhibit unruly and peculiar non-localities \cite{Metsaev:1991mt,Taronna:2011kt}.
In other words, a local Lagrangian description ceases to make much sense at the quartic order.
One might argue that HS gauge theories in Minkowski background essentially call for extended and possibly non-local objects
\cite{Fotopoulos:2010ay,Dempster:2012vw}, like the stringy Pomerons \cite{Brower:2006ea}.

Most importantly, the no-go theorems are not applicable in the presence of a cosmological constant. In AdS space, for example, there are no
asymptotic states so that the $S$-matrix itself is not defined. The Aragone-Deser obstruction, on the other hand, may be removed by the inclusion
of terms that contain inverse powers of the cosmological constant, which are therefore singular in the flat limit. In this context, the cosmological
constant can play a dual role: As an infrared cutoff and also as a dimensionful coupling constant, and can reconcile HS gauge symmetry and the equivalence principle.
As a result, in AdS one can write down the Fradkin-Vasiliev cubic action \cite{Fradkin:1987ks,Fradkin:1986qy,Vasiliev:2001wa,Alkalaev:2002rq,Vasilev:2011xf},
that consistently describes the cubic interactions of an infinite tower of HS gauge fields coupled to gravity according to the principle of equivalence.
One can also construct a set of fully non-linear gauge-invariant equations for these fields$-$Vasiliev's equations
\cite{Vasiliev:1990en,Vasiliev:1990vu,Vasiliev:1990bu,Vasiliev:1992av,Vasiliev:2003ev}. The formulation of these theories is rather technical, to which section \ref{sec:HS0} is devoted.

Let us stress that when expressed in terms of metric-like symmetric tensor(-spinor)s, the linearized
Vasiliev's equations take the standard Fronsdal form~(\ref{abba2}) and~(\ref{noname46}). However the interaction
terms do not stop at a finite number of derivatives, thus making the theory essentially non-local. Unlike in flat space,
this non-locality is weighted/controlled by inverse powers of the cosmological
constant (see Refs.~\cite{Kessel:2015kna,Boulanger:2015ova,Bekaert:2015tva,Skvortsov:2015lja} for a recent
discussion on the issues of locality). Such terms also forbid these theories to have any sensible flat limits, in accordance with the no-go theorems.

\subsection{Problems with Interacting Massive HS Fields}\label{sec:vz}

While interacting massive HS fields in flat space are immune from the grave inconsistencies their massless counterparts may exhibit, a local Lagrangian
description of the former cannot hold good up to an arbitrary energy scale. The reason is simple: the no-go theorems \cite{Weinberg:1964ew,Grisaru:1977kk,Grisaru:1976vm,Aragone:1979hx,Weinberg:1980kq,Porrati:2008rm,Porrati:2012rd} imply that such a Lagrangian
must be singular in the massless limit. An effective local field theory description would only ever make sense if the mass of the particle is well below
the cutoff scale. Luckily, the cutoff can be parameterically larger than the mass, the parameter being the inverse coupling constant
\cite{ArkaniHamed:2002sp,Porrati:2008ha,Rahman:2011ik}. Therefore, there may indeed exist a regime of energy scales in which the effective field theory
will be valid. However, even within this regime of validity the system may be plagued with pathologies, as we shall see below.

For simplicity, let us consider the case of a massive spin-2 field $\vf_{\m\n}$, of mass $m$ and charge $q$, coupled to electromagnetism (EM) in flat
space. If one introduces the coupling at the level of the Bargmann-Wigner equations~(\ref{kgb})--(\ref{trb}), by replacing ordinary derivatives with
covariant ones: $\partial_\m\rightarrow D_\mu=\partial_\mu+iqA_\mu$, the result is:
\beq \left(D^2-m^2\right)\vf_{\mu\nu}=0,\qquad D\cdot\vf_\mu=0,\qquad \vf'=0.\label{kutta0}\eeq
The mutual compatibility of the Klein-Gordon equation and the transversality condition requires that
\beq
\left[D^\mu,\,D^2-m^2\right]\vf_{\mu\nu}=0.\eeq
This requirement results in unwarranted constraints, owing to that fact that covariant derivatives do not commute. For example, for a constant EM field strength
$F_{\mu\nu}$, one obtains
\beq iqF^{\mu\rho}D_\mu\vf_{\rho\nu}=0.\label{kutta2}\eeq
This constraint disappears when the interaction is turned off, and so the system~(\ref{kutta0}) does not describe the same number of DoFs as the free theory.
As we already mentioned in the introduction, such difficulties prompted Fierz and Pauli \cite{Fierz:1939zz,Fierz:1939ix} to promote the Lagrangian formulation
as the right approach, which guarantees that the resulting EoMs and constraints are mutually compatible. The Lagrangian approach, however, is not
free of difficulties either, which we shall demonstrate in the following.

First, we take the spin-2 Lagrangian~(\ref{f1}) and make the field complex, and minimally couple it to a constant EM background. Due to the
non-commuting nature of covariant derivatives, the minimal coupling is ambiguous. We are therefore left with a one-parameter family of Lagrangians, with
the parameter $g$ being the gyromagnetic ratio (see, for example,~\cite{Deser:2001dt}):
\bea \mathcal L&=&-|D_\mu\vf_{\nu\rho}|^2+2|D\cdot\varphi^{\mu}|^2+|D_\mu \varphi'|^2+(\varphi^*_{\mu\nu} D^\mu D^\nu \vf'+\text{c.c.})
\nonumber\\&&-m^2(\vf^*_{\mu\nu}\vf^{\mu\nu}-\vf'^*\vf')-\,2iqg\,\text{Tr}(\vf\cdot F\cdot\vf^*),\label{vz1}\eea
where we have used the notation: $(A\cdot B)_{\m\n}\equiv A_\m{}^\r B_\r{}_\n$ and $\text{Tr}(A\cdot B)=(A\cdot B)^\m{}_\m$.
The EoMs resulting from this Lagrangian read:
\bea 0=\mathcal R_{\mu\nu}\equiv&&(D^2-m^2)\left(\vf_{\mu\nu}-\eta_{\mu\nu}\vf'\right)+\tfrac12 D_{(\mu} D_{\nu)}\vf'-D_{(\m}
D\cdot\vf_{\n)}+\eta_{\mu\nu}D\cdot D\cdot\vf\nonumber\\&&-iqgF_{\r(\m}\vf_{\n)}{}^\r.\label{vz2}\eea
\bex Combine the trace and the double divergence of Eq.~\eqref{vz2} and use the commutator: $[D_\m,\,D_\n]=iqF_{\m\n}$ to derive the
following expression:
\beq \left(\tfrac{d-1}{d-2}\right)m^4\vf'=-i(2g-1)q\,F^{\m\n}D_\m D\cdot\vf_{\n}+(g-2)q^2\,\text{Tr}(F\cdot\vf\cdot F)+\tfrac34\,q^2\,
\text{Tr}F^2\vf'.\label{vz6}\eeq\eex

If $g\neq\tfrac{1}{2}$, the first term on the right-hand side of Eq.~(\ref{vz6}) signals a breakdown of the DoF count, since it renders dynamical the
would-be trace constraint. The unique minimally coupled model that gives the correct DoF count therefore has $g=\tfrac{1}{2}$. This choice gives rise
to the following expressions:
\bea &D\cdot\vf_{\n}-D_\n\vf'=-\tfrac32(iq/m^2)\left[F^{\r\s}D_\r\vf_{\s\n}-F_{\n\r}(D\cdot\vf^{\r}-D^\r\vf')\right],&\label{vz7} \\ &D\cdot
D\cdot\vf-D^2\vf'=\tfrac32(q/m)^2\left[\,\text{Tr}(F\cdot\vf\cdot F)-\tfrac12\text{Tr}F^2\vf'\,\right]~,&\label{vz8}\\&\vf'=-\tfrac32
\left(\tfrac{d-2}{d-1}\right)(q/m^2)^2\left[1-\tfrac34\left(\tfrac{d-2}{d-1}\right)(q/m^2)^2\,\text{Tr}F^2\right]^{-1}\,\text{Tr}
(F\cdot\vf\cdot F).&\label{vz9}\eea
\bex Derive the expressions~(\ref{vz7})--(\ref{vz9}) from the divergence and double-divergence of Eq.~\eqref{vz2} and the trace constraint~\eqref{vz6}
with $g=\tfrac{1}{2}$.\eex

Note that in the free limit $q\rightarrow0$, the system has a vanishing trace: $\vf'=0$, which reduces Eq.~(\ref{vz7}) to the
divergencelessness condition: $\partial\cdot\vf_\n=0$, and as a result Eq.~(\ref{vz2}) yields the Klein-Gordon equation:
$(\Box-m^2)\vf_{\m\n}=0$, as expected. In the presence of interactions with the EM background, however,
the trace $\vf'$ does not vanish, as can be seen from Eq.~(\ref{vz9}). It would seem that the trace constraint~(\ref{vz9}) breaks down when
$\text{Tr}F^2=\tfrac43\left(\tfrac{d-1}{d-2}\right)(m^2/q)^2$, however this value of $\text{Tr}F^2$ is outside the regime of physical interest.
This is because if some EM field invariant is ${\mathcal O}(1)$ in units of $m^2/q$\,, a number of new phenomena would show up:
Schwinger pair production \cite{Schwinger:1951nm,Bachas:1992bh} and Nielsen-Olesen instabilities \cite{Nielsen:1978rm}.
With such instabilities, the idea of long-lived propagating particles ceases to have any physical meaning. Therefore,
an effective Lagrangian for a charged particle interacting with EM, even when explored well below its own cutoff scale, can be reliable
only for small EM field invariants.

Thus, in the regime of physical interest the interacting theory propagates the same number of DoFs as the free one. We still need to make sure
that the dynamical DoFs do not propagate outside the light cone. To investigate the causal properties of the system, one may employ the method
of characteristic determinant, which is outlined below (see \cite{Velo:1969bt,Velo:1970ur,Velo:1972rt} and references therein).
One replaces $i\partial_\mu$ with the vector $n_\mu$$-$the normal to the characteristic hypersurfaces$-$in the terms containing the highest
number of derivatives in the EoMs.  Essentially, the procedure makes an eikonal approximation:
$\vf_{\m\n}=\tilde{\vf}_{\m\n}\,e^{\,i\,t n\cdot x}$ with $t\rightarrow\infty$. Then one takes the resulting coefficient matrix and
computes its determinant $\Delta(n)$. The latter determines the causal properties of the system: The system is hyperbolic (i.e.,
describes a wave propagation in the first place)
if for any $\vec{n}$ the algebraic equation $\Delta(n)=0$ has real solutions for $n_0$, in which case the maximum
wave speed is given by $n_0/|\vec{n}|$. It is clear that if such a solution is time-like, $n^2<0$, it amounts to a
faster-than-light propagation.

To carry out the aforementioned procedure in the present case, let us first isolate the second-derivatives terms in the dynamical equations~(\ref{vz2}).
This gives
\bea \mathcal R_{\mu\nu}^{(2)}&=&D^2\vf_{\mu\nu}-\left[D_\m(D\cdot\vf_{\n}-D_\n\vf')+
(\m\leftrightarrow\n)\right]-\tfrac12 D_{(\mu} D_{\nu)}\vf'\nonumber\\&&+\eta_{\mu\nu}(D\cdot D\cdot\vf-D^2\vf').\label{vz11}\eea
The last term can be dropped in view of Eq.~(\ref{vz8}), while the second and third terms can respectively be substituted by the constraints~(\ref{vz7}) and~(\ref{vz9}). The result is
\bea \mathcal R_{\mu\nu}^{(2)}&=&\Box\vf_{\m\n}+\tfrac{3}{2}(iq/m^2)\left[F^{\r\s}\de_\r\de_{(\m}\vf_{\n)\s}
+F_{\r(\m}\de_{\n)}(\de\cdot\vf^{\r}-\de^\r\vf')\right]\nonumber\\&&+\tfrac32\left(\tfrac{d-2}{d-1}\right)(q/m^2)^2
\left[F^{\r\s}F_\s^{~\l}\de_\m\de_\n\vf_{\r\l}-\tfrac12\text{Tr}F^2\de_\m\de_\n\vf'\right].\label{vz12}\eea
With the substitution $\partial_\m\rightarrow-in_\m$, the coefficient matrix of Eq.~(\ref{vz12}) takes the form
\bea &&M_{(\m\n)}{}^{(\a\b)}(n)~=~-\tfrac12n^2\delta^{(\a}_\m\delta^{\b)}_\n-\,\tfrac32\left(\tfrac{d-2}{d-1}\right)
(q/m^2)^2\,n_\m n_\n\left[F^{\a\r}F_\r^{~\b}-\tfrac12\text{Tr}F^2\eta^{\a\b}\right]\nonumber\\&&~~~~~~~~~~~~~~~
-\tfrac34(iq/m^2)\left[n_\r F^{\r(\a}n_{(\m}\delta_{\n)}^{\b)}-n_{(\m}F_{\n)}^{~(\a}n^{\b)}+2n_{(\m}F_{\n)}^{~\r}n_{\r}
\eta^{\a\b}\right].\label{vz13}\eea
This expression is to be regarded as a $\tfrac12d(d+1)\times\tfrac12d(d+1)$ matrix, whose rows and columns are labeled by pairs
of Lorentz indices $(\m\n)$ and $(\a\b)$. To compute its determinant $\D(n)$, let us choose $d=4$.
\bex
Show that in 4d the $10\times10$ matrix~(\ref{vz13}) has the determinant:
\beq \Delta(n)=(n^2)^8\left[n^2-\left(\tfrac{q}{m^{2}}\right)^2\left(\tilde{F}\cdot n\right)^2\right]\left[n^2+\left(\tfrac{3q}{2m^{2}}\right)^2
\left(\tilde{F}\cdot n\right)^2\right],\label{vz14}\eeq where $\tilde{F}_{\m\n}\equiv\tfrac12\e_{\m\n\r\s}F^{\r\s}$, and so
$(\tilde{F}\cdot n)^2=(n_0\vec B+\vec n\times\vec E)^2-(\vec n\cdot\vec B)^2$.
\eex

Let us now consider the situation where the 4d EM field invariants are specified as: $\vec B\cdot\vec E=0,~\vec B^2-\vec E^2>0$, so that
the magnetic vector $\vec B$ is non-vanishing in all Lorentz frames. Let us choose a frame in which the EM field is purely magnetic.
Then, there exists a 3-vector $\vec n$, perpendicular to $\vec B$, for which the characteristic determinant~(\ref{vz14}) becomes zero
provided that $n_0$ obeys
\beq \frac{n_0}{|\vec n|}=\frac{1}{\sqrt{1-\left(\tfrac{3q}{2m^{2}}\right)^2\vec B^2}}\,.\label{vz15}\eeq
This is because the last factor appearing in the determinant vanishes for the above choice.
Therefore, in the given Lorentz frame the system is hyperbolic only below a critical value of the magnetic field:
$\vec{B}^2_{\text{crit}}=\left(\tfrac{2}{3}m^2/q\right)^2$. However, even for infinitesimally small values of $\vec B^2$ the ratio~(\ref{vz15})
exceeds unity, and so the propagation is superluminal. In this case, one has $n^2<0$. Because the latter is a Lorentz invariant statement
the pathology will show up in any Lorentz frame, even when $\vec B^2-\vec E^2$ is arbitrarily small but positive.
This is the so-called Velo-Zwanziger problem \cite{Velo:1969bt,Velo:1970ur,Velo:1972rt}.

The most disturbing aspect of the problem is that it persists, within the regime of validity of the effective field theory, for infinitesimally small values of the EM field invariants when all the instabilities \cite{Schwinger:1951nm,Bachas:1992bh,Nielsen:1978rm} are absent. Note that this pathology generically shows up for all charged massive HS particles with $s>1$. To make things worse, the problem is present for a wide class of non-minimal generalizations of the theory and for other interactions as well \cite{Shamaly:1972zu,Hortacsu:1974bm,Prabhakaran:1975yp,Seetharaman:1975dq,Kobayashi:1978xd,Kobayashi:1978mv,Deser:2000dz,Deser:2001dt}.
It is therefore quite a challenging task to write down consistent Lagrangians for interacting massive HS fields.

The problem originates from the very existence of the longitudinal modes of massive fields with high spin
\cite{Porrati:2008gv,Rahman:2011ik}. In the presence of a non-trivial background, some of these modes may acquire non-canonical kinetic terms that jeopardise their causal propagation. Let us emphasise that the Velo-Zwanziger problem is an inconsistency of the classical theory itself. Historically, what appeared earlier was the corresponding problem in the quantum theory for a massive charged spin-$\tfrac32$ field \cite{Johnson:1960vt}.
The common origin of these classical and quantum inconsistencies became clear from subsequent studies
\cite{Jenkins:1974dn,Takahashi:1978pi,Kobayashi:1987rt}.

\subsection{Causal Propagation of Massive HS Fields}\label{sec:open}

Despite being present not only for the minimally coupled theory but also for a wide class of non-minimal generalizations thereof,
the Velo-Zwanziger problem does have a cure: Addition of suitable non-minimal terms and/or new dynamical DoFs. A classic example
in this regard is $\mathcal{N}=2$ (broken) supergravity \cite{Ferrara:1976fu,Freedman:1976aw,Scherk:1978ta,Scherk:1979zr,deWit:1984px},
wherein the massive spin-$\tfrac{3}{2}$ gravitino propagates consistently, with or without cosmological constant, provided that under the graviphoton it has a charge $q=\tfrac{1}{\sqrt{2}}(m/M_\text{P})$ \cite{Deser:1977uq}. In this case, the remedy for the acausality lies in the presence of dynamical gravity along with suitable non-minimal couplings \cite{Deser:2001dt,Rahman:2011ik}.
In fact, one can do even without gravity: A set of non-minimal terms involving arbitrary powers of the EM field strength may suffice to causally propagate the physical modes of a massive spin-$\tfrac{3}{2}$ field in a constant EM background \cite{Porrati:2009bs}.

String Theory also provides a remedy for the Velo-Zwanziger problem. For a massive spin-2 field in a constant EM background,
for example, one can present an explicit string-field-theoretic Lagrangian \cite{Argyres:1989cu}, which gives rise to a consistent causal
set of dynamical equations and constraints in $d=26$ \cite{Porrati:2010hm}. A Lagrangian for non-critical dimensions can also be obtained
by a suitable dimensional reduction that keeps only the singlets of some internal coordinates. Starting with the string-theoretic spin-2
Lagrangian, in this way one ends up having a consistent model, say in $d=4$, that contains spin-2 field and a scalar with the same mass
and charge \cite{Porrati:2011uu}. Moreover, it was shown in Ref.~\cite{Porrati:2010hm} that in a constant EM background String Theory
propagates causally any field belonging to the first Regge trajectory. The consistent set of EoMs and constraints for these arbitrary-rank
symmetric tensor fields is guaranteed to result from a Lagrangian, but the latter may necessarily incorporate all the Regge trajectories
for a given mass level (see for example \cite{Klishevich:1998sr}).

\subsubsection*{Charged Open Strings in Constant EM Background}\label{sec:sigma}

In order to see how String Theory bypasses the Velo-Zwanziger problem, let us consider an open bosonic string, whose endpoints lie on
a space-filling $D$-brane, and carry charges $q_0$ and $q_\pi$ under a $U(1)$ gauge field $A_\mu$ living in the $D$-brane world-volume.
For a constant EM background, $F_{\mu\nu}=\text{constant}$, the $\sigma$-model is exactly solvable \cite{Fradkin:1985qd,Abouelsaood:1986gd}.
A careful analysis of the mode expansion results in commuting center-of-mass coordinates that have canonical commutation relations with the
covariant momenta \cite{Argyres:1989cu,Porrati:2010hm}. There appears the usual infinite set of creation
and annihilation operators, well defined in the physically interesting regimes \cite{Argyres:1989cu,Porrati:2010hm}:
\beq [\,a_m^\mu,a_n^{\dag\nu}\,]\,=\,\eta^{\mu\nu}\delta_{mn},\qquad [\,a_m^\mu,a_n^{\nu}\,]\,=\,[\,a_m^{\dag\mu},a_n^{\dag\nu}\,]\,=\,0
\qquad m,n\in\mathbb{N}_1.\label{r23}\eeq

In the presence of the EM background, the Virasoro generators get deformed but their commutation relations remain precisely the same
as in the free theory:
\beq [L_m,L_n]=(m-n)L_{m+n}+\tfrac{1}{12}\,d(m^3-m)\delta_{m,-n},\label{r53}\eeq
where $L_m$ with $m\in\mathbb Z$ are the deformed Virasoro generators. Only $L_0$, $L_1$ and $L_2$ are relevant for our subsequent
discussion. They are (with string tension set to $\alpha'=\tfrac{1}{2}$):
\bea L_0&=&-\tfrac{1}{2}\mathcal{D}^2+\sum_{m=1}^{\infty}(m-iG)_{\mu\nu}\,a_m^{\dag\mu}a_m^\nu+\tfrac{1}{4}\,\text{Tr}G^2,\label{b9.1}\\
L_1&=&-i\left[\sqrt{1-iG}\right]_{\mu\nu}\mathcal{D}^\mu a_1^\nu+\sum_{m=2}^{\infty}\left[\sqrt{(m-iG)(m-1-iG)}\right]_{\mu\nu}
a_{m-1}^{\dag\mu}a_m^\nu~,~~~~~~~~~\label{b9.2}\\L_2&=&-i\left[\sqrt{2-iG}\right]_{\mu\nu}\mathcal{D}^\mu a_2^\nu+\tfrac{1}{2}\left[\sqrt{1+G^2}
\right]_{\mu\nu}a_1^\mu a_1^\nu\nonumber\\&&~~~~~~~~~~~~~~~~~~~~~~~~~~+\sum_{m=3}^{\infty}\left[\sqrt{(m-iG)(m-2-iG)}\right]_{\mu\nu}a_{m-2}^{\dag\mu}
a_m^\nu~,\label{b9.3}\eea
where $\mathcal D^\m$ is the covariant derivative up to a rotation:
\beq \mathcal D^\mu\equiv\left(\sqrt{G/qF}\right)^\m{}_\n\,D^\n\,,\qquad [\,D_\m,\,D_\n\,]=iqF_{\m\n}.\label{fancyD}\eeq
with $q=q_0+q_\pi=\text{total charge}$, and $G_{\m\n}$ is an antisymmetric tensor given by
\beq G\equiv\frac{1}{\pi}\left[\,\text{arctanh}(\pi q_0F)+\text{arctanh}(\pi q_\p F)\,\right].\label{r13}\eeq

A generic string state $|\Phi\rangle$ is then constructed the same way as in the free theory: One takes the string vacuum $|0\rangle$
and apply creation operators to it,
\beq |\Phi\rangle=\sum_{s=1}^\infty\sum_{m_i=1}^\infty \phi_{\mu_1\cdots\mu_s}^{(m_1\cdots m_s)}(x)\,a^{\dag\mu_1}_{m_1}\cdots
a^{\dag\mu_s}_{m_s}\,|0\rangle,\label{kuttarbaccha}\eeq
where the rank-$s$ coefficient tensor $\phi_{\mu_1\cdots\mu_s}^{(m_1\cdots m_s)}$ is a string field, and as such a function of the string
center-of-mass coordinates $x^\m$.
\bex An eigenvalue $N$ of the number operator, $\mathcal{N}\equiv\sum_{n=1}^\infty n \, a_n^\dag\cdot a_n$, defines a string level.
Confirm that $N$ is a positive integer.\eex

\subsubsection*{Physical State Conditions}\label{sec:PSCEM}

Now we shall write down the physical state conditions for string states, which translate into a set of Fierz-Pauli conditions
in the language of string fields. A string state $|\Phi\rangle$ is called ``physical'' if it satisfies the following conditions
\cite{Abouelsaood:1986gd,Argyres:1989cu}:
\bea (L_0-1)|\Phi\rangle&=&0,\label{b11}\\ L_1|\Phi\rangle&=&0,\label{b12}\\L_2|\Phi\rangle&=&0,\label{b13}\eea
We will translate these conditions into the string field theory language level by level.

\subparagraph{Level $N=1$:}

The generic state at this level is given by: $|\Phi\rangle=A_\mu(x)\,a^{\dag\mu}_1\,|0\rangle$. In this case, the non-trivial equations
are~(\ref{b11}) and~(\ref{b12}), which give rise to
\beq\begin{aligned} \left(\mathcal{D}^2-\tfrac{1}{2}\text{Tr} G^2\right)A_\mu+2iG_\m{}^\n A_\n=0,\\
\mathcal{D}^\mu\left(\sqrt{\mathbf{1}-iG}\cdot A\right)_\mu=0,\end{aligned}\label{b18}\eeq
\bex Derive Eqs.~(\ref{b18}), and show that they reduce to the following form with the field redefinition
$\mathcal{A}_\m\equiv\left(\sqrt{\mathbf{1}-iG}\cdot A\right)_\m$:
\beq\left[\mathcal{D}^2-\tfrac{1}{2}\text{Tr}G^2\right]\mathcal{A}_\mu+2iG_\m{}^\n\mathcal{A}_\nu=0,\qquad
\mathcal{D}^\mu\mathcal{A}_\mu=0.\label{b20}\eeq\eex

\subparagraph{Level $N=2$:}

A generic state at this level is written as:
$|\Phi\rangle=h_{\mu\nu}(x)\,a^{\dag\mu}_{1}a^{\dag\nu}_{1}\,|0\rangle+\sqrt{2}iB_\mu(x)\,a^{\dag\mu}_{2}\,|0\rangle$.
The physical state conditions~(\ref{b11})--(\ref{b13}) take the following form after
the field redefinitions $\mathcal{H}_{\mu\nu}\equiv\left(\sqrt{1-iG}\cdot h\cdot\sqrt{1+iG}\,\right)_{\mu\nu},~
\mathcal{B}_\mu\equiv\left(\sqrt{1-iG/2}\cdot B\right)_\mu$:
\beq\begin{aligned}\left(\mathcal{D}^2-2-\tfrac{1}{2}\,\text{Tr}G^2\right)\mathcal{H}_{\mu\nu}+2i(G\cdot\mathcal{H}
-\mathcal{H}\cdot G)_{\mu\nu}=0,\\\left(\mathcal{D}^2-2-\tfrac{1}{2}\,\text{Tr}G^2\right)
\mathcal{B}_\mu+2iG_\mu{}^\nu\mathcal{B}_\nu=0,\\\mathcal{D}\cdot\mathcal{H}_\mu
-(1-iG)_\mu{}^\nu\mathcal{B}_\nu=0,\\\mathcal{H}'+2\mathcal{D}\cdot\mathcal{B}=0.\end{aligned}\label{b28}\eeq
The system~(\ref{b28}) enjoys an on-shell gauge invariance in $d=26$ with a vector gauge parameter
\cite{Porrati:2010hm}. Gauging away the vector field $\mathcal B_\mu$ one therefore obtains
\beq \left(\mathcal{D}^2-2-\tfrac{1}{2}\,\text{Tr}G^2\right)\mathcal{H}_{\mu\nu}-2i(G\cdot\mathcal{H}
-\mathcal{H}\cdot G)_{\mu\nu}=0,\qquad\mathcal D\cdot\mathcal H_{\mu}=0,\qquad\mathcal H'=0.\label{b37.3}\eeq

\subparagraph{Arbitrary Level $N=s$:}\label{sec:arbitrary}

For higher levels, fields belonging to subleading Regge trajectories show up. These string fields that do contract
with creation operators other than $a^{\dag\m}_1$, and therefore are not totally symmetric tensors like the leading (first) Regge
trajectory ones. At any given level, it turns out that a subleading Regge trajectory is not consistent on its own, whereas the leading
one is \cite{Porrati:2010hm}. The first Regge trajectory string field at level $N=s$ is symmetric rank-$s$ tensor $\vf_{\mu_1\cdots\mu_s}$,
whose equations can be derived from the physical state conditions~\eqref{b11}--\eqref{b13}. Modulo a field redefinition, they
are~\cite{Porrati:2010hm}:
\bea &\left[\mathcal{D}^2-2(s-1)-\tfrac{1}{2}\,\text{Tr}G^2\right]\vf_{\mu_1\cdots\mu_s}-2iG^\alpha{}_{(\mu_1}
\vf_{\mu_2\cdots\mu_s)\alpha}=0,&\label{b60}\\&\mathcal{D}\cdot\vf_{\mu_1\cdots\mu_{s-1}}=0,&\label{b61}\\
&\vf'_{\mu_1\cdots\mu_{s-2}}=0.&\label{b62}\eea
It is easy to see that Eqs.~(\ref{b60})--(\ref{b62}) are mutually compatible, i.e., they do not lead to any unwarranted constraint
like Eq.~(\ref{kutta2}). They comprise a consistent set of Fierz-Pauli conditions for a massive spin-$s$ field of
$\text{mass}^2=(1/\alpha')\left(s-1+\tfrac{1}{4}\text{Tr}G^2\right)$. It is manifest that the system produces the correct DoF
count of Eq.~(\ref{dofBm}). Now we will see that these equations indeed give causal propagation for the physical DoFs.

\subsubsection*{Causal Propagation of String Fields}\label{sec:causal}

To investigate whether the system~(\ref{b60})--(\ref{b62}) gives causal propagation, once again we employ the method of characteristic
determinants already discussed in Section~\ref{sec:vz}. The highest-derivative terms in the dynamical equation~(\ref{b60}) is
the scalar operator $\mathcal D^2$ acting on the field. Then, from the definition~(\ref{fancyD}) of $\mathcal D_\m$, one finds
that the vanishing the characteristic determinant is tantamount to the condition:
\beq \left(G/qF\right)^\m{}_\n\ n_\m \, n^\n=0.\label{c1}\eeq
Now let us make $F$ to be block skew-diagonal: $F^\m{}_\n=\text{diag}\left(\,F_1\,,F_2\,,F_3\,,\cdots\,\right)$, by choosing a particular
Lorentz frame. The blocks are of the form:
\beq F_i=a_i\left(\begin{array}{cc}
               0 &~~1 \\
            2\d_i^1-1 &~~0 \\
             \end{array}
           \right),\label{c3}\eeq
where $a_i$'s are real-valued functions of the EM field invariants. Clearly, $G$ will also be block skew-diagonal in the same
Lorentz frame.
\bex Prove that in our chosen frame, ($G/qF$) is the diagonal matrix:
\beq \left(G/qF\right)^\m{}_\n=\text{diag}\left[\,\frac{f_1(a_1)}{\p qa_1}~,\frac{f_1(a_1)}{\p qa_1}~,\frac{f_2(a_2)}{\p qa_2}~,\frac{f_2(a_2)}{\p qa_2}~,
\frac{f_3(a_3)}{\p qa_3}~,\frac{f_3(a_3)}{\p qa_3}~,\cdots\,\right].\label{c7}\eeq
\eex
where the functions $f_i(a_i)$'s are given by
\beq f_i(a_i)\equiv
\begin{cases}i=1:\quad \text{arctanh}(\pi q_0a_1)+\text{arctanh}(\pi q_\pi a_1)\\
i\neq1:\quad \arctan(\pi q_0a_i)+\arctan(\pi q_\pi a_i)\end{cases}.\label{c6}\eeq

Let us emphasise that in the regimes of physical interest the absolute values of the quantities $a_i$'s and the functions $f_i(a_i)$'s are
much smaller than unity. The latter, however, satisfy the inequalities: $f_1(a_1)\geq \p qa_1$, and $0<f_i(a_i)\leq \p qa_i$ for $i\neq1$.
These inequalities imply, in view of the diagonal matrix~(\ref{c7}), that any solution $n_\mu$ to the characteristic equation~(\ref{c1})
must be space-like: $ n^2\geq0$.

That $n_\m$ is space-like is a Lorentz invariant statement, and therefore must hold in all Lorentz frames. We conclude that the propagation of the massive
spin-$s$ field $\vf_{\mu_1\cdots\mu_s}$ is causal, since the maximum wave speed never exceeds unity:
\beq \frac{n_0}{|\vec n|}\leq 1.\label{c10}\eeq
\subparagraph{Remarks:}

\begin{itemize}
\item
The generalized Fierz-Pauli conditions~(\ref{b60})--(\ref{b62}) contain non-standard kinetic terms. It is important to check that
the flat-space no-ghost theorem extends to the present case. Indeed, in the presence of a constant EM background the no-ghost theorem continues to
be valid in the regimes of physical interest \cite{Porrati:2010hm}.\vspace{5px}
\item Open String Theory requires that the gyromagnetic ratio take the universal value of $g=2$ for all spin \cite{Ferrara:1992yc}.
This value is manifest in the dipole term appearing in the dynamical equation~(\ref{b60}). At the Lagrangian level, one may remove all
the kinetic-like cubic interactions by suitable field redefinitions, which leaves one with the value $g=2$. This is apparently in contradiction
with the unique choice of $g=\tfrac12$ made in section~\ref{sec:vz}. However, the conclusions of section~\ref{sec:vz} no longer hold when
non-minimal kinetic terms are present.\vspace{5px}
\item For $s>1$, the spin-$s$ string field becomes massless in the limit $\a'\rightarrow\infty$. This limit, however, makes sense only if one
simultaneously takes $qF\rightarrow0$, so that the eigenvalues of $\a'qF$ may remain finite and small to ensure absence of instabilities
\cite{Schwinger:1951nm,Bachas:1992bh,Nielsen:1978rm}. As a result, massless fields with $s>1$ cannot carry an electric charge. This is in complete
accordance the no-go theorems \cite{Weinberg:1980kq,Porrati:2008rm}.\vspace{5px}
\item String Theory does provide with a Lagrangian which in $d=26$ gives rise to the Fierz-Pauli system~(\ref{b60})--(\ref{b62}).
Such a Lagrangian naturally shows up with redundancies, i.e., there is a gauge invariance (St\"uckelberg symmetry) \cite{Argyres:1989cu,Klishevich:1998sr}.
Indeed, the gauge-invariant formulation is a consistent way of turning on interactions for massive HS fields. This approach allows one
to construct, order by order in number of fields, interacting Lagrangians in arbitrary dimensions
\cite{Klishevich:1997pd,Klishevich:1998ng,Klishevich:1998yt,Zinoviev:2008jz,Zinoviev:2009hu,Zinoviev:2010av}.
The gauge symmetry, however, is necessary but not sufficient for consistency. As it turns out, requirement of causal propagation may further
restrict otherwise allowed gauge-invariant couplings \cite{Buchbinder:2012iz,Henneaux:2013vca}.
\end{itemize}

\newpage
\section{Unfolded Formulation \& HS Equations}\label{sec:HS0}

We have seen in the previous sections that the explicit introduction of HS gauge potentials is possible in a Lagrangian framework, and it leads to the Fronsdal 
equations and the corresponding actions. Gauge potentials, however, can also be introduced in a different approach, which does not rely on an action principle. 
Known as the \textit{unfolded formalism}, this procedure is at the basis of Vasiliev's formulation of HS theories \cite{Vasiliev:1988sa}. To introduce the key 
ideas of this formalism, we choose to start with a different but instructive example: Riemann normal coordinates.

\subsection{From Riemann to Lopatin-Vasiliev}\label{sec:Unfolding}

This section starts with a review of Riemann normal coordinates. We then move to a derivation of Lopatin-Vasiliev's unfolded equations for free HS fields \cite{Lopatin:1987hz}.

\subsubsection{Riemann Normal Coordinates \& Riemannian Geometry}

Here we briefly review a slightly different description of Riemannian geometry, which will prove useful in understanding some basic ideas behind the HS case.
Consider a metric $g_{\mu\nu}$ defined on a smooth manifold $M$. At a given point $x$, one can always find a reference frame whose coordinates are measured 
along geodesics passing through the point $x$. Indeed, by choosing a point $y$ sufficiently close to $x$, it is always possible to ensure uniqueness of the 
geodesic passing across both of them. One can then define the coordinate $y$, in the reference system centered at $x$, to be the geodesic arc length measured 
from $x$ to $y$ on this geodesic. If $v^\mu$ is the tangent vector to the above geodesic,  and $t$ the corresponding geodesic length, one has
\begin{equation}
y^{\mu}=t \,v^\mu\,.
\end{equation}
In this choice of coordinates$-$known as the {\it Riemann normal coordinates}$-$the geodesic equations take the same simple form as in flat space:
\begin{equation}
\frac{d^2}{dt^2}\,y^\mu(t)=0\,.
\end{equation}
As a result, the corresponding Christoffel symbols in these coordinates satisfy:
\begin{equation}
y^\mu y^\nu \Gamma^\rho{}_{\mu\nu}(y)=0\,.\label{yGamma}
\end{equation}
\begin{svgraybox}
A key virtue of the Riemann normal coordinates is that in these coordinates the Taylor coefficients of a tensor around a point are Lorentz tensors themselves.
\end{svgraybox}
In order to appreciate this point more closely, it is useful to take a step back and consider the analog of this coordinate system in the context of Yang-Mills 
theory. In this case, the condition \eqref{yGamma} is written in terms of the gauge potential $A_\mu$ as:
\begin{equation}
i_y (A_{\mu}dx^{\mu})=y^\mu A_{\mu}(y)=0\,.\label{Schwinger-Fock}
\end{equation}
The above is in this context a gauge condition and goes under the name ``Schwinger-Fock"-gauge. In this gauge the gauge potential $A_\mu$ can be expressed formally 
in terms of the Yang-Mills curvature using the standard homotopy formula for the de-Rham differential (see Section~\ref{sec:conventions}):
\begin{multline}
dA=F-A\wedge A\\\longrightarrow A(y)=i_y\int_0^1 dt\, t\left[(F(y t))-A\wedge A(y t)\right]=y^\nu\int_0^1 dt\, t F_{\nu\mu}(t y)\,.
\end{multline}
The key is that the gauge condition \eqref{Schwinger-Fock} enabled us to drop the $A\wedge A$ term.

Most importantly, from the above expression relating the gauge potential to the curvature, one can deduce that in this gauge all symmetrized derivatives of the gauge 
potential vanish at the origin:
\begin{align}
\partial_{\nu}A_{\mu}(0)&=\int_0^1 dt\,t\,F_{\nu\mu}(0)\,,
\end{align}
due to the antisymmetry of the curvature tensor. Therefore
\begin{equation}
\partial_{\mu(s)}A_{\mu}(0)\equiv0\,.
\end{equation}
One then recovers that the above gauge choice makes the Taylor coefficients in the expansion of the gauge potential, covariant tensors:
\begin{multline}
A_\mu(\epsilon)=\sum_{n=0}^\infty\frac{(\epsilon\cdot\partial_x)^n}{n!}\,A_{\mu}(0)=\sum_{n=0}^{\infty}\frac{1}{n!}\frac{1}{n+2}(\epsilon\cdot\partial_x)^n\epsilon^\rho F_{\rho\mu}(0)\\=\sum_{n=0}^{\infty}\frac{1}{n!}\frac{1}{n+2}(\epsilon\cdot D)^n\epsilon^\rho F_{\rho\mu}(0)\,.
\end{multline}
In the last equality, we replaced ordinary derivatives with covariant derivatives by adding terms proportional to $\partial_{\mu(s)}A_{\mu}(0)$. This is possible because 
they vanish identically in this coordinate system, as shown above. We have thus shown that in the Schwinger-Fock gauge the Taylor expansion coefficients are manifestly 
Lorentz covariant, a property which does not hold in a generic gauge.

Similar conclusions can now be drawn in the gravity case where, in terms of the spin-connection $\omega$ and vielbein $e$, the normal coordinate gauge choice becomes:
\begin{equation}
y^\mu y^\nu\Gamma^{\rho}{}_{\mu\nu}=0\qquad\longleftrightarrow\qquad i_y\omega=0\,,
\end{equation}
and can be referred to again as the Schwinger-Fock gauge.
As with the spin-1 example, the virtue of this gauge choice is that the Taylor coefficients of the vielbein and spin-connection are Lorentz tensors \cite{Muller:1997zk}:
\begin{multline}\label{omeganormal}
\omega_\mu{}^a{}_b(\epsilon)=\sum_{n=0}^\infty\frac{(\epsilon\cdot\partial_x)^n}{n!}\,\omega_{\mu}{}^a{}_b(0)=\sum_{n=0}^{\infty}\frac{1}{n!}\frac{1}{n+2}(\epsilon\cdot\partial_x)^n\epsilon^\rho R^a{}_{b\rho\mu}(0)\\=\sum_{n=0}^{\infty}\frac{1}{n!}\frac{1}{n+2}(\epsilon\cdot \nabla)^n\epsilon^\rho R^a{}_{b\rho\mu}(0)=\int_0^1 dt\, t\epsilon^\rho R^a{}_{b\rho\mu}(t\epsilon)\,,
\end{multline}
with a similar but more complicated expansion for the vielbein.
\begin{exercise}
Integrate the torsion equation for the vielbein,
\be
de^a+\omega^{a}{}_b \wedge e^b=0\,,
\ee
 starting from $\omega_\mu{}^a{}_b$ given in \eqref{omeganormal}. Show that the following recursive formula for the vielbein holds \cite{Muller:1997zk}:
\be
e^a_{\mu}(x)=\delta_{\mu}^a+y^\ga y^\gb\int_0^1 dt \,t(1-t)\,R^a{}_{\ga\gb b}(ty)\,e^b_\mu(ty)\,,
\ee
which can be solved iteratively as an expansion in powers of the Riemann tensor.
\eex
The above formula can be further simplified by considering a covariant Taylor expansion of the Riemann tensor:
\begin{multline}\label{vieleq}
e^a_{\mu}(x)=\delta_{\mu}^a+y^\ga y^\gb\int_0^1 dt \,t(1-t)\,\sum_{k=0}^\infty (t\,y\cdot\nabla)^k R^a{}_{\ga\gb b}(0)\,e^b_\mu(ty)\\
=\delta_{\mu}^a+\sum_{k=0}^\infty (y\cdot\nabla)^k R^a{}_{b}(y|0)\int_0^1 dt \,t^{k+1}(1-t)\,e^b_\mu(ty)\,,
\end{multline}
where we have defined
\be
R^a{}_{b}(y|\,x)\equiv y^\ga y^\gb R^a{}_{\ga\gb b}(x)\,.
\ee
\bex
Integrate Eq. \eqref{vieleq} to show that the order-$k$ term in the expansion of the vielbein in powers of the Riemann tensor evaluated at the origin is given by
\be
e^{(k)}(y)=\sum_{n_1\cdots n_k=0}^{\infty}\,\prod_{l=1}^k\,\frac{(y\cdot\nabla)^{n_l}R(y|\,0)}{n_l!(n_l+\cdots+n_k+2k-2l+2)(n_l+\cdots+n_k+2k-2l+3)}\,,\label{vielsol}
\ee
where the tangent indices have been suppressed and matrix multiplication is assumed for $R(y|\,0)\equiv R^a{}_b(y|\,0)$.
\eex
These relations, when applied to the metric tensor $g_{\mu\nu}=\text{Tr}(e_{\mu} e_{\nu})$, give
\begin{equation}
g_{\mu\nu}(\epsilon)=\eta_{\mu\nu}-\tfrac{1}{3} R_{\mu\alpha\nu\beta}(0)\epsilon^\alpha\epsilon^\beta+O(\epsilon^3)\,.
\end{equation}
This well-known relation shows that the Riemann tensor evaluated at the point $x=0$ appears as Taylor coefficient at order $\epsilon^2$ in the expansion of the metric.

The general solution \eqref{vielsol} for the vielbein in normal coordinates also provides an arbitrary solution to the Einstein equations, when one considers a traceless 
decomposition of the derivatives of the Riemann tensor and sets to zero all components proportional to the Einstein equations. As we will clarify in the following section, 
what we observe here can be considered as a precursor to unfolding: dynamical equations like the Einstein equations are translated into the condition that certain terms 
in the Taylor expansion of the metric are absent:
\be
R_{\mu\nu\rho\sigma}=W_{\mu\nu\rho\sigma}\qquad\longleftrightarrow\qquad R_{\mu\nu}=0\,.
\ee
Note however that this cannot yet be considered as the complete unfolding of gravity. This is because the derivatives of the Riemann tensor are yet to be decomposed 
into traceless irreducible components.

To conclude this section, let us stress that the Schwinger-Fock gauge, alias Riemann normal coordinates, gives a convenient bridge between geometry and gauge theories. 
In what follows a generalization of the simple steps discussed above will lead us to the so-called unfolded formulation of HS theories, where linearized HS Riemann 
tensors will similarly appear as Lorentz covariant Taylor expansion coefficients of the solution to HS field equations.

\subsubsection{Lopatin-Vasiliev Formulation of Free Higher Spins}

In this section we present a very basic derivation of Lopatin-Vasiliev unfolded equations in flat space \cite{Lopatin:1987hz}, as well as a brief survey of the unfolded 
formalism \cite{Vasiliev:1980as,Vasiliev:1988xc,Vasiliev:1988sa,Vasiliev:1989yr,Vasiliev:1990en,Vasiliev:1992gr,Vasiliev:1995dn,Vasiliev:1999ba,Vasiliev:2000rn, Shaynkman:2000ts,Barnich:2004cr,Barnich:2006pc,Skvortsov:2008vs,Skvortsov:2009zu,Skvortsov:2009nv,Barnich:2010sw}. This derivation of the unfolded equations is centred on 
the idea introduced in the previous section, that in an appropriate normal-coordinate frame the Taylor expansion coefficients of the solution to the HS wave equations are 
manifestly Lorentz covariant. We begin the analysis by specialising to the flat space case, in which this feature is true in cartesian coordinates. As we shall demonstrate 
later, similar considerations will apply to other backgrounds. We will also comment on coordinate independence towards the end. Notably, the system of equations we will 
establish is completely coordinate independent.

In what follows we will consider in detail the spin-0, spin-1 and spin-2 cases, moving later on to the generic spin-s case. The starting point will be the Bargmann-Wigner 
equations discussed in Section \ref{sec:BW}, as opposed to the Fronsdal equations. The counterpart of the gauge potential in the Fronsdal case will be introduced later, 
as a consequence of the gauging of the rigid symmetries associated to HS fields.

\subparagraph{Scalar case:}

The scalar case is particularly simple, and might seem a bit trivial at first glance. We present this simple example to illustrate that the unfolded formulation is equivalent 
to the standard formulation of dynamics, and should be thought of as a change of variables for the dynamical system. In the following we focus on the example of the massless 
scalar satisfying the Klein-Gordon equation:
\begin{equation}\Box\phi(x)=0\,.\label{notunailam0}\end{equation}
Later, we will generalize this example to the massive case as well as to AdS background.
The starting point for the unfolded approach is to look at generic solutions to the wave equation (\ref{notunailam0})
in the form of Taylor expansions\footnote{This shows an interesting link with jet space that we detail in Section~\ref{sec:Jet}.} around a point $x_0$:
\begin{equation}
\phi(x)=\sum_{k=0}^\infty\frac1{k!}\,C_{a(k)}(x_0)(x-x_0)^{a(k)}\,.\label{Taylor}
\end{equation}
Here $C_{a(k)}(x_0)$ parameterize arbitrary Taylor coefficient functions defined on the tangent space at each point. We will collectively refer to them as
\textit{jet-vector}, owing to some nomenclature used in the context of jet-space. The choice of a Taylor expansion ansatz should be considered on the same 
footing as the plane wave ansatz:
\begin{equation}
\phi(x)=\int d^dp\,\tilde{\phi}(p)e^{ip\cdot x}\,,
\end{equation}
which can be used to solve the Klein-Gordon equation with a simple insertion of $\delta$-function:
\begin{equation}
\phi(x)=\int d^dp\,\theta(p_0)\delta(p^2)\tilde{\phi}(p)e^{ip\cdot x}\,.
\end{equation}
Above we have also inserted a Heaviside step function $\theta(p_0)$ enforcing energy positivity $p_0>0$, so that the Lorentz covariant measure 
$d^dp\,\theta(p_0)\delta(p^2)$ implies the mass-shell condition upon which the momentum lives on the upper light-cone.

One should note that an arbitrary choice of Taylor coefficients functions $C_{a(k)}(x)$ does not generically specify a scalar function $\phi(x)$ via \eqref{Taylor}. 
Therefore, it is necessary to ensure that the Taylor expansion coefficients as functions of the base point $x_0$ are compatible when moving the base point. To this 
effect, it is sufficient to consider two different expansions around the points $x_0$ and $x_0+\epsilon$ and require that their difference vanishes identically to 
linear order in $\epsilon$. This condition ensures that the above new variables $C_{a(k)}(x)$ define a compatible Taylor expansion at any point of space-time:
\begin{equation}
\sum_{k=0}^\infty\frac1{k!}\epsilon^\mu \left[\partial_{\mu}C_{a(k)}-\delta_{\mu}{}^b C_{b a(k)}\right](x-x_0)^{a(k)}=0\,.
\end{equation}
This condition can be also conveniently uplifted to a coordinate independent form:
\begin{equation}
\nabla C_{a(k)}-h^b C_{b a(k)}=0\,,\label{scalarUnf}
\end{equation}
upon introducing the vielbein $h^b=dx^\mu\delta_{\mu}{}^b$ encoding the local frame, together with the corresponding Lorentz covariant derivative $\nabla=dx^\mu\,\nabla_\mu$ 
acting as usual on tangent indices. Having worked out the above compatibility condition for the fields $C_{a(k)}$, one can now investigate the consequences of the Klein-Gordon 
equation on the above formal series and look for solutions. Exactly following the same logic that leads to the insertion of $\delta(p^2)$ in the Fourier representation, 
in terms of the fields $C_{a(k)}$ the on-shell condition becomes purely algebraic as it is tantamount to a tracelessness condition: $C^b{}_{ba(k)}=0$.

After the above change of variables, the dynamical equation involving time derivatives is replaced by the compatibility condition \eqref{scalarUnf}, while the initial 
dynamical equation is mapped into a purely algebraic statement about the dynamical variables which are bound to span a submanifold of the initial unconstrained jet-vector. 
Solving the dynamical equations would then reconstruct the on-shell Taylor expansion of the scalar field around a given point (see e.g. \cite{Shaynkman:2000ts}):
\begin{equation}
\sum_{k=0}^\infty\frac1{k!}\,C_{a(k)}(x_0)\,(x-x_0)^{a(k)}~\rightarrow~ \sum_{k=0}^\infty\frac1{k!}\,[\nabla_{a(k)}\phi(x_0)]\,(x-x_0)^{a(k)}\,.
\end{equation}
\vspace{5px}

At this point a few comments are in order:
\begin{itemize}
\item In considering the Taylor expansion ansatz we have replaced the field $\phi(x)$ with an infinite number of other fields $C_{a(k)}(x)$. 
These are the moments of $\phi(x)$ upon solving the unfolded equations. One should think of them as an equivalent but redundant set of variables that describe 
the same dynamical system$-$a scalar field. In the context of Hamiltonian formalism this set of variables is usually referred to as a vector in the jet space 
\cite{Barnich:2000zw} (see Section~\ref{sec:Jet} for more comments).\vspace{5px}
\item An advantage of the Fourier representation is that it naturally endows the space of solutions with a $L^2$-norm uplifting the corresponding space to a 
Hilbert space. Taylor representation of the solution hides this features. It is indeed easy to find polynomial solutions to the Klein-Gordon equation 
which however would correspond to non-normalizable solutions from the Fourier representation perspective.\vspace{5px}
\item A nice feature of the above choice of variables that we will investigate more in the following, is that it provides us with a description of the system 
which makes manifest the symmetries behind it. This makes this set of variables most suitable for the description of systems with a large amount of symmetry, 
like HS theories. It is mainly for this reason that these particular variables were most successful for this research direction, and lead to non-linear 
equations \cite{Vasiliev:1990en}.
\end{itemize}

It is useful to investigate a bit more the structure of the dynamical first order equations \eqref{scalarUnf} we have recovered. To this effect, 
it is instructive to study the transformation properties of a scalar field, under the action of the global isometry group $ISO(d-1,1)$, directly in terms 
of the new set of variables $C_{a(k)}(x)$. In other words, we consider the action of a translation and a Lorentz rotation of the scalar, and rewrite them 
in terms of the jet-vector components $C_{a(k)}(x)$. By considering first a translation $x\rightarrow x+\epsilon$ and then a Lorentz rotation around the 
base point $(x-x_0)_\mu\rightarrow (1+\Lambda)_{\mu}{}^\nu(x-x_0)_\nu$ in \eqref{Taylor}, one arrives at:
\begin{subequations}
\begin{align}
\delta_\epsilon C_{a(k)}(x)&=\epsilon^{\mu}h_{\mu}{}^b C_{ba(k)}\,,\\
\delta_\Lambda C_{a(k)}(x)&=\Lambda_{a}{}^b C_{ba(k-1)}(x)\,.
\end{align}
\end{subequations}
These can be considered as a definition of the action of translation operator $\hat{P}^a$ and Lorentz operator $\hat{L}^{ab}$ defined on the new dynamical 
variables $C_{a(k)}(x)$. Defining the infinite vector $C^{[0]}=\{C_{a(k)}(x)\}_{k=0,\cdots,\infty}$, whose components are given by the $C_{a(k)}(x)$, we can then write:
\begin{subequations}
\begin{align}
\delta_\epsilon C^{[0]}&=\epsilon^a \hat{P}_a C^{[0]}=\{\epsilon^{\mu}h_{\mu}{}^b C_{ba(k)}\}_{k=0,\cdots,\infty}\,,\\
\delta_\Lambda C^{[0]}&=\Lambda^{ab} \hat{L}_{ab} C^{[0]}=\{\Lambda_{a}{}^b C_{ba(k-1)}(x)\}_{k=0,\cdots,\infty}\,.
\end{align}
\end{subequations}
The above form of the action of isometries on the dynamical variables $C^{[0]}$ allows the dynamical EoM to be re-written in the following neat algebraic form:
\begin{equation}
\left(d-\omega^{ab}\hat{L}_{ab}-h^a\hat{P}_a\right)C^{[0]}\equiv \tadD C^{[0]}=0\,,
\end{equation}
which in turn defines a covariant derivative on the Poincar\'e-module:
\be C^{[0]}=\{C_{a(k)}(x)\}_{k=0,\cdots,\infty}\,.\ee
\bex
Using the explicit realization of $\hat{P}_a$ and $\hat{L}_{ab}$ given above, show that the derivative $\adD$ squares to zero in flat space.
\eex

A generic lesson we can learn from this example is that in terms of the above new variables the dynamical equations are mapped into covariant constancy conditions defined on 
a module of the isometry algebra, plus algebraic constraints on the infinite vector $C^{[0]}$ which restrict the field space to a submanifold of the initial jet.
Next we will extend the discussion to more general cases, and eventually to the generic case of Fronsdal fields.

\subparagraph{Maxwell case:}

Another instructive example to consider is the case of Maxwell equations. In particular, we want to unfold the following system:
\begin{subequations}\label{maxwelleq}
\begin{align}
\square F_{\mu\nu}&=0\,,\\
\partial^\mu F_{\mu\nu}&=0\,,\\
\partial_{[\rho} F_{\mu\nu]}&=0\,.
\end{align}
\end{subequations}
The logic we follow is the same as for the scalar. Keeping in mind the standard Fourier representation in terms of plane waves, we consider instead a Taylor expansion 
ansatz in Cartesian coordinates:
\be F_{\mu\nu}(x)=\sum_{k=0}^\infty \frac1{k!} \delta_{\mu}{}^m\delta_{\nu}{}^n C_{a(k);\,mn}(x_0)\,(x-x_0)^{a(k)}\,.\label{ansatzs1}\ee
Here we introduced the momenta $C_{a(k);\,mn}(x)$ as generic tensors totally symmetric in the first group of indices and antisymmetric in the last two. Notice that the 
$m$ and $n$ indices are tangent indices. They are related to the world indices $\mu$ and $\nu$ via the vielbein which in Cartesian coordinate is just a delta function.

As before, the first step is to ensure that the corresponding new variables are self-compatible when changing the base-point $x_0$. This ensures that the dependence on 
$x_0$ disappears so that $F_{\mu\nu}(x)$ is a function of $x$ only. The corresponding condition is a simple generalization of the condition found
in the scalar case and reads:
\be \sum_{k=0}^\infty\frac1{k!}\,\epsilon^\mu\delta_{\mu}{}^m\delta_{\nu}{}^n\left[\partial_{\mu}C_{a(k);\,mn}-\delta_{\mu}
{}^bC_{ba(k);\,mn}\right](x-x_0)^{a(k)}=0\,.\ee
This can be rewritten in a manifestly coordinate independent way, in terms of the standard Lorentz covariant derivative $\nabla=dx^\mu\,\nabla_\mu$ and frame 
$h^a=dx^\mu\delta_\mu{}^a$, as:
\be \nabla C_{a(k);\,mn}-h^bC_{ba(k);\,mn}=0\,.\ee
Packaging all the coefficient functions into an infinite dimensional jet vector: $$C^{[1]}\equiv\{C_{a(k);\,mn}\}_{k=0,\cdots,\infty}\,,$$ one can derive the structure 
constants for translation and Lorentz rotation generators from the transformation properties of the curvature $F_{\mu\nu}$ itself.
\bex Consider the action of the Poincar\'e group on the Maxwell tensor to show that the Poincar\'e generators are
represented on the infinite vector $C^{[1]}$ as:
\begin{subequations}
\begin{align}
\delta_\epsilon C^{[1]}&=\epsilon^a \hat{P}_aC^{[1]}=\{\epsilon^{\mu}h_{\mu}{}^b C_{ba(k);\,mn}\}_{k=0,\cdots,\infty}\,,\label{strS1_1}\\
\delta_\Lambda C^{[1]}&=\Lambda^{ab} \hat{L}_{ab}C^{[1]}=\{\Lambda_{a}{}^b C_{ba(k-1);\,mn}(x)+\Lambda_{[m}{}^c C_{a(k);\,c|n]}(x)\}_{k=0,\cdots,\infty}\,.\label{strS1_2}
\end{align}
\end{subequations}\eex

The above result allows again to represent the compatibility condition for the infinite dimensional jet vector $C^{[1]}=\{C_{a(k);\,mn}\}_{k=0,\cdots,\infty}$ directly in 
term of Lorentz and translation generators as a covariant constancy condition:
\be (d-\omega^{ab}\hat{L}_{ab}-h^a\hat{P}_a)C^{[1]}\equiv \tadD C^{[1]}=0\,.\ee
The compatibility condition takes the same form as for the scalar, but for a different (reducible) module of the Poincar\'e group given by $C^{[1]}=\{C_{a(k);\,mn}\}_{k=0,\cdots,\infty}$.

So far we have not yet imposed any on-shell condition, since all we have done is simply change variables from the off-shell Maxwell tensor $F_{\mu\nu}(x)$ to the corresponding infinite vector of momenta $C^{[1]}=\{C_{a(k);\,mn}\}_{k=0,\cdots,\infty}$. As for the scalar we should now impose the dynamical equations. Here there is a small complication with respect to the scalar case, which is related to the fact that we are describing a different module of the Poincar\'e group. Indeed the tensors $C_{a(k);\,mn}$, and with them the full infinite-dimensional jet-vector $C^{[1]}$, are not irreducible objects under the Lorentz subalgebra and can be decomposed as $gl(d)$ tensors as:
\be\begin{aligned}&\begin{tabular}{|c|}\hline\phantom{a1}\\\hline\end{tabular}\\[-4pt]
&\begin{tabular}{|c|}\phantom{a1}\\\hline\end{tabular}\end{aligned}~~\otimes~~\overbrace{\begin{tabular}
{|c|c|c|c|}\hline \phantom{a1}&\multicolumn{2}{|c|}{$~~~\cdots~~~$}&\phantom{a1}\\\hline
\end{tabular}}^{\displaystyle k}~~=~~\overbrace{\begin{aligned}&\begin{tabular}{|c|c|c|c|}\hline
$\phantom{a1}$&\multicolumn{2}{|c|}{$~~~\cdots~~~$}&\phantom{a1}\\\hline\end{tabular}\\[-4pt]
&\begin{tabular}{|c|}$\phantom{a1}$\\\hline\end{tabular}\end{aligned}}^{\displaystyle{k+1}}~~\oplus~~\overbrace{
\begin{aligned}&\begin{tabular}{|c|c|c|c|}\hline $\phantom{a1}$&\multicolumn{2}{|c|}{$~~~\cdots~~~$}
&\phantom{a1}\\\hline\end{tabular}\\[-4pt]&\begin{tabular}{|c|}$\phantom{a1}$\\\hline\end{tabular}\\[-4pt]&\begin{tabular}{|c|}$\phantom{a1}$\\\hline\end{tabular}\end{aligned}}^{\displaystyle k}~~~.\ee
Still, the above decomposition allows the Maxwell equations to be solved in a purely algebraic fashion. It is indeed easy to see that while the first Young tableaux $[2,1,\ldots,1]$ would manifestly solve the Bianchi identity, the second Young tableaux $[3,1,\ldots,1]$ and its traces would not -- the Bianchi identity is mapped into the Young projection condition on the tableaux $[2,1,\ldots,1]$ for each component $C_{a(k);\,mn}$. On top of that, any trace component would also be responsible for a violation of the Maxwell equations. This follows easily by plugging the ansatz \eqref{ansatzs1} into \eqref{maxwelleq}. The on-shell condition is then again mapped into algebraic $o(d)$ irreducibility conditions for the new dynamical variables.
\vspace{5px}

Summarizing:
\begin{itemize}\setlength\itemsep{0.5em}
\item Compatibility of the variable change in terms of the jet $C^{[1]}=\{C_{a(k);\,mn}\}_{k=0,\cdots,\infty}$ requires that a covariant constancy condition be imposed on $C^{[1]}(x)$:
\be\tadD C^{[1]}(x)=0\,.\ee
\item The on-shell condition for the system is translated into the vanishing of some of the irreducible components in $C^{[1]}(x)$. In the case at hand only one irreducible traceless
$o(d)$ component is allowed to be non-vanishing:
\be C^{[1]}~~\approx~~\left\{~\overbrace{\begin{aligned}&\begin{tabular}{|c|c|c|c|}\hline
$\phantom{a1}$&\multicolumn{2}{|c|}{$~~~\cdots~~~$}&\phantom{a1}\\\hline\end{tabular}\\[-4pt]
&\begin{tabular}{|c|}$\phantom{a1}$\\\hline\end{tabular}\end{aligned}}^{\displaystyle{k+1}}~
\right\}_{k=0,\cdots,\infty}~~~.\ee
\end{itemize}

It is useful to stress how Eqs.~\eqref{strS1_1} and \eqref{strS1_2} give the explicit form of the structure constants that parameterize the transformation properties of the given 
module under the Poincar\'e isometry. Although they look similar to the structure constants obtained in the scalar case, they are different. This can be appreciated by rewriting 
everything in the mostly symmetric convention for Young tableaux, which is also the most used in the literature.
To this effect, considering the antisymmetric projection of a manifestly symmetric tensor $C_{a(k+1),\,b}$:
\be C_{a(k);\,mn}=C_{a(k)m,\,n}-C_{a(k)n,\,m}\,,\ee
and using the Young projection condition in the manifestly symmetric basis:
\be C_{a(k)c,\,a}+ C_{a(k+1),\,c}=0\,,\ee
one readily arrives to:
\begin{equation}
C_{a(k);\,an}^{[2,1,\ldots,1]}=(k+2)\, C_{a(k+1),\,n}^{(k+1,1)}\,,
\end{equation}
where in the superscript we explicitly made manifest the difference in convention for the corresponding Young tableaux. By projecting Eq.~\eqref{strS1_1} 
in the manifestly symmetric basis, one then arrives to the following identities:
\begin{subequations}
\begin{align}
\delta_{\epsilon}C_{a(k-1);\,an}&=(k+1)\,\delta_{\epsilon}C_{a(k),\,n}\,,\\
\epsilon^{\mu}h_{\mu}{}^b C_{ba(k);\,an}&=\left(C_{a(k)b,\,n}-C_{a(k-1)nb,\,a}\right)\epsilon^b\,.
\end{align}
\end{subequations}
Using now the Young-projection condition for $C_{a(k-1)nb,a}$ that can be rewritten as:
\begin{equation}
C_{na(k-1)b,\,a}+C_{a(k)b,\,n}+C_{a(k)n,\,b}=0\,,
\end{equation}
and combining everything together, one finally arrives at the manifestly symmetric form of the same structure constants:
{\allowdisplaybreaks\begin{svgraybox}\vspace{-10px}
\begin{subequations}
\begin{align}
C^{[1]}(x)&\equiv\{C_{a(k+1),\,b}(x)\}_{k=0,\cdots,\infty}\,,\\[3pt]
\epsilon^c\hat{P}_cC^{[1]}(x)&=\epsilon^n\left\{C_{a(k+1)n,\,b}+\tfrac1{k+2}\,C_{a(k+1)b,\,n}\right\}_{k=0,\cdots,\infty}\,,\\[3pt]
\Lambda^{cd} \hat{L}_{cd}C^{[1]}(x)&=\left\{\Lambda_{a}{}^c C_{ca(k);\,b}(x)+\,\Lambda_{b}{}^c C_{a(k+1);\,c}(x)\right\}_{k=0,\cdots,\infty}\,.
\end{align}
\end{subequations}\vspace{-15px}
\end{svgraybox}
}
\subparagraph{Linearized gravity case:}

Following the spin-1 discussion, it is not too difficult to see how the spin-2 generalization would work. The main object to start with is the traceless tensor 
$R_{\mu\nu|\,\rho\sigma}$ in the manifestly antisymmetric convention, in terms of which the Bargmann-Wigner equations are written down (see Section~\ref{sec:BWmassless}).
The change of dynamical variables in this case is implemented by considering a Taylor expansion ansatz of the type:
\begin{equation}\label{S2comp}
R_{\mu\nu|\,\rho\sigma}(x)=\sum_{k=0}^\infty \frac1{k!} (\delta_{\mu}{}^{m_1}\delta_{\nu}{}^{n_1}\delta_{\rho}{}^{m_2}\delta_{\sigma}{}^{n_2})\,C_{a(k);\,m_1n_1|m_2n_2}
\end{equation}
where we have introduced an infinite-dimensional jet of dynamical variables:
%
\begin{equation}\label{Jets2}C^{[2]}(x)\equiv\left\{C_{a(k);\,m_1n_1|\,m_2n_2}(x)\right\}_{k=0,\cdots,\infty}
~\sim~\left\{~\text{\tiny$\begin{aligned}&\begin{tabular}{|c|c|}\hline\phantom{L}&\phantom{L}\\\hline
\end{tabular}\\[-4pt]&\begin{tabular}{|c|c|}\phantom{L}&\phantom{L}\\\hline\end{tabular}\end{aligned}$}
~\otimes~\overbrace{\text{\tiny{$\begin{tabular}{|c|c|c|c|}\hline \phantom{L}&\multicolumn{2}{|c|}
{$~~~\cdots~~~$}&\phantom{L}\\\hline\end{tabular}$}}}^{\displaystyle{k}}~\right\}_{k=0,\cdots,\infty}\,.
\end{equation}
The compatibility condition for \eqref{S2comp} is the same as for lower spins:
\begin{equation}
(d-\omega^{ab}\hat{L}_{ab}-h^a\hat{P}_a)C^{[2]}\equiv \tadD C^{[2]}=0\,,
\end{equation}
where again we have introduced Lorentz covariant derivative $\nabla=dx^\mu\,\nabla_\mu$ and frame $h^a=dx^\mu\delta_\mu{}^a$, with the following representation for Lorentz 
and translation generators as infinite dimensional matrices:
\begin{subequations}
\begin{align}
&\delta_\epsilon C^{[2]}=\epsilon^a \hat{P}_aC^{[2]}=\{\epsilon^{\mu}h_{\mu}{}^b C_{ba(k);\,m_1n_1|\,m_2,n_2}\}_{k=0,\cdots,\infty}\,,\label{strS2_1}\\[3pt]
&\delta_\Lambda C^{[2]}=\Lambda^{ab} \hat{L}_{ab}C^{[2]}\nonumber\\[3pt]
&~~=\left\{\Lambda_{a}{}^b C_{ba(k-1);\,m_1n_1|\,m_2,n_2}(x)+\Lambda_{[m_1}{}^c C_{a(k);\,c|\,n_1;\,m_2n_2]}(x)+\cdots\right\}_{k=0,\cdots,\infty}\,.\label{strS2_2}
\end{align}
\end{subequations}
The last equation is simply the statement that each tensor transforms as a Lorentz tensor.
The on-shell condition requires the analysis of Bianchi identities and in particular how they are mapped into irreducibility conditions for the jet $C^{[2]}$. Analogous to the spin-1 case, plugging the ansatz \eqref{S2comp} into Bargmann-Wigner equations, one recovers that on-shell some of the irreducible components of $C^{[2]}(x)$ are forced to vanish. The on-shell condition can be shown to be equivalent to:
%
\be C^{[2]}~~\approx~~\left\{~\overbrace{\begin{aligned}&\begin{tabular}{|c|c|c|c|c|}\hline
$\phantom{a1}$&$\phantom{a1}$&\multicolumn{2}{|c|}{$~~~\cdots~~~$}&\phantom{a1}\\\hline\end{tabular}\\[-4pt]
&\begin{tabular}{|c|c|}$\phantom{a1}$&$\phantom{a1}$\\\hline\end{tabular}\end{aligned}}^{\displaystyle{k+2}}
~\right\}_{k=0,\cdots,\infty}~~~,\ee
in terms of $o(d)$ traceless Young diagrams. This restricts the original jet \eqref{Jets2} to the corresponding on-shell subspace. Otherwise, all other components and traces violate the on-shell conditions.


\begin{exercise}
Starting from the manifestly antisymmetric basis and following the $s=1$ example and exercise \ref{Ex2}, prove that the action of the translation 
generator in the linearized spin-2 case in the symmetric basis takes the following form:
\begin{subequations}
\begin{align}
C^{[2]}(x)&\equiv{C_{a(k+2)\,,b(2)}}_{k=0,\cdots,\infty}\,,\\[3pt]
\hat{P}_nC^{[2]}(x)&=\left\{C_{a(k+2)n,\,b(2)}+\tfrac{1}{k+2}\,C_{a(k+2)b,\,bn}\right\}_{k=0,\cdots,\infty}\,.
\end{align}
\end{subequations}
\end{exercise}

\subparagraph{Generic Fronsdal field:}
The above examples make it all very clear how to proceed for generic spin. One starts by recalling the Bargmann-Wigner equations \eqref{w-curvature1}--\eqref{w-curvature3}:
\begin{subequations}
\begin{align}
\square R_{\mu_1\nu_1|\,\cdots\,|\,\mu_s\nu_s}&=0\,,\\
\partial^{\mu_1} R_{\mu_1\nu_1|\,\cdots\,|\,\mu_s\nu_s}&=0\,,\\
\partial_{[\rho_1} R_{\mu_1\nu_1]\,|\,\cdots\,|\,\mu_s\nu_s}&=0\,,
\end{align}
\end{subequations}
written in terms of a traceless two-row tensor $R_{\mu_1\nu_1|\,\cdots\,|\,\mu_s\nu_s}$ (HS Weyl tensor).
One then introduces an infinite-dimensional jet of dynamical variables:
%
%
\begin{equation}\label{Weyl_s}
C^{[s]}(x)\equiv\left\{C_{a(k)|\mu_1\nu_1|\,\cdots\,|\,\mu_s\nu_s}(x)\right\}_{k=0,\cdots,\infty}
~\sim~\left\{~\overbrace{\text{\tiny$\begin{aligned}&\begin{tabular}{|c|c|c|c|}\hline\phantom{L}
&\multicolumn{2}{|c|}{$~~~\cdots~~~$}&\phantom{L}\\\hline\end{tabular}\\[-4pt]&\begin{tabular}{|c|c|c|c|}\phantom{L}&\multicolumn{2}{|c|}{$~~~\cdots~~~$}&\phantom{L}\\\hline\end{tabular}\end{aligned}$}}
^{\displaystyle{s}}~\otimes~\overbrace{\text{\tiny{$\begin{tabular}{|c|c|c|c|}\hline\phantom{L}
&\multicolumn{2}{|c|}{$~~~\cdots~~~$}&\phantom{L}\\\hline\end{tabular}$}}}^{\displaystyle{k}}
~\right\}_{k=0,\cdots,\infty}\,.\end{equation}
They reconstruct the Taylor expansion of the HS Weyl tensors at any point $x_0$:
\begin{equation}\label{Sscomp}
R_{\mu_1\nu_1|\,\cdots\,|\,\mu_s\nu_s}=\sum_{k=0}^\infty \frac1{k!} \left(\delta_{\mu_1}{}^{m_1}\delta_{\nu_1}{}^{n_1}\cdots \delta_{\mu_s}{}^{m_s}\delta_{\nu_s}{}^{n_s}\right)\,C_{a(k);\,m_1n_1|\,\cdots\,|\,m_sn_s}(x_0)\,(x-x_0)^{a(k)}\,,
\end{equation}
provided the following compatibility condition is enforced:
\begin{equation}
(d-\omega^{ab}\hat{L}_{ab}-h^a\hat{P}_a)C^{[s]}\equiv \tadD C^{[s]}=0\,.
\end{equation}

The Bargmann-Wigner on-shell conditions become, in terms of the infinite dimensional vector $C^{[s]}(x)$, completely algebraic allowing on-shell 
only the following non-zero $o(d)$ traceless components:
%
\be C^{[s]}~~\approx~~\left\{~\begin{aligned}&\overbrace{\begin{tabular}{|c|c|c|}\hline\multicolumn{3}
{|c|}{$~~~\cdots~~~\cdots~~~\cdots~~~$}\\\hline\end{tabular}}^{\displaystyle{k+s}}\\[-4pt]
&\underbrace{\begin{tabular}{|c|c|}\multicolumn{2}{|c|}{$~~~\cdots~~~\cdots~~~$}\\\hline\end{tabular}
}_{\displaystyle{s}}\end{aligned}~\right\}_{k=0,\cdots,\infty}~~~.\ee
\begin{svgraybox}
To summarize, the unfolded version of the Bargmann-Wigner equations for arbitrary symmetric spins can be recast as:
\begin{align}
\tadD C^{[s]}\equiv(d-\omega^{ab}\hat{L}_{ab}-h^a\hat{P}_a)C^{[s]}=0,
\end{align}
in terms of the spin-$s$ module:
\begin{equation}
C^{[s]}(x)=\left\{C_{a(k);\,m_1n_1|\,\cdots\,|\,m_sn_s}(x)\right\}^{[2,\ldots,2,1,\ldots,1]}_{k=0,\cdots,\infty}\,,
\end{equation}
each component of which transforms as a Lorentz tensor. The translation generator acts on the indecomposable module as:
\begin{equation}
\hat{P}_bC^{[s]}=\{C_{ba(k);\,m_1n_1|\,\cdots\,|\,m_s,n_s}\}_{k=0,\cdots,\infty}\,,\label{strS2_s}
\end{equation}
in the antisymmetric basis. In the symmetric basis, the corresponding module reads:
\begin{equation}
C^{[s]}(x)=\left\{C_{a(s+k),\,b(s)}(x)\right\}^{(s+k,\,s)}_{k=0,\cdots,\infty}\,,
\end{equation}
while the action of the translation generator can be expressed as:
\begin{equation}
\hat{P}_nC^{[s]}(x)=\left\{C_{a(s+k)n,\,b(s)}+\tfrac{1}{k+2}\,C_{a(s+k)b,\,b(s-1)n}\right\}_{k=0,\cdots,\infty}\,.
\end{equation}\vspace{-15px}
\end{svgraybox}

As anticipated, the above equations make manifest how the modules $C^{[s]}(x)$ transform as infinite dimensional unitary indecomposable representations
of the Poincar\'e group. It might be useful to stress here that unfolding is at this level just a change of dynamical variables exactly on the same 
footing as the more standard Fourier transform, which replaces field variables with plane waves. The above jet vectors $C^{[s]}$ are usually referred to 
as Weyl modules, because they encode purely gauge-invariant information of the system--the degrees of freedom or moduli.

\subsubsection{Unfolding the Killing Equation: The Gauge Module}\label{sec:gaugemod}

The introduction of gauge potentials was a crucial ingredient in order to write consistent interacting theories beyond the free case. Therefore, it is useful 
to introduce gauge potentials also in the unfolded setting. In the following we will describe how this can be achieved. The way we choose to this effect is to 
first clarify the HS rigid symmetries behind each Weyl module. One can then introduce the gauge potentials as linearized connections of the corresponding HS 
rigid symmetries.

In analogy with Yang-Mills theory, in order to introduce gauge potentials it is convenient to first identify the corresponding global/rigid symmetries. These can 
be promoted to a local gauge-invariance principle by introducing appropriate gauge connections. In the case of Yang-Mills theory, for instance, these rigid 
symmetries are usually compact groups like $U(N)$ or simply $U(1)$ in the case of Maxwell electrodynamics that we have described above.

A convenient way to extract the information about rigid symmetries within a gauge theory setting is to solve the so called Killing equations. In the following 
we will study the Killing equations of the Fronsdal theory. The logic we are going to follow is to gauge the corresponding{\it unfolded} rigid symmetries.

\subparagraph{Spin 1:}
The simplest, although slightly trivial, example is the spin-1 case where the corresponding Killing equation written in arbitrary coordinates takes the form:
\begin{equation}
\nabla \xi=0\,.
\end{equation}
The above equation admits as solutions constant functions $\xi$. This is in agreement with the fact that a spin-1 field is gauging an internal symmetry whose 
parameters are space-time scalars. A set of scalar generators parameterizes internal symmetries of the theory like Chan-Paton factors \cite{Paton:1969je}.

\subparagraph{Spin 2:}
Already in the spin-2 case the discussion becomes more interesting. The Killing equation now takes the following form:
\be \nabla_{(\m}\xi_{\n)}=0\,,\ee
while the gauge parameter carries a Lorentz index.
Therefore, the gauge parameter transforms non-trivially under the space-time isometry. Furthermore, the solutions of the above equation are in one-to-one 
correspondence with the generators of the isometry algebra of the vacuum.

What we would like to do in the following is to consider an change of variables analogous to that considered in the previous section, introducing a Taylor expansion 
ansatz for the tensor $\xi_{\m}$:
\be \xi_{\m}=\sum_{k=0}^\infty\frac1{k!}\delta_{\m}{}^m\xi_{a(k);\,m}(x_0)(x-x_0)^{a(k)}\,.\label{notsoimportant0}
\ee
The decomposition of the above Taylor coefficient is given by:
%
\be\label{kx1}\begin{tabular}{|c|}\hline\phantom{a1}\\\hline\end{tabular}~~\otimes~~
\overbrace{\begin{tabular}{|c|c|c|c|}\hline \phantom{a1}&\multicolumn{2}{|c|}{$~~~\cdots~~~$}
&\phantom{a1}\\\hline\end{tabular}}^{\displaystyle{k}}~~=~~\overbrace{\begin{aligned}&\begin{tabular}
{|c|c|c|c|}\hline$\phantom{a1}$&\multicolumn{2}{|c|}{$~~~\cdots~~~$}&\phantom{a1}\\\hline\end{tabular}
\\[-4pt]&\begin{tabular}{|c|}$\phantom{a1}$\\\hline\end{tabular}\end{aligned}}^{\displaystyle{k}}~~
\oplus~~\overbrace{\begin{tabular}{|c|c|c|c|}\hline \phantom{a1}&\multicolumn{2}{|c|}{$~~~\cdots~~~$}
&\phantom{a1}\\\hline\end{tabular}}^{\displaystyle{k+1}}~~~.\ee
while the Killing equation is mapped into an algebraic condition on the variables:
\be \xi_{a(k-1)(b;\,m)}=0\,,\qquad k\geq1\,.\label{Ks1}\ee
The $k=0$ term satisfies the equation trivially, being the constant piece of the Taylor expansion. For $k>0$
the above condition can have non-trivial solutions only if $k=1$, due to the fact that this is the only case
in which Eq.~\eqref{Ks1} is mapped into a proper Young-symmetry projection condition. In this case, it requires
that the symmetric component in \eqref{kx1} be set to zero. We can then conclude that the most general solution to
the Killing equation is just a polynomial:
\be\xi_{\m}(x)=\delta_{\m}{}^m\left(\xi_{m}(x_0)+\tfrac{1}{2}\xi_{[a;\,m]}(x_0)(x-x_0)^{a}\right)\,.\ee
This 
also demonstrates that the Taylor coefficients (\ref{notsoimportant0}) are in one-to-one
correspondence with the isometry generators of the vacuum$-$in this case translations $\xi_{m}(x_0)$ and the
antisymmetric rotation generators $\xi_{[a;\,m]}(x_0)$.

Having restricted the structure of the Taylor coefficients (which we have analyzed first in order to restrict the
attention to a finite number of Taylor coefficients) the next step is to ensure the above polynomial
actually defines the same function when changing the base point $x_0$. This condition amounts in this case to:
\be\epsilon^c\left(\nabla_c\xi_a(x_0)-\xi_{a,c}(x_0)+\nabla_c\xi_{a,\,b}(x_0)(x-x_0)^b\right)=0\,.\ee
This leads, as for the Weyl module, to a covariant constancy condition for $\xi^{[2]}$:
\be\adD\xi^{[2]}=(d-\omega^{ab}\hat{L}_{ab}-h^a\hat{P}_a)\xi^{[2]}=0\,,\ee
where:
\be\xi^{[2]}(x)=\left(\begin{matrix}\xi_a(x)\\[2pt]\xi_{a,\,b}(x)\end{matrix}\right)\,.\ee
The explicit realization of the Lorentz generators is again such that each component is a Lorentz tensor. On the other hand, the
explicit realization of the translation generator $\hat{P}_a$ is now different with respect to what we have obtained in
the case of the Weyl module in the previous section. This is indeed expected, since the modules are different: the Weyl module is 
unitary and infinite-dimensional while Killing tensor module is finite dimensional. Looking explicitly at the
transformation properties of the polynomial solution above, we find upon shifting $x\rightarrow x+\epsilon$ and keeping
only the linear order in $\epsilon$:
\be\hat{P}_b\left(\begin{matrix}\xi_a(x)\\[2pt]\xi_{a,b}(x)\end{matrix}\right)=\left(\begin{matrix}
\xi_{b,a}(x)\\[2pt]0\end{matrix}\right)\,.\ee

\subparagraph{Spin $s$:}

Everything we have done so far can be easily generalized to HS Killing equations, starting from the HS
Killing equation of the Fronsdal theory:
\begin{subequations}
\begin{align}
\nabla_{\m}\xi_{\m(s-1)}&=0\,,\\
\xi_{\m(s-3)\n}{}^{\n}&=0\,.
\end{align}
\end{subequations}
In this case, the unfolding is achieved by the following Taylor expansion ansatz:
\be \xi_{\m(s-1)}(x)=\sum_{k=0}^\infty\frac1{k!}\underbrace{\delta_{\m}{}^b\cdots \delta_{\m}{}^b}_{s-1}\xi_{a(k);\,b(s-1)}(x_0)(x-x_0)^{a(k)}\,.\ee
By substituting it into the Killing equation, the differential equation is mapped into an algebraic condition that this time reads:
\be \xi_{b(k-1)a;\,a(s-1)}=0\,.\ee
For arbitrary $s$ it admits finitely many non-trivial solutions given by
\be\xi_{b(k);\,a(s-1)}=\xi^{(s-1,\,k\leq s-1)}_{a(s-1),\,b(k)}\,,\ee
where on the right hand side the tensor is an irreducible traceless tensor with Young tableaux $(s-1,\,k\leq s-1)$ in the symmetric basis:
%
%
\be \xi^{[s]}~~\sim~~\left\{~\begin{aligned}&\overbrace{\begin{tabular}{|c|c|c|}\hline\multicolumn{3}
{|c|}{$~~~\cdots~~~\cdots~~~\cdots~~~$}\\\hline\end{tabular}}^{\displaystyle{s-1}}\\[-4pt]
&\underbrace{\begin{tabular}{|c|c|}\multicolumn{2}{|c|}{$~~~\cdots~~~\cdots~~~$}\\\hline\end{tabular}
}_{\displaystyle{k}}\end{aligned}~\right\}_{k=0,\cdots,\,s-1}~~~.\ee
Working out the action of the translation generator one gets, similarly as for the lower spin cases:
\begin{equation}
\hat{P}_c\xi^{[s]}(x)=\left\{\xi_{a(s-1),\,b(k)c}(x)\right\}_{k=0,\cdots,\,s-1}\,,\qquad \xi_{a(s-1),\,b(s-1)c}(x)\equiv0\,,\end{equation}
which is the action of the translation generators on the Killing tensor module. This module carried by the HS algebra generators generalizes
the isometry generators obtained for $s=2$.

The above analysis is tantamount to a classification of HS generators which underline Fronsdal HS fields. In algebraic terms we have been
able to fix the structure constants carried by HS generators when branched with respect to the background isometry generators, namely:
%
%
\be\left[~T_2~,~T_2~\right]~\sim~T_2~,\qquad\Big[~T_2~,~T_{\includegraphics[width=0.35cm]{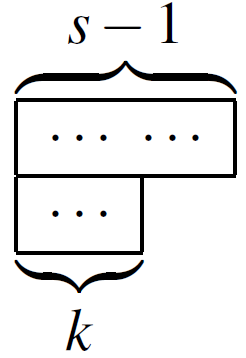}}
~\Big]~\sim~T_{\includegraphics[width=0.35cm]{index.png}}~.\ee
%
%
The non-trivial commutation relation between the HS generators and the isometry generators is a key ingredient behind any HS theory.
The missing ingredients are however the remaining commutation relations:
\be \Big[~T_{\includegraphics[width=0.35cm]{index.png}}~,~T_{\includegraphics[width=0.35cm]{index.png}}
~\Big]~\sim~\cdots\,,\ee
%
%
which will allow one to define corresponding HS algebras (see e.g. \cite{Vasiliev:2004cm,Boulanger:2013zza,Joung:2013nma,Govil:2013uta,Govil:2014uwa,Joung:2014qya,Fernando:2015tiu}).

\subparagraph{Introducing gauge fields:}
Having in detail the structure of the Killing module, one can introduce gauge fields along the lines of the Maxwell case by considering a $1$-form gauge connection 
transforming in the same module:
\begin{equation}
\omega(x)=\omega_\mu dx^\mu=\left\{\omega_{a(s-1),\,b(k)}(x)\right\}_{k=0,\cdots, s-1}\,.
\end{equation}
In more complicated cases, when the global-symmetry generators have gauge-for-gauge components, one would also be forced to introduce higher-form degree connections \cite{Alkalaev:2003qv,Alkalaev:2003hc,Skvortsov:2009zu,Skvortsov:2009nv}.

The remaining construction just follows the YM case, so that the associated ``linearized'' two-form curvature can be defined exactly as in YM theory by:
\begin{equation}
R=\adD\omega\,,
\end{equation}
in terms of the adjoint covariant derivative $\adD$.

On-shell the curvature $R$ must be related to the Weyl module described in the previous section. Indeed, the linearized curvature should give rise on-shell to the
linearized HS-Weyl tensor. This mimics the gravitational case where Einstein equations are tantamount to the tracelessness of the Riemann tensor. More explicitly,
in components the most general linear equations that can be written respecting form degree read:
\begin{subequations}
\begin{align}
\adD\omega_{a(s-1),\,b(k)}&=0\,,\qquad k<s-1\,,\\
\adD\omega_{a(s-1),\,b(s-1)}&=h^{c}\wedge h^d C_{a(s-1)c,b(s-1)d}\,.
\end{align}
\end{subequations}
The above can be interpreted as the most general coordinate independent solution to the Bianchi identity equation in terms of gauge potentials. In the unfolded
language it encodes the non-trivial cohomology in form degree 2 of the adjoint derivative $\adD$ (see e.g. \cite{Kessel:2015kna,Boulanger:2015ova} for a detailed account).

Notice that the Bianchi identity and the existence of the above non-trivial right-hand side is a necessary condition to link the gauge potential to the curvature!
A zero-curvature condition would otherwise give only pure gauge solutions with no propagating degree of freedom stored into the Weyl-module.

To summarize, the Weyl tensors encoding the local degrees of freedom/moduli of the system appear here as non-trivial cohomologies that can source the curvature,
resulting in propagating degrees of freedom.
\bex Check that the system of unfolded equations:
{\allowdisplaybreaks
\begin{subequations}
\begin{align}
(\adD\omega)_{a(s-1),\,b(k)}&=0\,,\qquad k<s-1\,,\\
(\adD\omega)_{a(s-1),\,b(s-1)}&=h^{c}\wedge h^d C_{a(s-1)c,\,b(s-1)d}\,,\\
(\tadD C)_{a(s+k),\,b(s)}&=0\,,
\end{align}
\end{subequations}}

\noindent are compatible with $d^2=0$. Recall that $\adD=d-\omega^{ab}\hat{L}_{ab}-h^a \hat{P}_a$ and the Lorentz and translation generators act on each module as computed above.\eex
\begin{exercise}
Check that the compatibility of the above system of equations with $d^2=0$ implies the following gauge symmetries:
\begin{subequations}
\begin{align}
\delta_{\xi} \omega(x)&=\adD\xi(x)\,,\\
\delta_{\xi} C(x)&=0\,,
\end{align}
\end{subequations}
compatible with the interpretation of HS gauge potentials as the gauge fields gauging the rigid symmetries described above. The Weyl module is gauge invariant at the linear level. Notice however that it transforms covariantly under the isometry action. The gauge-module is gauge dependent. All in all, the above is nothing but a decomposition of the HS jet space associated to a Fronsdal field into: 1) On-shell components which are set to zero algebraically, 2) gauge covariant components which carry the degrees of freedom/moduli of the system and 3) the gauge dependent components that are naturally described by one-forms as connections gauging the HS global symmetries (see e.g. \cite{Barnich:2004cr,Barnich:2010sw,Grigoriev:2012xg} for further details).
\end{exercise}

\subsection{Unfolded Equations in Anti-de Sitter Space}\label{sec:AdS_Unf}

In this section we discuss the AdS extension of the above unfolded equations describing HS fields, pointing out the key aspects.
The basic difference between the AdS and flat space can be appreciated by looking at the generic Taylor expansion for a scalar field in normal-coordinates:
\be
\phi(x)=\sum_{k=0}^\infty\frac1{k!}\left(\nabla_{(a_1}\cdots\nabla_{a_k)}\phi(x_0)\right)(x-x_0)^{a(k)}\,.
\ee
The compatibility condition for the Taylor coefficients:
\be
\tilde{\phi}(x)=\{\nabla_{a(k)}\phi(x)\}\,,
\ee
does not change and takes exactly the same form as in flat space:
\be
\left(\nabla-h^a\hat{P}_a\right)\tilde\phi(x)=0\,,
\ee
with the following representation for the translation generator:
\be
\hat{P}_b\left(\nabla_{(a_1}\cdots\nabla_{a_k)}\phi(x)\right)=\nabla_{(a_1}\cdots\nabla_{a_k}\nabla_{b)}\phi(x)\,.
\ee
However, the difference is that the Taylor coefficients $\tilde{\phi}(x)$ do not correspond in this case to the proper unfolded variables:
\be
C_{a(k)}\neq\left(\nabla_{(a_1}\ldots\nabla_{a_k)}\phi(x)\right)\,.
\ee
The reason for this is that $\nabla_{a(k)}\phi(x)$ is not a traceless irreducible tensor. One should consider a traceless projection, which in this case reads:
\bea \nabla_{a\{k\}b}\phi(x)&=&\nabla_{\{a\{k\}b\}}\phi(x)+f_{k}^{\,m}\eta_{\{a|b}\nabla_{a\{k-1\}\}}\phi(x)\nonumber\\[3pt]
&=&C_{a(k)b}+f_{k-1}^{\,m}\left(\eta_{ab}C_{a(k-1)}-\tfrac{2}{d+2k-4}\,\eta_{a(2)}C_{ba(k-2)}\right)\,.\label{whatacartoon}\eea
Here, the $\{\}$ bracket notation implies that the corresponding indices are projected into their traceless components. The very reason this needs to be done is for instance:
\be \h^{a_1a_2}\nabla_{(a_1}\nabla_{a_2}\nabla_{a_3)}\phi=-\tfrac{2}{3}\Lambda(d-1)\nabla_{a_3}\phi\,,\qquad\Box\phi(x)=0\,.\ee
The above gives an example of how, in the massless case, on curved backgrounds symmetrized derivatives have non-vanishing trace components on-shell contrary to the flat space case. To take into account this feature, one has to introduce coefficients $f_{k}^{\,m}$ which will depend on the mass-shell of the scalar and that parameterize the most general traceless decomposition. Again the main logic/feature of the unfolded-formalism is to translate dynamical information into purely algebraic conditions. These are related in this case to the traceless decomposition of the Taylor-coefficient dynamical variables. One can then see that the condition:
\be\label{scalarAdS}
\left(\nabla-h^a\hat{P}_a\right)C^{[0]}(x)=0\,,
\ee
can be translated in terms of the unfolded variables $C_{a(k)}$ by specifying the following action of the translation generator:
\be\hat{P}_bC^{[0]}=\left\{C_{a(k)b}+f_{k-1}^{\,m}\left(\eta_{ab}C_{a(k-1)}-\tfrac{2}{d+2k-4}\,\eta_{a(2)}C_{ba(k-2)}\right)\right\}_{k=0,\cdots,\infty}\,.\ee
It is now sufficient to use:
\be
\nabla^2C_{a(k)}=\Lambda H_{a}{}^bC_{ba(k-1)}\,,\qquad H_{ab}=h_a\wedge h_b\,,
\ee
together with
\bea
\nabla C_{a(k)b}&=&h^c\Big[C_{a(k)bc}+f_{k}^{\,m}\left(\eta_{bc}C_{a(k)}+\eta_{ac}C_{ba(k-1)}\right)\nonumber\\[3pt]
&&~~~~~-\tfrac{2}{d+2k-2}f_{k}^{\,m}\left(\eta_{ab}C_{ca(k-1)}+\,\eta_{a(2)}C_{cba(k-2)}\right)\Big]\,,\nonumber\\[3pt]
\nabla C_{a(k-1)}&=&h^c\left[C_{a(k-1)c}+f_{k-2}^{\,m}\left(\eta_{ac}C_{a(k-2)}-\tfrac2{d+2k-6}\,\eta_{a(2)}C_{a(k-3)c}\right)\right]\,,\nonumber
\eea
to obtain the following recursion relation:
\be f_k^m=\tfrac{d+2(k-1)}{d+2k}\left(f^m_{k-1}+\Lambda\right)\,,\ee
whose solution can be expressed as
\be f_k^m=\tfrac{1}{d+2k}\left[\Lambda k(k+d-1)+m^2\right]\,.\ee
Considering the unfolded equations \eqref{scalarAdS} for $k=0,1$ it is easy to see that only $f_k^m$ for $k=0$ contributes and one can extract the Klein-Gordon 
equation which justifies the interpretation of $m^2$ as mass term:
\be
(\Box-m^2)C(x)=0\,.
\ee
The conformally coupled scalar is picked out by the choice $m^2=-\frac{\Lambda}4 d(d-2)$. Notice also that on AdS one recovers non-unitary discrete points for 
which degeneracies appear in the coefficients $f_k^m$ (see \cite{Shaynkman:2000ts}).
\bex
Unfold the Maxwell equations on AdS by using the appropriate traceless projectors for the corresponding two-raw young tableaux. First show that the most general 
structure constants for the action of the translation generator on the spin-1 massless module read:
\beq
h^a \hat{P}_{a}(C^{[1]})=\sigma_-^1(C^{[1]})+f^{m}_{+,k}\sigma_+^1(C^{[1]})\,,\label{ansatzA}
\eeq
with
\begin{subequations}\label{sigma1}
\begin{align}
&\sigma_-^1(C_{a(k),\,b})=C_{a(k-1)m,b}\,h^m+\tfrac{1}{k} C_{a(k-1)b,\,m}\,h^m\,,\\[3pt]
&\sigma_+^1(C_{a(k),\,b})=h_a C_{a(k),\,b}-\tfrac{2}{d+2k-2} \eta_{aa}C_{a(k-1)m,\,b}\,h^m\nonumber\\[3pt]
&~~~~~~~~~~-\tfrac{1}{d+k-2}\,\eta_{ab}C_{a(k),\,m}h^m+\tfrac{2}{(d+2k-2)(d+k-2)}\eta_{aa}C_{a(k-1)b,\,m}\,h^m\,.
\end{align}
\end{subequations}
Hint: consider the part of the decomposition of the tensor product of a vector with a $(k,1)$ tensor which keeps the structure of the spin-1 massless module.

Fix the AdS mass and the coefficients $f^{m}_{+,\,k}$ following similar steps as in the scalar case. Notice that to describe a massive vector one 
would need to enlarge \eqref{ansatzA} by considering also a scalar modulxe on top of the massless spin-1 module. The scalar module would play the role 
of St\"uckelberg component for the massive vector field.
\eex
\bex
The $\sigma$-operators appearing above are sometimes referred to as ``cell-operators'', and parameterize the most general action of the translation generator on a given jet. 
The origin of the name comes from the fact that the translation generator has a single index usually depicted as a cell in the Young-diagram notation. The cell-operators 
in \eqref{sigma1} are those two which do not change the length of the second row while increase or reduce by one unit the length of the first row. In general, one can 
define the cell operators $\sigma^{i}_{\pm}$ which add/subtract a cell to the $i$-th row of the corresponding Young tableaux in a way compatible with irreducibility. 
For instance, in the two row case there are $4$ cell-operators associated to the following decomposition of the tensor product of a two-row tensor with the translation 
generator:
%
%
\be\text{\tiny$\begin{tabular}{|c|}\hline\phantom{L}\\\hline\end{tabular}$}
~~\otimes~~
\text{\tiny$\begin{aligned}&\overbrace{\begin{tabular}{|c|c|c|}\hline\multicolumn{3}
{|c|}{$~~~\cdots~~~\cdots~~~$}\\\hline\end{tabular}}^{\displaystyle{k}}\\[-4pt]
&\underbrace{\begin{tabular}{|c|c|}\multicolumn{2}{|c|}{$~~~\cdots~~~$}\\\hline\end{tabular}}
_{\displaystyle{l}}\end{aligned}$}
~~=~~\underbrace{
\text{\tiny$\begin{aligned}&\overbrace{\begin{tabular}{|c|c|c|}\hline\multicolumn{3}
{|c|}{$~~~\cdots~~~\cdots~~~$}\\\hline\end{tabular}}^{\displaystyle{k+1}}\\[-4pt]
&\underbrace{\begin{tabular}{|c|c|}\multicolumn{2}{|c|}{$~~~\cdots~~~$}
\\\hline\end{tabular}}_{\displaystyle{l}}\end{aligned}$}
}_{\sigma^1_+}~~\oplus~~\underbrace{
\text{\tiny$\begin{aligned}&\overbrace{\begin{tabular}{|c|c|c|}\hline\multicolumn{3}
{|c|}{$~~~\cdots~~~\cdots~~~$}\\\hline\end{tabular}}^{\displaystyle{k-1}}\\[-4pt]
&\underbrace{\begin{tabular}{|c|c|}\multicolumn{2}{|c|}{$~~~\cdots~~~$}
\\\hline\end{tabular}}_{\displaystyle{l}}\end{aligned}$}
}_{\sigma^1_-}~~\oplus~~\underbrace{
\text{\tiny$\begin{aligned}&\overbrace{\begin{tabular}{|c|c|c|}\hline\multicolumn{3}
{|c|}{$~~~\cdots~~~\cdots~~~$}\\\hline\end{tabular}}^{\displaystyle{k}}\\[-4pt]
&\underbrace{\begin{tabular}{|c|c|}\multicolumn{2}{|c|}{$~~~\cdots~~~$}
\\\hline\end{tabular}}_{\displaystyle{l+1}}\end{aligned}$}
}_{\sigma^2_+}~~\oplus~~\underbrace{
\text{\tiny$\begin{aligned}&\overbrace{\begin{tabular}{|c|c|c|}\hline\multicolumn{3}
{|c|}{$~~~\cdots~~~\cdots~~~$}\\\hline\end{tabular}}^{\displaystyle{k}}\\[-4pt]
&\underbrace{\begin{tabular}{|c|c|}\multicolumn{2}{|c|}{$~~~\cdots~~~$}
\\\hline\end{tabular}}_{\displaystyle{l-1}}\end{aligned}$}
}_{\sigma^2_-}~~\oplus~~\cdots~~.\ee

By imposing tracelessness and ensuring the appropriate Young symmetry property, show that the cell operators for two-row Young tableaux read:
\begin{subequations}
\begin{align}
&\sigma_-^1\left(C_{a(k),\,b(l)}\right)~=~C_{a(k-1)m,\,b(l)}\,h^m+\tfrac{1}{k-l+1}\,C_{a(k-1)b,\,b(l-1)m}\,h^m\,,\\[5pt]
&\sigma_+^1\left(C_{a(k),\,b(l)}\right)~=~h_a C_{a(k),\,b(l)}-\tfrac{2}{d+2k-2}\eta_{aa}C_{a(k-1)m,\,b(l)}\,h^m\nonumber\\[5pt]
&-\tfrac1{d+k+l-3}\eta_{ab}C_{a(k),\,b(l-1)m}\,h^m+\tfrac{2}{(d+2k-2)(d+k+l-3)}\eta_{aa}C_{a(k-1)b,\,b(l-1)m}\,h^m\,,\\[5pt]
&\sigma_-^2\left(C_{a(k),\,b(l)}\right)~=~C_{a(k),\,b(l-1)m}\,h^m\,,\\[5pt]
&\sigma_+^2\left(C_{a(k),\,b(l)}\right)~=~h_b C_{a(k),\,b(l)}-\tfrac{1}{k-l}\,h_a C_{a(k-1)b,\,b(l)}-\tfrac2{d+2l-4}\eta_{b(2)}C_{a(k),\,b(l-1)m}\,h^m\nonumber\\[5pt]
&~~~~~~~~~~~-\tfrac{k-l-1}{(k-l)(d+k+l-3)}\eta_{ab}C_{a(k-1)m,\,b(l)}h^m+\tfrac2{(k-l)(d+k+l-3)}\eta_{a(2)}C_{a(k-2)mb,\,b(l)}\,h^m\nonumber\\[5pt]
&~~~~~~~~~~~+\tfrac{d+2k-4}{(k-l)(d+2l-4)(d+k+l-3)}\eta_{ab}C_{a(k-1)b,\,b(l-1)m}\,h^m\nonumber\\[5pt]
&~~~~~~~~~~~-\tfrac{4}{(k-l)(d+k+l-3)(d+2l-4)} \eta_{a(2)}C_{a(k-1)b(2),\,b(l-1)m}\,h^m\,.
\end{align}
\end{subequations}
Conclude the exercise proving that the above cell operators are nilpotent. The most general ansatz for the action of translation generators on spin-$s$ particles 
involves all the above terms which do play a role for HS massive particles. For instance, the St\"uckelberg fields couple among each other via $\sigma^2_{\pm}$.\label{cell}
\eex
To summarize, the covariant derivative defined on a given jet vector splits into different pieces, which increase/decrease or leave invariant the rank of the tensors 
on which they act upon. They are the Lorentz derivative $\nabla$ which preserves the rank of the tensor and encodes the action of Lorentz generators on the jet components, 
plus the ``cell-operator'' which select irreducible components appearing in the action of the translation generator on the jet itself (see exercise \ref{cell}):
\be
\mathcal{D}=\nabla+\sigma_-+\sigma_+\,.\label{decomp}
\ee
Above we have labelled collectively all various cell-operators which decrease or increase the rank of the tensors on which they act upon. More explicitly we have 
introduced the notation $\sigma_{\pm}=\sum_i f_i\,\sigma^i_{\pm}$, for some constants $f_i$ specifying the action of the translation generator on the module.

Given a generic form degree-$n$ jet vector $\omega$ taking values in the jet bundle $[\omega]$, and decomposing it as in \eqref{decomp} according to the total number 
of indices: $\omega=\sum_i\omega_i$, one recovers the following system of unfolded equations:
\begin{subequations}
\begin{align}
\nabla\omega_k+\sigma_-\omega_{k+1}+\sigma_+\omega_{k-1}&=0\,,\label{1formeq}\\
\delta\omega_k=\nabla\xi_k+\sigma_-\xi_{k+1}+\sigma_+\xi_{k-1}&=0.\label{gaugetransf}
\end{align}
\end{subequations}
Physical fields should neither be auxiliary (expressible as derivatives of other fields) nor St\"uckelberg. Now, in the jet space context it is natural to consider an iteration starting from\footnote{In the jet space theory, truncations to finitely many derivatives ($k<\infty$ in our language) are at the basis of the notion of inverse limit. See e.g.~\cite{Krasil'shchik:2010ij}.} $k=0$ so that the $\sigma_-$ operator assumes a distinguished role with respect to $\sigma_+$, which vanishes for $k=0$: $\xi_{k<0}\equiv0$. One can then see that:
\begin{itemize}
\item A component of the jet is not auxiliary if $\sigma_-\omega_k=0$. In all other cases $\sigma_-$ can be inverted and one can express $\omega_k$ in terms of lower jet components:
\be
\omega_{k}=-(\sigma_-)^{-1}\left(\nabla\omega_{k-1}-\sigma_+\omega_{k-2}\right)\,.
\ee
Analogously, a jet component is not St\"uckelberg if $\omega_k\neq\sigma_-\xi_{k+1}$. Indeed, otherwise $\omega_k$ may be iteratively removed by a proper choice of $\xi_{k+1}$. Therefore, physical components of the $p$-form jet are non-vanishing elements of $\mathbb{H}^p(\sigma_-,[\omega])$ with coefficient on the jet $[\omega]$.

\item While we have already shown that if $\sigma_-\xi_k\neq0$ we can fix the corresponding gauge parameter by removing St\"uckelberg components, one should also take into account possible reducibility of gauge parameters: $\delta \xi=\mathcal{D}\epsilon$. Hence any $\xi_k=\sigma_-\epsilon_{k+1}$ can be iteratively removed by appropriate choices of $\epsilon_{k+1}$, analogously as for St\"uckelberg components $\omega_{k}$ above. Non-trivial gauge symmetries of the $p$-form jet $\omega$ are then parameterized by $\mathbb{H}^{p-1}(\sigma_-,[\omega])$.

\item As for fields and gauge parameters, also differential equations are not all independent. Only some components encode the non-trivial conditions while all others are differential consequences thereof. In order to clarify which one are the non-trivial differential conditions one must take into account Bianchi identities encoded into the compatibility condition $\mathcal{D}^2=0$. In components one obtains:
\be
\sigma_-(\mathcal{D}\omega)_{k+1}=-\nabla(\mathcal{D}\omega)_{k}-\sigma_+(\mathcal{D}\omega)_{k-1}\,.
\ee
Hence, any component of $(\mathcal{D}\omega)_{k+1}$ which is not annihilated by $\sigma_-$ is a differential consequences of lower $k$ equations $(\mathcal{D}\omega)_{k}$ and $(\mathcal{D}\omega)_{k-1}$:
\be
(\mathcal{D}\omega)_{k+1}=-(\sigma_-)^{-1}\left[\nabla(\mathcal{D}\omega)_{k}+\sigma_+(\mathcal{D}\omega)_{k-1}\right].
\ee
Non-trivial differential consequences must then be closed: $\sigma_-(\mathcal{D}\omega)_{k+1}=0$, while $\sigma_-$ exact terms can be removed by appropriate 
redefinitions of $\omega_k\rightarrow \omega_k+f_k$. Hence, the physical differential conditions for a $p$-form jet sits into $\mathbb{H}^{p+1}(\sigma_-,[\omega])$. 
Notice that when the latter cohomology group is empty, no non-trivial differential condition is imposed while unfolded equations simply express different components 
as derivatives of others.
\end{itemize}

\bex
Compute $\mathbb{H}^{0,1}(\sigma_-,[C^{[s]}])$, $\mathbb{H}^{0,1,2}(\sigma_-,[\omega^{[s]}])$, for the adjoint and twisted adjoint massless modules of arbitrary spin, 
using the explicit formulas for the $\sigma_-$ operators on the two modules:
\begin{align*}
\sigma_-C^{[s]}(x)&=\left\{h^n C_{a(s+k)n,\,b(s)}+\tfrac{1}{k+2}\,h^n C_{a(s+\,k)b,\,b(s-1)n}\right\}_{k=0,\cdots,\infty}\,,\\[3pt]
\sigma_-\omega^{[s]}(x)&=\left\{h^m \omega_{a(s-1),\,b(k-1)m}\right\}_{k=1,\cdots,\,s-1}\,.
\end{align*}
A key step is to consider the irreducible projection of the form indices together with the tangent ones. Keep in mind that what one should recover are doubly traceless 
Fronsdal fields and the Fronsdal equations in the corresponding cohomologies.
\eex

The above discussion clarifies why the $\sigma_-$ operator is universal and indeed does not depend on flat or AdS backgrounds, as one can see in the $s=0$ case:
\be
\sigma_-:T_l^p\rightarrow T_{k-1}^{p+1}\,,\qquad (\sigma_- C)_{a(k-1)}=h^bC_{a(k-1)b}\,.
\ee
On the other hand, on AdS or in general for massive fields and possibly other backgrounds, the positive grade component is not universal. In the $s=0$ case above:
\begin{subequations}
\begin{align}
&\sigma_+:T_l^p\rightarrow T_{k+1}^{p+1}\,,\\[3pt]
&(\sigma_+ C)_{a(k)}=f_{k-1}^m\left(h_aC_{a(k-1)}-\tfrac1{d-2k-4} \eta_{aa}C_{a(k-2)b}h^b\right)\,.
\end{align}
\end{subequations}

\subparagraph{The 4d spinorial language:}

So far we have seen in various examples how the change of dynamical variables at the basis of the unfolded formalism translates dynamical conditions into algebraic 
tensorial relations. Free dynamics is translated into structure constants of the modules under isometry generators. However, in order to perform the unfolding of 
spin-$s$ fields one should consider the most general ansatz for the structure constants among the momentum generator and the spin-s module which is given by the 
cell-operators introduced in the previous section. This becomes more and more cumbersome, and it turns out to be much more convenient to find some explicit realization
of the structure constants.

This is possible thanks to oscillator realizations of the relevant isometry algebras. For instance in 4d, due to the isomorphism between the isometry algebra $so(3,2)$ 
and $sp(4)$ one can simplify the tensorial operations, in particular traceless projections. The whole idea is based on the following matrix representation of 4-vectors:
\begin{align}
p_{\ga\gad}=p_\mu \sigma^\mu_{\ga\gad}=\begin{pmatrix}
p_0+p_3&~~p_1-i p_2\\
p_1+ip_2&~~p_0-p^3
\end{pmatrix}\,,
\end{align}
which maps a 4-vector into a Hermitian matrix.
The charge conjugation tensor:
\be
\epsilon_{\ga\gb}=\begin{pmatrix}
~0&~1\\
-1&~0
\end{pmatrix}=\epsilon^{\ga\gb}\,,
\ee
is then introduced, which allows one to raise and lower indices:
\begin{subequations}
\begin{align}
\bar{\sigma}^\mu{}^{\gad\ga}\equiv\sigma^{\mu}{}^{\ga\gad}&=\epsilon^{\ga\gb}\epsilon^{\gad\gbd}\sigma^\mu_{\gb\gbd}\,,\\
\sigma^\mu_{\ga\gad}\sigma_{\nu}^{\ga\gad}&=-2\delta^\mu_\nu\,,\\
\sigma^\mu_{\ga\gad}\sigma_\mu^{\gb\gbd}&=-2\,\delta_{\ga}^\gb\delta_{\gad}^\gbd\,.
\end{align}
\end{subequations}
We recall that in our conventions the signature of the metric is mostly positive.
The dictionary between spinorial and vectorial notation can be obtained by using the above $\sigma$-matrices, like for example:
\be
\sigma^{a\ga\gad}\sigma^{b\gb\gbd}F_{[ab]}=\epsilon^{\ga\gb}\bar{F}^{\gad\gbd}+\epsilon^{\gad\gbd}F^{\ga\gb},
\ee
where $F^{\ga\gb}$ and $\bar{F}^{\gad\gbd}$ are the (anti-)self-dual components of the Maxwell tensor, and so on for $s>1$. As a result, it turns out that one 
can rewrite the jets of the HS gauge fields in the spinorial language, in terms of the following 1-forms and 0-forms:
\begin{subequations}\label{d4set}
\begin{align}
\omega^{(s)}(x)&=\left\{\omega_k\equiv\omega^{\ga(s-1+k)\gad(s-1-k)}\right\}_{k=-(s-1),\cdots,s-1}\,,\label{gauge}\\
C^{(s)}(x)&=\left\{C^{\ga(s+k)\gad(k)},\,\bar{C}^{\ga(k)\gad(s+k)}\right\}_{k=0,\cdots,\infty}\,.\label{Weyl}
\end{align}
\end{subequations}
\bex Show, by using standard $\sigma$-matrix identities to map vector indices into spinorial ones, that the above sets of spinorial-language variables 
match their vectorial counterparts derived in the previous sections, in particular for $s=0,1,2,3$.\eex
In spinorial language one can perform from scratch the same manipulations considered in vector notation in the previous sections.
The explicit form of unfolded $AdS_4$ equations in spinorial language will be presented later after introducing other relevant ingredients. However, as it will be 
explained more in detail in the following, the main simplification achieved by the spinorial language is to allow a simpler treatment of the on-shell vanishing 
components of the jet space. These are associated to trace components of the off-shell jet vector and are encoded for instance into the coefficients $f_m$ parameterizing 
the momentum structure constants. Further key simplifications arise thanks to the oscillator realizations for the corresponding structure constants inherited from 
the construction of HS algebras. These oscillator realizations turn out to encode all the structure coefficients $f_m$, which would depend on the mass-shell and 
spectrum of the theory, into relatively simple $\star$-product operations.

\subsubsection{HS Algebra}\label{sec: HS Algebra}
The set of one-forms $\omega$ and zero-forms $C$ appearing in Eqs.~\eqref{d4set} can be embedded into a HS algebra. The latter is the global symmetry algebra 
of a free conformal boson on the boundary of $AdS_4$. The $AdS_4$ HS algebra is the Weyl algebra with two pairs of canonical oscillators \cite{Vasiliev:1986qx}, 
thanks to the isomorphism $so(3,2)\simeq sp(4,\mathbb{R})$.

While so far we have always given an explicit realization of the action of translation and Lorentz generators on the relevant modules by giving the explicit 
index form of the relevant structure constants, in 4d it is convenient to rely on a realization of the latter in terms of a quartet $\hat{Y}^A$, $A=1,...,4$ 
of operators. The latter are assumed to obey canonical commutation relations:
\begin{align}
[ \hat{Y}^A , \hat{Y}^B ] & = 2 i C^{AB}\,,
\end{align}
where $C^{AB}$ is the $sp(4)$ charge conjugation matrix:
\be
C_{AB}=\begin{pmatrix}
\epsilon_{\ga\gb}&0\\
0&\epsilon_{\gad\gbd}
\end{pmatrix}\,.
\ee
The bilinears give rise to an oscillator realization of $sp(4)\sim so(3,2)$:
\begin{align}
T^{AB} &=-\tfrac{i}{4} \{\hat{Y}^A ,\hat{Y}^B \} \, , && [T^{AB} ,T^{CD}]= T^{AD}C^{BC}+\text{3 terms}\,.
\end{align}
The bosonic HS algebra is then defined as the algebra of even functions $f(\hat{Y})$ in $\hat{Y}^A$. This is an associative algebra with a product that can be 
realized in terms of the Moyal (Weyl-ordered) star-product:
\begin{align}\label{expform}
(f \star g)(Y)=\exp\Big[i\Big({\overleftarrow{\pl}_A  C^{AB}\overrightarrow{\pl}_B}\Big)\Big]\,,
\end{align}
allowing the replacement of the operators $\hat{Y}^A$ with symbols $Y^A$. By definition,
\be
\pl^{(Y)}_AY_B=C_{AB}\,.
\ee
The $Y$'s are ordinary commuting variables to be multiplied with the Moyal-product, and whose symplectic indices\footnote{This in particular implies that 
$\pl^{(Y)}{}^A\neq\frac{\pl}{\pl Y_A}$.} are raised and lowered as $Y^A=C^{AB}Y_B$, $Y_B=Y^AC_{AB}$, $\pl^{(Y),A}=C^{AB}\pl^{(Y)}_B$, $\pl^{(Y)}_A=\pl^{(Y),B}C_{BA}$\,. 
For completeness we also give a useful integral representation of the star-product
\begin{equation}\label{Ystarproduct}
(f\star g)(Y)=\int dU dV f(Y+U) g(Y+V) e^{iU_A V^A}\,,
\end{equation}
from which the exponential formula \eqref{expform} can be derived, dropping the boundary terms. Some useful formulas are also:
\begin{align}
Y_A\star&=Y_A+i\partial^{(Y)}_{A}\,,&Y_A\star&=Y_A-i\partial^{(Y)}_{A}\,.
\end{align}
The integral form extends beyond polynomials and is used extensively. However, it is always important to keep in mind that non-polynomial extensions of the 
$\star$-product considered for different ordering prescriptions (Weyl ordering, normal ordering, etc.) are inequivalent.

All in all, the main simplification of introducing the HS algebra as above is that one can pack all one- and zero-form modules of \eqref{d4set} into master 
fields $\omega=\omega_\mm (Y|\,x)\,dx^\mm$ and $C = C (Y|\,x)$.
\begin{svgraybox}
The bosonic components of the master fields $\omega$ and $C$, even in $Y^A$, read
\begin{subequations}
\begin{align}
&\omega^{(s)}(y,\bar{y}|\,x)\,=\,\sum_{k=0}^{s-1}\tfrac{1}{(s-1+k)!(s-1-k)!}\,\omega_{\ga(s-1+k)\gad(s-1-k)}(x)\,y^{\ga(s-1+k)}\,\bar{y}^{\gad(s-1-k)}\,,\\
&C^{(s)}(y,\bar{y}|\,x)\,=\,\sum_{k=0}^{\infty}\tfrac{1}{(s+k)!k!}\,C_{\ga(2s+k)\gad(k)}(x)\,y^{\ga(2s+k)}\,\bar{y}^{\gad(k)}\nonumber\\[3pt]
&~~~~~~~~~~~~~~~~~~~~~~~~~~~~~~+\sum_{k=0}^{\infty}\tfrac{1}{(s+k)!k!}\,C_{\ga(k)\gad(2s+k)}(x)\,y^{\ga(k)}\,\bar{y}^{\gad(2s+k)}\,,\\[3pt]
&C^{(0)}(y,\bar{y}|\,x)\,=\,\sum_{k=0}^{\infty}\tfrac{1}{k!k!}\,C_{\ga(k)\gad(k)}(x)\,y^{\ga(k)}\,\bar{y}^{\gad(k)}\,.
\end{align}
\end{subequations}\vspace{-15px}
\end{svgraybox}
As anticipated, having embedded the field content into the HS algebra, the unfolded equations can be simply expressed in terms of $\star$-commutators of the 
translation and Lorentz generators as realized in the HS algebra. The adjoint and Weyl-module covariant derivatives take then the suggestive form:
\begin{subequations}\label{derivatives}
\begin{align}
(\adD f)(y,\bry|\,x)&=\nabla f(y,\bry|\,x)-h^{\ga\gad}[P_{\ga\gad},f(y,\bry|\,x)]_{\star}\,,\\
(\tadD f)(y,\bry|\,x)&=\nabla f(y,\bry|\,x)-h^{\ga\gad}\{P_{\ga\gad},f(y,\bry|\,x)\}_{\star}\,,
\end{align}
\end{subequations}
where the $sp(4)$ algebra has been split in an $AdS_4$-covariant way identifying $AdS$ translations and Lorentz generators as:
\be
T_{AB}=\begin{pmatrix}
L_{\ga\gb}&P_{\ga\gbd}\\
P_{\gb\gad}&L_{\gad\gbd}
\end{pmatrix}=-\tfrac{i}{2}\,\begin{pmatrix}
\ y_{\ga}y_{\gb}\ &\ y_{\ga}\bry_{\gbd}\ \\
\ y_{\ga}\bry_{\gbd}\ &\ \bry_{\gad}\bry_{\gbd}\
\end{pmatrix}
\ee
Notice that the action of the translation generator on the Weyl module is nicely rewritten as an anti-commutator in the HS algebra.
\bex Extract the index form of the above derivatives by performing the $\star$-(anti)commutators above on the respective adjoint module $\omega^{(s)}(y,\bry|\,x)$ 
and Weyl module $C^{(s)}(y,\bry|\,x)$. Compare the formulas obtained with those obtained in vector notation in the previous section.\eex
The action of the translation generator on the Weyl module is usually called after the embedding into the HS algebra, twisted-adjoint action. This is mainly due 
to the fact that one can rewrite the anticommutator in terms of the $\pi$ involution of the HS algebra defined by its action on the oscillators $\pi(y)=-y$, 
$\pi(\bry)=\bry$. This involution is nothing but the usual Chevalley involution flipping in the conformal basis boundary translations and conformal boost generator. 
It allows to rewrite the Weyl module derivative as:
\be(\tadD f)(y,\bry|\,x)=\nabla f(y,\bry|\,x)-h^{\ga\gad}\left[P_{\ga\gad}\star f(y,\bry|\,x)-f(y,\bry|\,x)\star\pi(P_{\ga\gad})\right]\,,\ee
from which the name ``twisted-adjoint''. The above way of rewriting the $\tadD$ derivative is the most useful to extend the theory beyond the linear order.

In terms of these ingredients we can finally write down the linear unfolded equations describing all symmetric massless HS fields as:
\begin{subequations}
\begin{align}
\adD\omega(y,\bry|\,x)&=-\tfrac{1}{2}\,H^{\ga\ga}\pl_{\ga}\pl_{\ga}C(y,0|\,x)-\tfrac{1}{2}\,H^{\gad\gad}\bar{\pl}_{\gad}
\bar{\pl}_{\gad}C(0,\bry|\,x)\,,\label{OnShell1}\\[3pt]\tadD C(y,\bry|\,x)&=0\,,
\end{align}
\end{subequations}
where $H^{\ga\ga}=h^{\ga}{}_{\gnd}\wedge h^{\ga\gnd}$, and $H^{\gad\gad}=h_{\gamma}{}^{\gad}\wedge h^{\gamma\gad}$ are a basis of two forms. As in vector notation, 
the right hand side of \eqref{OnShell1} can be interpreted as the most general solution to the Bianchi identity in Bargmann-Wigner equations. It links one-form and 
zero-form modules together ensuring that the gauge fields are not flat connections and propagate physical degrees of freedom stored into the zero-forms.

\subsubsection{Non-linear Unfolded Equations}
Abstracting from the example above, it is useful to write down the most general structure of unfolded equations \cite{Vasiliev:1988sa} which, as we have seen, take the 
form of first order differential equations of the following type
\begin{subequations}
\label{xpaceseq}
\begin{align}
&d\omega=F^{\omega}(\omega,C)\label{xpaceseqXA}\,,\\
&dC=F^C(\omega, C)\,.\label{xpaceseqXB}
\end{align}
\end{subequations}
Here the power of $\omega$ is fixed by the form degree and we have introduced unknown structure functions $F^{\omega,C}$
admitting an expansion in powers of the zero-forms $C$:
\besubeqs
\label{Cexpansion}
\begin{align}
&F^{\omega}(\omega,C)=\mathcal{V}(\omega,\omega)+\mathcal{V}(\omega,\omega,C)+
\mathcal{V}(\omega,\omega,C,C)+\cdots\,,\\
&F^C(\omega, C)=\mathcal{V}(\omega,C)+\mathcal{V}(\omega,C,C)+\mathcal{V}(\omega,C,C,C)+\cdots\,.
\end{align}
\esubeqs
The latter are constrained by the requirement of being compatible with $d^2=0$.
The first vertices can be considered as the initial data for the deformation problem, and are given by the HS algebra
\begin{align}\label{trivialcocycles}
\mathcal{V}(\omega,\omega)&=\omega\star\omega\,,
&\mathcal{V}(\omega, C)&=\omega\star C-C\star \pi(\omega)\,,
\end{align}
generalizing the free theory discussion to include HS generators on top of the isometry of the background. This step is almost straightforward once the HS algebra 
is given in terms of an oscillator realization. We recall that $\star$ denotes the (associative) product in the HS algebra and $\pi$ is the automorphism of the HS 
algebra that is induced by the reflection of $AdS$ translation generators $P_a\rightarrow -P_a$. The latter distinguishes the gauge module $\omega=\sum_s\omega^{(s)}$ 
from the Weyl-module $C=\sum_s C^{(s)}$, which is referred to in this context as the twisted adjoint module. In the oscillator realization described above, $\pi$ is 
given explicitly as the operation which flips the sign of $y:\,y\rightarrow -y$.

To summarize, Eq.~\eqref{trivialcocycles} extends the linear unfolded equations which encode the representation theory of the
background isometry to the full HS algebra.
The corresponding unitary irreducible representations involve all symmetric tensors with multiplicity one.

Before concluding this section we would like to stress that the interaction terms in \eqref{trivialcocycles} can be considered as an initial condition for the 
deformation problem which requires the higher-order cocycles to be fixed by the compatibility condition $d^2=0$. The HS problem is mapped to the problem of 
determining a full non-linear completion of the structure constants \eqref{trivialcocycles}.

\subsubsection{Unfolding, Jet Space \& BRST-BV Formalism}\label{sec:Jet}

Before concluding this introductory review of the unfolded formalism and also before moving to a more detailed analysis of 4d HS unfolded equations, we would like 
to point out some generic features of unfolding that relate it with the BRST-BV formalism \cite{Barnich:2004cr,Barnich:2006pc,Barnich:2009jy,Barnich:2010sw,Grigoriev:2010ic,Grigoriev:2012xg,Alkalaev:2013hta,Alkalaev:2014nsa}. 
The interested reader may also find in this link some possible bridges with string field theory and various related ideas.

The structure of unfolded equations we have detailed so far are naturally that of a cohomological problem, encoding dynamics
into nilpotent operators which, in this set of variables, acquire a plain geometric interpretation.

The basic ingredients of the BRST-BV formalism are very similar, since the basic common object is the jet space$-$an infinite dimensional manifold constructed as 
the tangent bundle with coordinates given by fields (a scalar in the example below) and all of their derivatives:
\be
\pi^\infty: J^\infty(E)=M\times V^\infty\rightarrow M\,,\qquad V^\infty\equiv\{\phi(x),\pl_{\mm}\phi(x),\cdots,\pl_{\mm(s)}\phi(x),\cdots\}\,.
\ee
In the following we will not consider any issues related to topologically non-trivial manifolds or to the appearance of gauge symmetries and just state the main ideas. 
Inside this space, which may include also ghost fields, one usually defines a so called
\textit{stationary surface}, which is the submanifold $\Sigma^\infty$ of the jet space where all equations of motions and their differential consequences hold. Namely, 
their zero-locus. In the case ghosts are included, the only constraints will be placed on the physical fields when dealing with the ungauge-fixed theory. Another key 
ingredient is given by the Bianchi identities, usually referred to as ``Noether identities'' in this context. The latter encode the redundancy among equations of motion 
related to the underlying gauge invariance of the system. More in detail, denoting the physical fields by $\varphi^i$ and the extended set including ghosts and antighosts 
by $\phi^I=\{\varphi^i,C^\ga,\ldots\}$, one can define a non-Lagrangian system by specifying a BRST operator $Q$ squaring to zero as a(n evolutionary) vector field on the 
jet bundle. One can then consider the graded Lie algebra generated by evolutionary vector fields under the Lie Bracket on the jet bundle:
\be
[\bullet,\bullet]:H^{g_1}\left([Q,\bullet]\right)\times H^{g_2}\left([Q,\bullet]\right)\rightarrow H^{g_1+g_2}\left([Q,\bullet]\right)\,,
\ee
where we have also considered the corresponding restriction to BRST cohomologies. The ghost degree zero cohomology is then endowed with the structure of a graded Lie algebra, 
while any other ghost degree cohomology group forms modules with respect to the ghost degree zero cohomology.

In this language one may say that the idea of the unfolded formalism is to promote the above cohomology classes to main dynamical variables through a change of dynamical 
variables. Background independence is furthermore achieved by trading ghost number with form degree, while requiring the BRST operator to have a plain geometric meaning 
with respect to the ghost degree zero graded Lie algebra. The form degree zero adjoint action is given by reducibility parameters or global symmetries, and all other modules 
correspond to various cohomologies of the corresponding BRST operator. Hence, while the language of unfolding is closely related to that of BRST-BV formalism, the main 
difference and virtue is a choice of dynamical variables which makes geometry of the jet space, rather than space time, manifest.

\newpage
\section{From Higher Spins to Strings}\label{sec:fromHS}

In this section we specialise the general setting above to the 4 dimensional case and develop the formalism in order to provide the reader with some basic examples. 
At the end of this section we are going to use the results obtained to consider a linear analysis of asymptotic symmetries and their implications. The aim is to link 
symmetries of Vasiliev's theory and Superstring Theory.

\subsection{Vacuum Solutions, Flat Connections \& AdS Space}

The unfolded equations above \eqref{xpaceseq} have a natural family of maximally symmetric vacua given by flat connections $\Omega$ of the HS algebra:
\begin{align}
d\Omega& =\Omega\star\Omega\,,& C&=0\,.\label{adsbckgrnd}
\end{align}
It is straightforward to show that any solution of the above type posses the full HS algebra as a global symmetry, which is usually broken by $C$ as we will explicitly 
see in some simple examples below. Among all flat connections of the HS algebra the simplest vacuum solution is $AdS_4$:
\begin{align}\label{Omegaconn}
\Omega=\tfrac{1}{2}\varpi^{\ga\ga}L_{\ga\ga}+ h^{\ga\gad} P_{\ga\gad}+\tfrac{1}{2}\varpi^{\gad\gad}\bar{L}_{\gad\gad}\,,
\end{align}
where we have identified explicitly among the $sp(4)$ generators $T^{AB}$, the Lorentz generators $L_{\ga\ga}$ and $\bar{L}_{\gad\gad}$ and the translation 
generators $P_{\ga\gad}$\,:
\begin{subequations}
\begin{align}
L_{\ga\ga}&=-\tfrac{i}{4}\{y_\ga,y_\ga\}\,, \\[3pt]
P_{\ga\gad} &= - \tfrac{i}{4} \{y_\ga,\bry_\gad\} \,, \\[3pt]
\bar{L}_{\gad\gad} & = -\tfrac{i}{4}\{\bry_\gad,\bry_\gad\}\,,
\end{align}
\end{subequations}
with $Y_A=(y_\ga,\bry_\gad)$. The (anti)self-dual components of the $4d$ spin-connection are $\varpi^{\ga\ga}=\varpi^{\ga\ga}_\mm\, dx^\mm$ and 
$\varpi^{\gad\gad}=\varpi^{\gad\gad}_\mm\, dx^\mm$, while $h^{\ga\gad}=h^{\ga\gad}_\mm\, dx^\mm$ is the invertible vierbein of $AdS_4$. In the 
following$-$and also in the view of applications to the AdS/CFT correspondence$-$it is useful to specialise the above coordinate independent setting to a particular 
choice of coordinates given by Poincar\'e coordinates,  which can be conveniently defined by a conformal foliation of $AdS_4$. To this end it is useful to 
introduce a different splitting of the $sp(4)$ generators, that are natural from the perspective of the 3d boundary \cite{Fradkin:1989yd,Vasiliev:2012vf}. 
This can be achieved by introducing a different pair of oscillators related to $y$ and $\bry$ as follows
\be y^+_\ga=\tfrac12(y_\ga-i\bry_\ga),\qquad y^-_\ga=\tfrac12(\bry_\ga-iy_\ga)\,.\ee
From the bulk perspective, one breaks AdS covariance as:
\be x^{\ga\gad}=\text{x}^{\ga\ga}\epsilon_{\ga}{}^{\gad}-i\epsilon^{\ga\gad}z\,,\qquad \sigma_2^{\ga\gad}\equiv
-i\epsilon^{\ga\gad}\,,\ee
and defines $\bry_\ga=\epsilon_{\ga\gad}\bry^\gad$. In this way $\text{x}^{\ga\ga}$ is a 3d coordinate on which the
conformal group acts via the following realization given by Fradkin and Linetski of the $sp(4)$ conformal algebra:
\begin{align}
P_{\ga\ga}&=iy_{\ga}^-y_{\ga}^-\,,& K^{\ga\ga}&=-iy^{+\ga}y^{+\ga}\,,\\L^\ga{}_\gb&=y^{+\ga}y^{-}_\gb-\tfrac{1}{2}\,
\delta^\ga_\gb\, y^{+\gamma}y^-_{\gamma}\,,& D&=\tfrac{1}{2} y^{+\gamma}y^-_{\gamma}\,.\end{align}
The above can be checked easily by noticing that
\be
\left[y^-_\ga\,,\,y^{+\gb}\right]_{\star}=\delta_\ga^\gb\,.
\ee

In order to write down a conformal foliation of $AdS_4$ space, one can consider a dilatation of a 3d flat connection:
\be \Omega=e_\star^{\log z \,D}\star\left(\tfrac{i}{2}\,P_{\ga\ga} d\text{x}^{\ga\ga}\right)\star e_\star^{-\log z \,D}
+e_\star^{\log z \,D} d e_\star^{-\log z \,D}\,,\ee
where $e_\star$ is a $\star$-exponential. The following commutation relations hold:
\be \left[D,P_{\ga\ga}\right]_{\star}=-P_{\ga\ga},\qquad \left[D,K^{\ga\ga}\right]_{\star}=K^{\ga\ga},\qquad
\left[D,L^{\ga}{}_{\gb}\right]_{\star}=0\,,\ee
which can be checked using the general expression:
\be [D,f]_{\star}=\tfrac{1}{2}\left(y^{+\ga}\pl_\ga^+-y^{-\ga}\pl^-_{\ga}\right) f\,.\ee
One can then easily perform the above $\star$-products just using the commutation relations above and arriving to the following conformally foliated $AdS_4$ connection:
\be
W=\frac{1}{z}\left[P_{\ga\ga}d\text{x}^{\ga\ga}-D\,dz\right]\,,
\ee
or in terms of $y$ and $\bry$:
\bea
W&=&-\frac{1}{8z}\left(d\text{x}^{\gad\gad}\bry_{\gad}\bry_{\gad}-d\text{x}^{\ga\ga}y_\ga y_\ga\right)+\frac{i}{4z}\left(d\text{x}^{\ga\gad}-i\epsilon^{\ga\gad} dz\right)
y_\ga\bry_\gad\nonumber\\&=&\tfrac{i}{4}\omega^{\ga\ga}y_\ga y_\ga+\tfrac{i}{4}\varpi^{\gad\gad}\bry_\gad \bry_\gad+\frac{i}{2}h^{\ga\gad}y_\ga\bry_\gad\,.\label{whatsupbaby}
\eea
The above conveniently gives:
\begin{subequations}
\begin{align}
\omega^{\ga\ga}&=-\frac{i}{2z}\,d\text{x}^{\ga\ga}\,,\\
\varpi^{\gad\gad}&=\frac{i}{2z}\,d\text{x}^{\gad\gad}\,,\\ h^{\ga\gad}&=\frac{1}{2z}\,\left(d\text{x}^{\ga\gad}-i\epsilon^{\ga\gad}dz\right)=\frac{1}{2z}\,dx^{\ga\gad}\,,
\end{align}
\end{subequations}
that can be checked to be a constant curvature solution of the Einstein equations:
\besubeqs\label{flatadsconnection}
\begin{align}
R_{\ga\ga}&=d\varpi^{\ga\ga}-\varpi^{\ga}_{\gamma}\wedge\varpi^{\gamma\ga}-h^{\ga}_{\dot\gamma} \wedge h^{\ga\dot\gamma}=0\,,\\
\bar R_{\gad\gad}&=d\bvarpi^{\gad\gad}-\bvarpi^{\gad}{}_{\dot\gamma}\wedge\bvarpi^{\dot\gamma\gad}-h_{\gamma}{}^{\gad} \wedge h^{\gamma\gad}=0\,,\\
T_{\ga\gad}&=dh^{\ga\gad}-\varpi^{\ga}_{\gamma}\wedge h^{\gamma\gad}-\bvarpi^{\gad}_{\dot\gamma} \wedge h^{\ga\dot\gamma}=0\,,
\end{align}
\esubeqs
with
\begin{align}
h_\mm^{\ga\gad}h^\mm_{\gb\gbd}&= \delta_{\gb}{}^{\ga}\delta_{\gbd}{}^{\gad}\,, &&
h_\mm^{\ga\gad} h^\nnn_{\ga\gad}=\delta^\nnn_\mm\,,
\end{align}
and
\begin{align}
h_\mm^{\ga\gad}\,dx^\mm&=\frac1{2z} \sigma_\mm^{\ga\gad} dx^\mm\,, && h^\mm_{\ga\gad}=-z \sigma^\mm_{\ga\gad}\,, && g_{\mm\nnn}=\frac1{2z^2} \eta_{\mm\nnn}dx^\mm dx^\nnn\,.
\end{align}
Here we use a non-canonical normalisation for the cosmological constant to simplify some factors of $2$.

\subsection{Linear Order} Once a vacuum solution/flat connection, such as \eqref{Omegaconn} which describes $AdS_4$, is identified, we can consider linearized fluctuations around it \cite{Vasiliev:1988xc}. The generic structure of the equations reads:
\besubeqs\label{firstordereqs}
\begin{align}
&d\omega=\{\Omega,\omega\}_\star+\mathcal{V}(\Omega,\Omega,C)\label{xpaceseqEA}\,,\\
&dC=\Omega\star C-C\star \pi(\Omega)\label{xpaceseqEB}\,,
\end{align}
\esubeqs
which reproduces the unfolded equations of the gauge and Weyl modules.
As already stated, the appearance of $\mathcal{V}(\Omega,\Omega,C)$ corresponds to a deviation from a flat connection, and parameterizes the non-vanishing components of the HS curvature tensor compatibly with Bargmann-Wigner equations and the Bianchi identities.

As we described in the first part of this chapter, to each gauge field there corresponds a jet-bundle, namely the space of all its derivatives, each treated as an independent coordinate. This includes an infinite-dimensional subspace of \underline{gauge covariant} components, a subspace of which is set to zero on the equations of motion. The remaining non-vanishing components parameterize the Weyl module $C$. This includes the HS Weyl tensors $C^{a(s),\,b(s)}$, given by the order-$s$ curl of the Fronsdal field:\footnote{The link between world and tangent tensors is performed via the vierbein $h^{\ga\gad}_\mm$ and its inverse.}
\begin{align}
\begin{aligned}
C^{ \ga (2s)} & \\
C^{ \gad (2s)}
\end{aligned}
&: && C^{a(s),b(s)}\sim\,\nabla^{\{b_1}\cdots\nabla^{b_s}\Fron^{a_1\cdots a_s\}}\,.\label{LinWeylTensors}
\end{align}
Above we used the spinorial $4d$ language to decompose the HS Weyl tensors into their (anti)self-dual components $C_{\ga(2s)}$ and $C_{\gad(2s)}$, that can be found in the expansion of  $C(y,\bry=0)$ and $C(y=0,\bry)$ respectively. Other components of $C(y,\bry)$ contain on-shell nontrivial derivatives of the Weyl tensors:
\begin{align}
\begin{aligned}
C^{\ga(2s+k),\gad(k)}&\\
C^{\ga(k),\gad(2s+k)}
\end{aligned}&: && \, \nabla^{(a_{s+1}}\nabla^{a_{s+k}} C^{a_1\cdots a_s),\,b_1\cdots b_s}-\text{traces}\,, &&k=0,\cdots,\infty\,.
\end{align}
The zero-form $C$ includes also a scalar field, $\Fron_0=C(y=0,\bry=0)$, together with all of its on-shell nontrivial derivatives encoded as $C^{\ga(k),\gad(k)}\sim \nabla\cdots\nabla \Fron_0$.

The gauge connection $\omega$ includes the Fronsdal field $\Phi_{\mm(s)}$, which in $4d$ is embedded as the following component with equal number of dotted and undotted indices:
\begin{align}\label{FronVielLinear}
\Phi_{\mm(s)}&=\omega^{\ga(s-1),\gad(s-1)}_\mm\, h_{\mm|\ga\gad}\cdots h_{\mm|\ga\gad}\,.
\end{align}
The other components of $\omega(y,\bry)$ describe \underline{gauge-dependent} derivatives of the Fronsdal field up to order-$(s-1)$:
\begin{align}
\begin{aligned}
\omega_{-k}\equiv\omega^{\ga(s-1-k),\gad(s-1+k)}&\\
\omega_{k}\equiv\omega^{\ga(s-1+k),\gad(s-1-k)}
\end{aligned}&: && \,\nabla^{b_1}\cdots\nabla^{b_k}\Fron^{a_1\cdots a_s}\,, &&k=0,\cdots,s-1\,.
\end{align}
Notice that there is no traceless projection here, due to the fact that a one-form with traceless tangent indices describes also trace component when translated in the metric-like language (the Fronsdal field is only doubly traceless).

We can finally rewrite the free equations and gauge transformations as:
\begin{subequations}
\begin{align}
&\adD \omega=\mathcal{V}(\Omega,\Omega,C)\,,\label{xpaceseqQBA}&&\delta \omega=\adD\xi\,,\\
&\tadD C=0\,, && \delta C=0\,,
\end{align}
\end{subequations}
where we have used the background covariant derivatives $\adD$ and $\tadD$ already introduced. Note that $C$ is gauge invariant to the lowest order,
but it transforms covariantly under global HS transformations:
\be
\delta_\epsilon C=\epsilon\star C-C\star\pi(\epsilon)\,.
\ee
The corresponding transformation of $\omega$ under HS ``global'' symmetries is given by the adjoint action of the HS algebra and reads:
\be
\delta_\epsilon \omega=\epsilon\star \omega-\omega\star\epsilon\,.
\ee
These transformations extend to the full HS algebra the structure constants previously derived for the isometry generators.
The background nilpotent derivatives take, as anticipated, the following form:
\begin{subequations}
\begin{align}
\adD\bullet&=d\bullet -\Omega\star \bullet\pm\bullet\star \Omega=\nabla-h^{\ga\ga}(y_\ga\bpl_\gad+\bry_\gad\pl_\ga)\,,\label{AdjointDer}\\
\tadD\bullet&=d\bullet -\Omega\star \bullet\pm\bullet\star \pi(\Omega)=\nabla+ih^{\ga\gad}(y_\ga\bry_\gad-\pl_\ga\bpl_\gad)\,,\label{TwistedDer}\\
\nabla&=d-\varpi^{\ga\ga}y_\ga \pl_\ga-\varpi^{\gad\gad}\bry_\gad \bpl_\gad\,,\label{LorentzDer}
\end{align}
\end{subequations}
where $\pm$ accounts for a graded commutator, $\nabla$ is the Lorentz-covariant derivative on $AdS_4$ and we also give the explicit $y$-operatorial form. The link between $C$ and $\omega$ is realized via the following coupling:
\begin{align}\label{OMSTcocycle}
\mathcal{V}(\Omega,\Omega,C)&=-\tfrac{1}{2} H^{\ga\ga}\pl_\ga\pl_\ga C(y,0|\,x)-
\tfrac{1}{2} H^{\gad\gad}\bar{\pl}_\gad\bar{\pl}_\gad C(0,\bry|\,x)\,,
\end{align}
where $H^{\ga\ga}=h^{\ga}{}_{\gnd}\wedge h^{\ga\gnd}$, and $H^{\gad\gad}=h_{\gamma}{}^{\gad}\wedge h^{\gamma\gad}$. Setting $y=0$ or $\bry=0$ selects the (anti)self-dual components of the HS Weyl tensors.

The above form is the one which makes manifest $AdS_4$ covariance. It is however interesting to investigate briefly the transformation properties of the above modules from a conformal-algebra perspective. In particular, considering the twisted-adjoint commutator of the dilatation generators one arrives at:
\be\left\{D\,,\, C(y,\bry|\,x)\right\}_{\star}=\left(y_\ga\bry^\ga-\pl_\ga\bar{\pl}^\ga\right)C(y,\bry|\,x)\,,\ee
which shows how the zero-form $C$ does not diagonalise the dilatation generator. To go to a conformal basis, one has to diagonalise the dilatation generator \cite{Vasiliev:2012vf}. This change of basis is achieved by the boundary-to-bulk propagator, which we discuss in the next sections.

\subsection{The Boundary-to-Bulk Propagator Solution}

Here we review how to find the solution for the boundary-to-bulk propagator of the linear HS Weyl tensors following the work of
\cite{Giombi:2009wh,Giombi:2010vg,Vasiliev:2012vf,Didenko:2012tv,Didenko:2013bj}. For this, we need to solve the following equation:
\be \tadD C(y,\bry|\,x)=0\,,\ee
with a $\d$-function boundary condition for the HS field. This is equivalent to solving the Bargmann-Wigner equations in AdS
space. To this end, the first step is to rewrite the above equations in Poincar\'e coordinates:
{\allowdisplaybreaks
\besubeqs
\begin{align}
\left[d_{\text{x}}+\frac{i}{2z}\,d\text{x}^{\ga\ga}\left(y_\ga\pl_\ga-\bry_\ga\bar{\pl}_\ga+y_\ga\bry_\ga-\pl_\ga\bar{\pl}_\ga\right)
\right]C(y,\bry|\,x)&=0\,,\\\left[d_z+\frac{1}{2z}\,dz\left(y_\gamma\bry^\gamma-\pl_\gamma\bar{\pl}^\gamma\right)\right]C(y,\bry|\,x)&=0\,.
\end{align}
\esubeqs}
It is convenient to make the following change of variables \cite{Vasiliev:2012vf}:
\be C(y,\bry|\,x)=z \exp\left(y_\gamma \bry^\gamma\right) J(yz^{1/2},\,\bry z^{1/2}|\,\text{x},z)\,,\label{ansatz}\ee
which results in simpler unfolded equations in terms of $J(w=yz^{1/2},\,\bar{w}=\bry z^{1/2}|\,\text{x},z)$:
\begin{subequations}
\begin{align}
\left[d_{\text{x}}-\tfrac{i}{2}\,d\text{x}^{\ga\ga}\pl^w_\ga\bar{\pl}^w_\ga\right]J(w,\bar{w}|\,\text{x},z)&=0\,,\label{currcons}\\[3pt]
\left[d_z-\tfrac{1}{2}\,dz\,\pl^w_\gamma\bar{\pl}^{w\gamma}\right]J(w,\bar{w}|\,\text{x},z)&=0\,.\label{currz}
\end{align}
\end{subequations}
The first equation is a particular type of unfolded equation, tantamount to current conservation for some components of $J(w,\bar{w}|\,x)$.
These two equations can be combined nicely into a single covariant one:
\be
\left[d_{x}-\tfrac{i}{2}\,dx^{\ga\gad}\pl^w_\ga\bar{\pl}^{w}_\gad\right]J(w,\bar{w}|\,x)=0\,.
\ee
\bex
Show that the above unfolded equations \eqref{currcons} imply the following conservation equations:
\begin{align}
\frac{\partial}{\partial \text{x}^{\ga\ga}}\frac{\partial^2}{\partial w_\ga\partial {w}_\ga}J(w,0|\,\text{x},z)&=0\,,&\frac{\partial}{\partial \text{x}^{\ga\ga}}
\frac{\partial^2}{\partial \bar{w}_\ga\partial \bar{w}_\ga}J(0,\bar{w}|\,\text{x},z)&=0\,,&
\end{align}
while the other components with both $w\neq0$ and $\bar{w}\neq0$ are descendants of such conserved currents, apart from the primary operators:
\begin{align}
J(w,\bar{w}|\,\text{x},z)&\sim J_{\text{even}}(\text{x},z)\,,& J(w,\bar{w}|\,\text{x},z)&\sim  w_\ga\bar{w}^\ga J_{\text{odd}}(\text{x},z)\,,
\end{align}
which are dual to a bulk scalar field with $\Delta=1$ and $\Delta=2$ respectively.
\eex
The above equations admit a very simple ansatz in terms of the following $AdS_4$ covariant bispinor:
\be
\Pi^{\ga\gad}=\frac{z}{(\text{x}-\text{x}_0)^2+z^2}\,\left[(\text{x}-\text{x}_0)^{\ga\gad}-i\epsilon^{\ga\gad} z\right]\equiv z\,\widetilde{\Pi}^{\ga\gad}\,,
\ee
where for convenience we have also introduced  $\widetilde{\Pi}^{\ga\gad}$, as it plays the role of covariant bispinor for $w$ and $\bar{w}$. Introducing a
boundary constant spinors $\eta_\ga$ we then have\footnote{Reality conditions for the boundary constant spinor can be fixed directly at the level of $\eta$ adding
appropriate factors of $i$. We leave this arbitrary in the following.}:
\beq J(w,\bar{w}|\,\text{x},z)=K f(u) \left[e^{\xi^\ga w_\ga}+e^{\brxi^\gad\bar{w}_{\gad}}\right]\,,\qquad
K=\frac{z}{(\text{x}-\text{x}_0)^2+z^2}\,,\label{JKappear}\eeq
where
\be u=\widetilde{\Pi}^{\ga\gad}w_\ga\bar{w}_{\gad}\,,\qquad
\xi^\ga=\widetilde{\Pi}^{\ga\gad}\eta_{\ga}\,,\qquad
\brxi^\gad=\widetilde{\Pi}^{\ga\gad}\eta_{\gad}\,.\ee
One can now solve the equations by just looking at the $z$ dependence of \eqref{currz}, by choosing for convenience $\text{x}=\text{x}_0$. With this simplification
and plugging the ansatz into \eqref{currz}, one gets a very simple differential equation for the arbitrary function $f(u)$:
\be
2\left(f-\tfrac{i}{2} f'\right)+u\left(f'-\tfrac{i}{2}f''\right)+q\left(f-\tfrac{i}{2} f'\right)=0\,,
\ee
where $q=-i z^{-1}\eta^\ga w_\ga$. The above has only one analytic solution at $u\sim 0$, given by
\be
f(u)=e^{-2i u}\,.
\ee
Combining the above solution with \eqref{ansatz}, one can finally write down the unfolded solution for the boundary-to-bulk
propagator for the Weyl module:
\begin{multline}
C(y,\bry|\,x) = \frac{z}{(\text{x}-\text{x}_0)^2+z^2} \\\times\exp\left[- i y_\alpha \bar{y}_{\ad} \left( i\epsilon + \frac{2z\left( (\text{x}-\text{x}_0)-i\epsilon z\right)}
{(\text{x}-\text{x}_0)^2+z^2} \right)^{\alpha\ad} \right]\left(e^{ \xi^\alpha y_\alpha+i\theta} + e^{ \bar{\xi}^{\ad} \bar{y}_{\ad}-i\theta} \right),
\end{multline}
where
\be \xi^\alpha = \frac{z^{1/2}}{(\text{x}-\text{x}_0)^2+z^2} \lb (\text{x}-\text{x}_0)-i \epsilon\,
z \rb^{\alpha\beta} \eta_\beta\,,\label{xi}\ee
and we have also considered a phase parameterizing parity breaking.
\begin{exercise}
Show that the $y$-dependent Gaussian part of the boundary-to-bulk propagator can be rewritten as
\be e^{-iy_\ga\bry_{\gad}F^{\ga\gad}}\,,\ee
in terms of the wave vector defined as:
\begin{align}
F_{\ga\gad} &= \nabla_{\ga\gad}\ln K=-z \,\frac{\partial}{\partial x^{\alpha\ad}} \tln \frac{z}{(\text{x}-\text{x}_0)^2+z^2} \\
&= \left( i\epsilon_{\alpha\ad} + \frac{2z}{(\text{x}-\text{x}_0)^2+z^2}\left( (\text{x}-\text{x}_0)_{\alpha\ad}
-i\epsilon_{\alpha\ad} z \right) \right)\,,\nonumber\\
&= i\epsilon_{\alpha\ad} + 2\Pi_{\ga\gad}\,.\nonumber
\end{align}
Show that the wave vector so defined squares to the identity:
\be F_{\ga}{}^\gad F_{\gb\gad}=\epsilon_{\ga\gb}\,.\ee
\end{exercise}

It is also useful to compute the covariant derivative of $\xi^\ga y_\ga$, which comprises the following exercise:
\bex
Using the equation for the zero-forms by restricting the attention to a linear dependence in $\eta$:
\begin{multline}
-i\,h^{\ga\gad}\left[\left(y_\ga\bry_{\gad}-\pl_\ga\bpl_{\gad}\right)Ke^{-iy_\ga\bry_{\gad}F^{\ga\gad}+\xi^{\ga}y_\ga}\right]_{\bry=0,\eta\text{-linear}}\\=\nabla(K\xi^\ga y_\ga)=K \,\left[(\xi^\ga y_\ga)d\ln K+\nabla(\xi^\ga y_\ga)\right]\,,
\end{multline}
prove that:
\begin{subequations}\label{nablax}
\begin{align}
\nabla(\xi^\ga y_\ga)&=h^{\ga\gad}y_{\gb}F^{\gb}{}_{\gad}\xi_\ga\,,\\
\nabla(\xi^\gad \bry_\gad)&=h^{\ga\gad}\bry_{\gbd}F_{\ga}{}^{\gbd}\xi_\gad\,.
\end{align}
\end{subequations}
\eex
\bex
Use the above property and the integral definition of the Moyal product to show that, given:
\begin{align}
P= \exp \lb - i y_\beta\bar{y}_{\bd} F^{\beta\bd} + \xi^\beta y_\beta \rb\,,
\end{align}
the following relation holds (see e.g. \cite{Vasiliev:2012vf,Didenko:2012tv}),
\be
P\star P=P\,.
\ee
We have then recovered that the embedding of the boundary-to-bulk propagator in the unfolded language is achieved through a projector in the star-product algebra.
\eex
\bex
Show that (see e.g. \cite{Vasiliev:2012vf,Didenko:2012tv}):
\begin{itemize}\setlength\itemsep{0.5em}
\item $P\star\pi(P)$ is ill defined.
\item Given $y^{\pm}_\ga\equiv F_{\ga}{}^\gad \bry_\gad\pm y_\ga-i\xi_\ga$, derive the relation:
$y^+_\ga\star P=0=P\star y^-_\ga$\,. The latter defines a Fock vacuum for any $F_{\ga\gad}$.
\end{itemize}
\eex

\bex
Considering the other possible primary given by $w_\ga\bar{w}^\ga J_{\text{odd}}$, solve the above unfolded equations with an ansatz of the type:
\be
J(w,\bar{w}|\,\text{x},z)=K^2\left[f_1(u)+w_\ga\bar{w}^\ga f_2(u)\right]\,,
\ee
and show that the boundary-to-bulk propagator for a scalar with boundary condition $\Delta=2$ is given by (see e.g. \cite{Giombi:2009wh,Giombi:2010vg,Didenko:2013bj})
\be
C(y,\bry|\text{x},z)=\left(\frac{z}{(x-x_0)^2+z^2}\right)^2\left(1-i y_\ga\bry_\gad F^{\ga\gad}\right)e^{-i y_\ga\bry_\gad F^{\ga\gad}}\,,
\ee
with $F^{\ga\gad}$ the same wave-vector as for the $\Delta=1$ case.\eex

\subsection{The 1-Form Sector}

In the previous section we have studied the boundary-to-bulk propagator solution in the Weyl module sector of the unfolded equations. Once the solution for the Weyl
module is found, one can also use the equations for $\omega$ to find the corresponding solution for its bulk-to-boundary propagator. It is equivalent to integrating
the Bianchi identity for gauge connections. This step is analogous to solving for torsion in the frame-like formulation of General Relativity, but it requires an
iterative procedure due to the fact that the gauge module for HS fields involves more fields than gravity does. The solution for the $\omega$-propagator was originally
found in \cite{Giombi:2009wh}. In this section we derive it using techniques based on \cite{Didenko:2012tv,Boulanger:2015ova}.

The equations we want to analyze are the following:
\begin{align}
\adD\omega(y,\bry|\,\text{x},z)&=-\tfrac{1}{2}H^{\ga\ga}\pl_\ga\pl_\ga C(y,0|\,\text{x},z)-\tfrac{1}{2}H^{\gad\gad}\bpl_\gad\bpl_\gad C(y,\bry|\,\text{x},z)\,,\\[3pt]
\delta\omega(y,\bry|\,\text{x},z)&=\adD\epsilon(y,\bry|\,\text{x},z)\,,
\end{align}
where the parity violating phase has been included into the definition of $C$. Plugging the explicit form of the solution for $C(y,\bry|\,\text{x},z)$, we get the
following set of equations:
\begin{align}
\adD\omega(y,\bry|\,\text{x},z)&=-\tfrac{1}{2}KH^{\ga\ga}\pl_\ga\pl_\ga e^{i\theta+\xi^\ga y_\ga}-\tfrac{1}{2}KH^{\gad\gad}\bpl_\gad\bpl_\gad
e^{-i\theta+\xi^\gad\bry_\gad}\,,\\[3pt]\delta\omega(y,\bry|\,\text{x},z)&=\adD\epsilon(y,\bry|\,\text{x},z)\,.\end{align}
Our strategy is to first solve the general equation for arbitrary $C$, and then apply the result to the case of the
boundary-to-bulk propagator in the end.

First of all it is necessary to fix the freedom up to Fierz identities by picking a canonical basis for differential forms. This can be always done with the help
of the following identity:
\be\label{Fierz} f_{\ga}(y)=\pl_{\ga}\left(\tfrac{1}{N}\,y^\gb f_\gb\right)-y_{\ga}\left(\tfrac{1}{N+2}\,\pl^\gb f_{\gb}\right)\,,\ee
with
\beq N=y^\ga\pl^y_\ga\,,\qquad\brN=\bry^\gad\bar{\pl}^\bry_\gad\,,\label{whatisthis1}\eeq
being number operators. One can then find the following 4-dimensional canonical basis for one-forms given by:
\be\label{1formB} \omega=h^{\ga\gad}\pl_\ga\bpl_{\gad}\omega^{\pl\bpl}+Q_+\omega^{Q_+}+Q_-\omega^{Q_-}+h^{\ga\gad}y_\ga\bry_\gad\omega^{y\bry}\,,\ee
where we have defined for later convenience:
\beq Q_+=-h^{\ga\gad}y_\ga\bpl_\gad\,,\qquad Q_-=-h^{\ga\gad}\bry_\gad\pl_\ga\,.\label{whatisthis2}\eeq
We also have the following 6-dimensional canonical basis for 2-forms:
\begin{multline}\label{2formB}
H^{\ga\ga}\pl_\ga\pl_\ga J^{\pl\pl}+H^{\ga\ga}y_\ga\pl_\ga J^{y\pl}+ H^{\ga\ga}y_\ga y_\ga J^{yy}
\\+H^{\gad\gad}\bpl_\gad\bpl_\gad {J}^{\bpl\bpl}+H^{\gad\gad}\bry_\gad\bpl_\gad {J}^{\bry\bpl}+ H^{\gad\gad}\bry_\gad \bry_\gad {J}^{\bry\bry}\,.
\end{multline}
One can then easily show that:
\bea
&(\adD\epsilon)_k=h^{\ga\gad}\pl_\ga\bpl_\gad\left[\tfrac{1}{N\brN}\,y^\gb\bry^\gbd\nabla_{\gb\gbd}\epsilon_k\right]
+h^{\ga\gad}y_\ga\bry_{\gad}\left[\tfrac{1}{(N+2)(\brN+2)}\,\pl^\gb\bpl^\gbd\nabla_{\gb\gbd}\epsilon_k\right]~~~~~~~~~~~~\nonumber\\
&+Q_+\left[\tfrac{1}{\brN(N+2)}\,\bry^\gbd\pl^\gb\nabla_{\gb\gbd}\epsilon_k+\epsilon_{k-1}\right]
+Q_-\left[\tfrac{1}{(\brN+2)N}\,y^\gb\bpl^\gbd\nabla_{\gb\gbd}\epsilon_k+\epsilon_{k+1}\right].\label{whatanequation}
\eea
The above allows to gauge fix all $Q_{\pm}$ components of $\omega$ to zero, by tuning the gauge parameters $\epsilon_k$ with $k\neq0$. In this way we only
fixed the St\"uckelberg symmetries, and we are left with the degrees of freedom of a doubly-traceless Fronsdal field and all derivatives thereof:
\begin{align}\label{doublytraceless}
\omega_0=h^{\ga\gad}\pl_\ga\pl_\gad\Phi+h^{\ga\gad}y_\ga\bry_{\gad}\Phi'\,,\\
\omega_k=h^{\ga\gad}\pl_\ga\pl_\gad\omega_k+h^{\ga\gad}y_\ga\bry_{\gad}\omega_k'\,.
\end{align}
On-shell we can however use the leftover non-St\"uckelberg gauge symmetries to also gauge fix to zero the trace of the Fronsdal field, as can be seen from the
above equation for $k=0$ by looking at the $h^{\ga\gad}y_\ga \bry_\gad$ component. In the following we will hence set $\Phi'=0$, assuming to have chosen a
traceless on-shell gauge.

In this gauge it is then easy to solve unfolded equations for $\omega_k$. The first step is to decompose the equation according to their components ending up with:
\be
Q\left(\begin{matrix}
\omega_{k-1}\\
\omega_{k+1}
\end{matrix}\right)\equiv Q_+\omega_{k-1}+Q_-\omega_{k+1}=-\nabla\omega_k\,.
\ee
The above needs then to be inverted. To this end, first of all we should note that:
\bea Q_+\omega&=&\tfrac{1}{2}NH^{\gad\gad}\partial_\gad\partial_\gad\omega^{\pl\pl}-\tfrac{1}{2}(\bar{N}+2)H^{\ga\ga}y_\ga y_\ga\,\omega^{y\bry}
\nonumber\\&&+\tfrac{1}{2}\left[-N H^{\gad\gad}\bry_\gad\partial_\gad+(\bar{N}+2)H^{\ga\ga}y_\ga\partial_\ga\right]\omega^{Q_-}\,,\label{weirdname1}\\[5pt]
Q_-\omega&=&\tfrac{1}{2}\bar{N}H^{\ga\ga}\partial_\ga\partial_\ga\omega^{\pl\pl}-\tfrac{1}{2}(N+2)H^{\gad\gad}\bry_\gad\bry_\gad\,\omega^{y\bry}\,
\nonumber\\&&+\tfrac{1}{2}\left[-\bar{N}H^{\ga\ga}y_\ga\partial_\ga+(N+2)H^{\gad\gad}\bry_\gad\partial_\gad\right]\omega^{Q_+}\,.\label{weirdname2}\eea
%
Hence, one can invert the equation $Q\binom{\omega_-}{\omega_+}=S$, by decomposing it into the canonical basis above and inverting a simple linear system:
\begin{subequations}
\begin{align}
\begin{pmatrix}
\omega_{+}^{\pl\pl}\\[3pt]
\omega_{-}^{\pl\pl}
\end{pmatrix}&=
2\begin{pmatrix}
\tfrac{1}{\bar{N}}\,S^{\pl\pl}\\[3pt]
\tfrac{1}{N}\,S^{\bpl\bpl}
\end{pmatrix}\,,\\
\begin{pmatrix}
\omega_{+}^{y\bry}\\[3pt]
\omega_{-}^{y\bry}
\end{pmatrix}&=
-2\begin{pmatrix}
\tfrac{1}{N+2}\, S^{\bry\bry}\\[3pt]
\tfrac{1}{\bar{N}+2}\,{S}^{yy}
\end{pmatrix}\,,\\
\begin{pmatrix}
\omega_{+1}^{Q_+}\\[3pt]
\omega_{-1}^{Q_-}
\end{pmatrix}&=
\tfrac{1}{N+\bar{N}+2}\begin{pmatrix}
N & \bar{N}+2\\[3pt]
N+2 & \bar{N}
\end{pmatrix}
\begin{pmatrix}
S^{y\pl}\\
{S}^{\bry\bpl}
\end{pmatrix}\,.
\end{align}
\end{subequations}
It is evident that the solution for $\omega_+$ is independent from that for $\omega_-$, and involves a different component of
the decomposition of $S$. It is now useful to define a ladder operator increasing the degree $k$ of the one forms. Thus we arrive at a simple relation,
which links together the non-vanishing components of $\omega$:
\be \omega_k=-(Q_-)^{-1}\nabla\omega_{k-1}\,.\ee
Finally, we need the decomposition of $\nabla\omega$ in the basis \eqref{2formB}. This decomposition is again straightforward to achieve using \eqref{Fierz}, and reads:
\begin{subequations}
\begin{align}
(\nabla\omega)^{\pl\pl}=&\tfrac{1}{2N}\left(y^\gb\bpl^\gbd \nabla_{\gb\gbd}\omega^{\pl\bpl}\right)\,,\\[3pt]
(\nabla\omega)^{y\pl}=&\tfrac{1}{2}\left(\tfrac{1}{N}\,y^\gb\bry^\gbd\nabla_{\gb\gbd}\omega^{y\bry}-\tfrac{1}{N+2}\,\bpl^\gbd\pl^\gb\nabla_{\gb\gbd}\omega^{\pl\bpl}\right)\,,\\[3pt]
(\nabla\omega)^{yy}=&-\tfrac{1}{2(N+2)}\left(\pl^\gb\bry^{\gbd}\nabla_{\gb\gbd}\omega^{y\bry}\right)\,,\\[3pt]
(\nabla\omega)^{\bar{\pl}\bar{\pl}}=&\tfrac{1}{2\bar{N}}\left(\bry^\gbd\pl^\gb\nabla_{\gb\gbd}\omega^{\pl\bpl}\right)\,,\\[3pt]
(\nabla\omega)^{\bry\bar{\pl}}=&\tfrac{1}{2}\left(y^\gb\bry^\gbd\nabla_{\gb\gbd}\omega^{y\bry}-\tfrac{1}{\bar{N}+2}\pl^\gb\bpl^\gbd\nabla_{\gb\gbd}\omega^{\pl\bpl}\right)\,,\\[3pt]
(\nabla\omega)^{\bry\bry}=&-\tfrac{1}{2(\bar{N}+2)}\left(y^{\gb}\bpl^\gbd\nabla_{\gb\gbd}\omega^{y\bry}\right)\,,
\end{align}
\end{subequations}
where we have restricted the attention for simplicity to the components in Eq. \eqref{doublytraceless} only. Starting from the $k=0$ component and using that in our gauge $\Phi'=0$,
one gets first of all that $\omega_k^{y\bry}=0$ for any $k$, and the following recursive formula for $\omega_k^{\pl\bpl}$:
\be\label{recomega}
\omega_k^{\pl\bpl}=-\tfrac{1}{N\brN}\,y^\gb\bpl^\gbd\nabla_{\gb\gbd}\,\omega_{k-1}^{\pl\bpl}\,.
\ee
Now we only need the link between $\omega_{s-1}$ and the Weyl tensor, which can be found in a similar way as
\be\label{recC}
C(y,0)=-\tfrac{1}{N}\,y^\gb\bpl^\gbd\nabla_{\gb\gbd}\,\omega^{\pl\bpl}_{s-1}\,,
\ee
where again we assume the $\theta$ dependence to be included into the definition of $C$. The above system of equations is also easy to integrate once $C(y,0)$ is specified.

Let us consider an ansatz for the various components of $\omega$ of the following type:
\be \omega_k^{\pl\bpl}=\gamma_k\, K (\xi^\ga y_\ga)^{s+k}(\xi^\gad\bry_{\gad})^{s-k}\,,\ee
where $K$ is again the scalar boundary-to-bulk propagator, $K=\tfrac{z}{(\text{x}-\text{x}_0)^2+z^2}$, and $\gamma_k$ are unknown coefficients. One obtains from
Eq.~\eqref{nablax} the following identity:
\beq y^\gb\bpl^{\gb}\nabla_{\gb\gb}\omega_{k-1}^{\pl\bpl}\,=\,-\gamma_{k-1}\,\left(s^2-(k-1)^2\right)\,y_\gb F^{\gb\gbd}\xi_{\gbd}\,K(\xi y)^{s+k-1}(\xi\bry)^{s-k}\,.
\label{whatisthis3}\eeq
Setting $\text{x}-\text{x}_0=0$ (i.e., $F^{\ga\gad}=-i \epsilon^{\ga\gad}$) for simplicity and substituting into \eqref{recomega} and \eqref{recC} gives the following
recurrence relations:
\beq \gamma_{k-1}=-i\gamma_{k}\,\frac{(s+k)(s-k)}{(s+k-1)(s-k+1)}\,,\qquad \gamma_{s-1}=e^{i\theta}\frac{1}{2s-1}\frac{1}{(2s-1)!}\,,\label{whatisthis4}\eeq
which can be solved as
\be
\gamma_{q}=(-i)^{s-1-q} \,e^{i\theta}\frac{1}{(2s-1)!}\,\frac{1}{(s+q)(s-q)}\,.
\ee
Combining the above coefficients we get the boundary-to-bulk propagator for $\omega^{(s)}$:
\be
\omega^{\pl\bpl(s)}(y,\bry|x)=\sum_{k=-s+1}^{s-1}\gamma_k\omega_k +\text{c.c}\,.
\ee
\bex Upon using the identity \be (-i)^{s-1-k}\frac{(s+k-1)!(s-k-1)!}{(2s-1)!}=\int_0^1dt (-it)^{s-k-1}(1-t)^{s+k-1}\,,\ee
exponentiate the boundary-to-bulk propagator for $\omega$ by introducing an auxiliary integration variable. The final result takes the form:
\be \omega^{\pl\bpl}(y,\bry|x)=i K\,e^{i\theta}\int_0^1 \frac{dt}{t(1-t)}\,e^{-it \xi^\ga y_\ga+(1-t)\xi^\gad\bry_{\gad}}+\text{c.c.}~.
\footnote{We thank S.~Didenko and Z.~Skvortsov for useful discussions and communications on the exponentiated form \eqref{genOmega} of the propagator.}\label{genOmega}
\ee\eex

\subsection{Asymptotic Symmetries of the AdS Theory}

This section is devoted to the analysis of the asymptotic symmetries of HS theories at the free level. This relies on the transformation properties of $\delta$-function
sources at the boundary. In the simplest bosonic example, we will give the details of the argument that leads to the identification of boundary conditions that preserve
or break HS symmetry. Furthermore, we will present a generalization of this analysis to the supersymmetric case where the matching of symmetries among supersymmetric HS
theories in the bulk and ABJM type boundary duals has led to interesting conjectures relating String Theory to HS theories. Although this analysis is carried out entirely
at the level of symmetries, the corresponding matching is quite striking and it would be exciting to test such dualities beyond the linear regime. It is however important
to stress that the current status of these conjectures does not go beyond the matching of symmetries for both theories.

We begin with the analysis of boundary HS symmetries, which can be easily studied in the oscillator realization by considering arbitrary powers of $y^{\pm}$ oscillators
in the manifestly conformal basis. A convenient basis for the boundary HS-conformal algebra is given by an exponential generating function of the type:
\be \epsilon(y^{\pm})~=~e^{\,\Lambda_+\,^\ga y^+_\ga\,+\,\Lambda_-^\ga\,y^-_\ga}\,,\ee
in terms of constant spinors $\Lambda_+^\ga$.
The $z$-dependence can be reinstated by considering the conformal foliation of $AdS_4$ and acting with the dilatation generator:
\be \epsilon(y^\pm)~=~e_\star^{\ln z\,D}\star e^{\Lambda_+^\ga\,y^+_\ga\,+\,\Lambda_-^\ga\,y^-_\ga}\star e_\star^{-\ln z\,D}
~=~e^{\Lambda^\ga(z)\,y_\ga+\bar{\Lambda}^\gad(z)\,\bry_\gad}\,,\ee
where for convenience we have set $\text{x}-\text{x}_0=0$, since we will be mostly interested in the $z$-dependence. We also defined
\begin{align}
\Lambda^\ga&=\tfrac{1}{2}\left(z^{-1/2}\Lambda_-^\ga-iz^{1/2}\Lambda_+^\ga\right)\,,& \bar{\Lambda}^\gad&=\tfrac{1}{2}\left(z^{1/2}\Lambda_+^\gad-iz^{-1/2}\Lambda_-^\gad\right)\,.
\end{align}
The above parameterizes the most general solution to the $AdS_4$ killing equation $\adD\epsilon=0$.
The main reason for defining above HS charges, is because we want to study the action of such charges on linearized solutions to the free equations given by propagators. This analysis will be instrumental in order to match the asymptotic symmetries of the bulk and boundary theories, at least to the linear order. The spin-s global symmetry generating parameter is obtained by restricting the above generating function to its degree $2s-2$ subsector, obtaining exactly a one to one correspondence with the module \eqref{gauge}.
Having obtained the explicit solution to the HS killing equations in $AdS_4$, we can use the corresponding charges on delta-function boundary sources and see how they transform. The HS global transformation of the Weyl module are just twisted-adjoint transformations, given by
\be \delta_\epsilon C(y,\bry|\,x)=\epsilon\star C-C\star\pi(\epsilon)\,.\ee
The $\star$-products are easy to perform, thanks to the simple exponential form of $\epsilon$:
\be \delta_\epsilon C(y,\bry|\,x)=-\epsilon(y,\bry)C(y+i\Lambda,\bry+i\bar{\Lambda})
-\epsilon(y,-\bry)C(y-i\Lambda,\bry+i\bar{\Lambda})\,.\ee
Using the explicit form of the boundary-to-bulk propagator for the Weyl module, and restricting the attention to spin-$s$
Weyl tensors by setting $\bry=0$, we arrive at
\bea \delta_\epsilon C^{(s)}(y,0|\,\text{x},z)&=&-\frac{z^{s+1}}{[(\text{x}-\text{x}_0)^2+z^2]^{2s+1}}\;e^{y^\ga\left(F_{\ga}{}^\gbd\bar{\Lambda}_\gbd-\Lambda_\ga\right)}
~~~~~~~~~~~~~~~~~~~~~~~~~~~~~~\nonumber\\&&~~~~~~~~\times\left[e^{i\left(\Lambda^\ga F_{\ga}{}^\gbd\bar{\Lambda}_{\gbd}+\theta\right)}\mathcal{X}_+
-e^{-i\left(\Lambda^\ga F_{\ga}{}^\gbd\bar{\Lambda}_{\gbd}-\theta\right)}\mathcal{X}_-\right],\label{bigequation0}\eea
where
\bea
\mathcal{X}_\pm&\equiv&\left[\left(y\pm i\Lambda\right)_\ga\left((\text{x}-\text{x}_0)^{\ga\gb}-i\epsilon^{\ga\gb} z\right)\eta_\gb\right]^{2s}~~~~~~~~~~~~~~~~~~\nonumber\\
&&~~~~~~~~~~~~~~~~~~~+e^{-2i\theta}\left[i\bar{\Lambda}_\gad\left((\text{x}-\text{x}_0)^{\gad\gbd}-i\epsilon^{\gad\gbd} z\right)\eta_\gbd\right]^{2s}.\label{bigequation1}
\eea
This quantity (\ref{bigequation0}) has a $z^{s+1}$-like asymptotic behaviour, which is expected for a HS boundary source. This also
shows how HS global symmetries preserve the boundary condition for a $\delta$-function HS source.
The situation is different for scalar-field boundary sources of the HS multiplet, for which one can choose between
two possible boundary behaviors:
\be C^{(0)}\sim a z+b z^2\,.\ee
Setting also $\bry=0$ in order to restrict the attention at the HS transformation of a scalar, we readily see the following behavior:
\bea \delta_\epsilon C^{(0)}(\text{x},z)&=&-\frac{2 i z^{s+1}}{[(\text{x}-\text{x}_0)^2+z^2]^{2s+1}}\,\sin\left(\Lambda^\ga F_{\ga}{}^\gbd\bar{\Lambda}_{\gbd}\right)\,
e^{i\theta}\,\mathcal{X}_+(y=0)\nonumber\\[5pt]&\sim& z \cos\theta (\cdots)+iz^2 \sin\theta(\cdots)+\mathcal{O}(z^3)\,.\label{scalarA}\eea
Now we see that the boundary condition $\Delta =1$ is preserved for $\theta=0$; this is the so-called \textit{A-type model}. The other boundary condition $\Delta=2$, on the other hand, is preserved for $\theta=\frac{\pi}{2}$ in the so-called \textit{B-type model}. Any other choice of $\theta$ breaks HS symmetries. This linearized symmetry analysis gives complete agreement with free-boson and free-fermion respectively.
\bex Work out the explicit form of the ellipses in Eq.~\eqref{scalarA} to show that the broken symmetries appear only for $s>2$.\eex

The lesson is that the boundary conditions for $s=0$ and $\frac12$ may be responsible for HS symmetry breaking. Even when the vacuum is endowed with full HS symmetry, a boundary source may break it through the mechanism discussed above. In fact, any flat connection of the HS algebra preserves all HS symmetries, but background zero forms may break them in general.

Let us emphasize that the global symmetry analysis performed so far turns out to be powerful enough to match symmetries between various CFTs on the boundary and HS theories in the bulk with different boundary conditions (double- or triple-trace boundary operators turned on from the boundary perspective). An example of this analysis is the Sezgin-Sundell-Klebanov-Polyakov conjecture \cite{Sezgin:2002rt,Klebanov:2002ja} between the free boson at the boundary and the bosonic minimal Vasiliev's theory in the bulk. A consequence of this analysis is also the duality between double trace deformation of free boson in 3d, and the same Vasiliev's theory with $\Delta=2$ boundary conditions for the scalar field \cite{Klebanov:2002ja}.

\subsection{ABJM Triality}

The whole analysis performed in the previous section is based on the unfolded form of linear HS equations in the bulk, associated to some HS algebra. Although until now we have restricted our attention to the simplest bosonic HS algebra, it is possible to consider Chan-Paton dressings thereof \cite{Paton:1969je} and include fermionic generators as well.
The analysis of symmetries goes exactly along the same lines, although it is more involved as one needs to keep track of more complicated reality conditions for the spinorial generators. Generalizing the above analysis, one can see that in such cases one must consider the action of asymptotic super-HS symmetries on spin-s boundary sources when in the bulk such transformations generate $s=0,1/2,1$ bulk fields. In the case of $s=1$ fields it is easy to convince ourselves that the appearance of the parity-violating phase would translate into mixed electric-magnetic boundary conditions for the spin-1 bulk field for instance.

For describing some of the supersymmetric results along the lines of the previous section, a key ingredient is the extension of the Moyal HS algebra to include spinorial or internal generators. The simplest way to achieve this extension is to introduce Chan-Paton factors into Moyal $\star$-product and consider appropriate reality condition on the corresponding associative algebras \cite{Vasiliev:2004cm,Chang:2012kt}. In the case at hand one can introduce $U(M)$ Chan-Paton factors, promoting the $s=1$ bulk gauge field to a $U(M)$ gauge field while all other bulk gauge field transform in the adjoint of $U(M)$. More concretely, such an extension can be achieved by introducing Clifford oscillators $\psi_i$, $i=1,\cdots,2n$ satisfying a Clifford algebra $\{\psi_i,\psi_j\}=2\d_{ij}$ on top of the $Y_A$ oscillators of the HS algebra of Section \ref{sec: HS Algebra}. The end result is exactly associated to $U(2^{n/2})$ Chan-Paton factors, but it is more convenient for the analysis of supersymmetry in this context.

On such HS algebras one can realize the same anti-involution used for the bosonic algebra which can be used to define appropriate reality conditions:
\begin{align}
(y_\ga)^\dagger&=\bry_\gad\,,&(\bry_\gad)^\dagger&=y_\ga\,,\\
(f\star g)^\dagger&=(-1)^{|g||f|}\,g^\dagger\star f^\dagger\,,& (\ga f+\gb g)^\dagger&=\ga^* f^\dagger+\gb^* g^\dagger\,,
\end{align}
where the notation $|\bullet|$ is related to form-degree and $\psi_i$ are Hermitian.
The reality conditions in the bosonic case then read simply:
\begin{align}
\omega^\dagger&=-\omega\,,& C^\dagger&=\pi(C)\,,
\end{align}
the first being the standard one in the context of gauge theories, the second using the automorphism $\pi$ compatibly with HS symmetries while reproducing the correct reality condition for the translation generator.

Let us briefly recall that the $\pi$ operation flips the sign of $y$, extending to the full HS algebra the operation which exchanges the sign of the translation generator and which defines the Weyl module. It can be realized within the HS algebra in terms of a pair of Kleinian operators $\kappa$ and $\bar{\kappa}$ as
\begin{align}
\kappa\star f(y,\bry)&=f(-y,\bry)\star\kappa\,,& \kappa\star\kappa&=1\,,\\
\bar\kappa\star f(y,\bry)&=f(y,-\bry)\star\bar{\kappa}\,,& \bar\kappa\star\bar\kappa&=1\,.
\end{align}
Once the HS algebra has been extended with Clifford elements, one is required to extend both the kinematical constraints (that were previously projecting the theory onto its bosonic components) and the reality conditions. The kinematical constraints are extended by considering the total-Kleinian $K=\kappa\star\bar\kappa \Gamma$, combining both $\kappa$ and $\bar\kappa$ with the corresponding Kleinian in the Clifford algebra defined as:
\be
\Gamma\equiv i^{n(n-1)/2}\,\psi_1\cdots\psi_{2n}\,.
\ee
They read
\begin{align}\label{Kinem}
[K,\omega]_\star&=0\,,& [K,C]_\star&=0\,,&[K,\epsilon]_\star&=0\,,
\end{align}
where $\epsilon$ is the generic element of the HS algebra, while $\omega$ and $C$ are the usual gauge and Weyl modules. The latter reduces to the condition that all fields are even functions of the oscillator when no Clifford element is present.

Also the extension of the reality conditions for the fields is quite simple and reads:
\begin{align}\label{Reality}
\omega^\dagger&=-\omega\,,& C^\dagger&=\Gamma\pi(C)=\bar{\pi}(C)\Gamma\,.
\end{align}
Notice that with our choices
\be
\Gamma^\dagger=\Gamma^{-1}=\Gamma\,,
\ee
while the extra factor of $\Gamma$ in Eq.~\eqref{Reality} must be introduced for compatibility with the kinematical constraint \eqref{Kinem}.

Having described the extension of the HS algebra to include Clifford elements, it is easy to study the corresponding spectra. Looking at the scalar components by setting $y=0$ and $\bry=0$, but keeping the Clifford elements and imposing the kinematic condition \eqref{Kinem}, one arrives to
\be \Gamma \Phi\Gamma=0\,,\ee
which implies that the scalar components must be even in $\psi_i$. The overall number of scalars is $2^{2n-1}$, while half of them are parity odd and
the other half are parity even upon decomposing the corresponding space according to the projectors $\frac12(1\pm\Gamma)$. Following \cite{Chang:2012kt}
one can then write down the most general boundary conditions for the parity odd and parity even scalar, by first selecting a half-dimensional subspace
of the scalars and then acting with appropriate projectors:
\be C^{(0)}=\left(e^{i\gamma}+\Gamma e^{-i\gamma}\right)f_1(\psi)\,z+\left(e^{i\gamma}-\Gamma e^{-i\gamma}\right)f_2(\psi)\,^2+\cdots\,,\ee
where $\gamma$ is an arbitrary Hermitian operator acting on the half-dimensional space in which $f_1$ and $f_2$ live. The standard boundary conditions
for which parity-odd scalars all have boundary condition $\Delta=2$ and parity-even scalars have instead $\Delta=1$ is recovered with $\gamma=0$.

The most general boundary condition for spin-$\frac12$ fermions is recovered as
\be C^{(1/2)}=z^{\frac{3}{2}}\left(e^{i\alpha}y^\ga\eta_\ga(\psi)-\Gamma e^{-i\alpha}\bry^\gad\eta_\gad(\psi)\right)+\cdots~,\ee
where $\alpha$ is some Hermitian operator on the Clifford-algebra acting on $\eta$, which is itself an element of the Clifford algebra.
For $\alpha=0$ all fermions are parity odd. Analogously one can proceed with most general spin-1 boundary conditions:
\be C^{(1)}=z^2\left(e^{i\beta}y^\ga y^\ga C_{\ga\ga}(\psi)+\Gamma e^{-i\beta}\bry^\gad \bry^\gad C_{\gad\gad}(\psi)\right)+\cdots\,.\ee
Choosing $\beta=\theta$, where $\theta$ is the parity violating phase, one recovers the standard magnetic boundary conditions for an un-gauged flavour group.
More general choices of $\beta$ as an $\psi$-operators acting on $C_{\ga\ga}(\psi)$ give other boundary conditions.

An analysis similar to the one performed in the bosonic case allows us to find among the boundary conditions described above those which match the asymptotic
symmetries of ABJ $\mathcal{N}=6$ theory. First of all, one fixes $n=3$ which constrains the number of Clifford elements to be $6$. When the theory is dual to
a free theory on the boundary, one can have up to $2^6=64$ supersymmetries.

The boundary conditions that can be shown to preserve $\mathcal{N}=6$ supersymmetries necessary to match a dual of ABJ type must involve a non-vanishing parity
violating phase $\theta\neq0$. They can be expressed as in \cite{Chang:2012kt} by using appropriate projectors on some subspaces spanned by certain Clifford algebra
elements $e$: $P_{e}$. In the following we give the explicit form of $\alpha$, $\beta$ and $\gamma$ that solve the problem:
\begin{align} \alpha&=\theta(1-2 P_{\psi_i \Gamma})\,,&\beta&=\theta(1-2 P_{\Gamma})\,,&\gamma=\theta P_{\{1,\psi_i\psi_j\}}\,.\end{align}
Moreover, we must restrict the choice of functions $f_{1,2}(\psi)$ imposing the condition:
\be P_{\Gamma,\psi_i\psi_j\Gamma}f_i=0\,,\ee
which projects out half of the components of $f_i$. Finally, $\theta$ can be compared with the Chern-Simons level by matching the spin-1 boundary condition, obtaining
in this way a non-trivial mapping of parameters following from symmetry matching.

Having described a non-trivial matching of symmetries between $\mathcal{N}=6$, $U(N)_k\times U(1)_{-k}$ ABJ model and parity violating HS theories in the
bulk \cite{Chang:2012kt}, one can then argue that the duality between the same ABJ model and type IIA string theory on $AdS_4 \times \mathcal{C}P^3$ gives a bulk-to-bulk
duality between parity violating HS theories and String Theory. The regime of Vasiliev's HS theory would correspond to the regime with a small bulk 't Hooft coupling
$\lambda_{\text{bulk}}\sim\frac1{N}$ for $N,k\rightarrow\infty$, while the stringy regime would be recovered for $\lambda_{\text{bulk}}\sim 1$.

To conclude this section, we would like to emphasize the power of the unfolded formalism in facilitating the important matching of symmetries between bulk and boundary,
by making the action of such symmetries manifest. Unfolding turns out to be a convenient slicing and choice of dynamical variables that are tuned to the symmetry behind the theory.

\begin{acknowledgement}
We are grateful to K.~Alkalaev, S.~Didenko, M.~Grigoriev, E.~Joung, C. Sleight, Z.~Skvortsov for valuable comments on the draft and for discussions. RR and MT would
like to thank the organizers of the ``International Workshop on Higher Spin Gauge Theories'' and Nanyang Technological University, Singapore for its hospitality
during the course of this work. MT is partially supported by the Fund for Scientific Research-FNRS Belgium and by the Russian Science Foundation grant 14-42-00047 in
association with Lebedev Physical Institute.
\end{acknowledgement}

\newpage
\bibliographystyle{ieeetr}
\bibliography{megabib}

\end{document}